\documentclass[aps,prx,twocolumn,superscriptaddress,nofootinbib,10pt]{revtex4-2}
\usepackage[utf8]{inputenc}
\usepackage[english]{babel}
\usepackage[T1]{fontenc}
\usepackage{amsmath}
\usepackage{amsfonts}
\usepackage{bbm}
\usepackage{amssymb}
\usepackage{bm}
\usepackage{amsthm}
\usepackage{braket}
\usepackage[dvipsnames, table]{xcolor}
\usepackage{hyperref}
\hypersetup{
  colorlinks  = true, 
  urlcolor    = Blue, 
  linkcolor   = BrickRed, 
  citecolor   = PineGreen 
}
\usepackage{graphicx}
\usepackage[caption=false]{subfig}
\usepackage{diagbox}
\usepackage{tikz}
\usetikzlibrary{tikzmark}

\usepackage{enumitem}

\usepackage{chngcntr} %

\usepackage{array}

\usepackage{wasysym}
\usepackage[only,llbracket,rrbracket,Lbag,Rbag]{stmaryrd}
\newcommand{\com}[2]{\left\llbracket{#1},{#2}\right\rrbracket}

\def\paddedtext#1#2{\leavevmode\hbox to#1{#2\hss}\ignorespaces}

\begin{document}

\title{Spacetime Spins: Statistical mechanics for error correction with stabilizer circuits}

\author{Cory T. Aitchison}
\affiliation{DAMTP, University of Cambridge, Wilberforce Road, Cambridge, CB3
	0WA, UK}
\author{Benjamin Béri}
\affiliation{DAMTP, University of Cambridge, Wilberforce Road, Cambridge, CB3
	0WA, UK}
\affiliation{T.C.M. Group, Cavendish Laboratory, University of Cambridge, J.J. Thomson Avenue, Cambridge, CB3 0HE, UK\looseness=-1}%

\begin{abstract}
	A powerful method for analyzing quantum error-correcting codes is to map them onto classical statistical mechanics models. Such mappings have thus far mostly focused on static codes, possibly subject to repeated syndrome measurements. Recent progress in quantum error correction, however, has prompted new paradigms where codes emerge from stabilizer circuits in spacetime---a unifying perspective encompassing syndrome extraction circuits of static codes, dynamically generated codes, and logical operations. We show how to construct statistical mechanical models for stabilizer circuits subject to independent Pauli errors, by mapping logical equivalence class probabilities of errors to partition functions using the spacetime subsystem code formalism. We also introduce a modular language of spin diagrams for constructing the spin Hamiltonians, which we describe in detail focusing on independent circuit-level $X$-$Z$ error channels. With the repetition and toric codes as examples, we use our approach to analytically rank logical error rates and thresholds between code implementations with standard and dynamic syndrome extraction circuits, describe the effect of transversal logical Clifford gates on logical error rates, and perform Monte Carlo simulations to estimate maximum likelihood thresholds. Our framework offers a universal prescription to analyze, simulate, and compare the decoding properties of any stabilizer circuit, while revealing the innate connections between dynamical quantum systems and noise-resilient phases of matter. 
\end{abstract}

\maketitle

\section{Introduction}

Quantum error-correction (QEC) enables quantum computers to operate
reliably despite the presence of noise \cite{kitaevQuantum1997,shorFaulttolerant1997,bravyiQuantum1998,knillResilient1998,dennisTopological2002,
	kitaevFaulttolerant2003, aharonovFaultTolerant2008, gottesmanStabilizer1997,
	steaneError1996, preskillFaultTolerant1998, 
	poulinStabilizer2005}.
The simplest QEC setting is that of a quantum memory, robustly storing logical
information via repeated rounds of syndrome extraction. Quantum computation then
evolves this with fault-tolerant logical operations that limit the accumulation  
of errors.

Quantum memories exemplify noise-resilient phases of
matter~\cite{dennisTopological2002,bombinTopological2010,bombinStrong2012,
 kubicaThreeDimensional2018,chubbStatistical2021,baoMixedstate2023,fanDiagnostics2024,leeMixedState2025,sangMixedState2024,vennCoherentError2023,behrendsStatistical2025,behrendsSurface2025,baoPhases2024,
sangStability2025, wichettePartition2025,
englishIsing2025, lavasaniStability2025,chengEmergent2025,jingIntrinsic2025,
huangRobust2025,liuCoherent2025,xuPhenomenological2025,
wangDecoherenceinduced2025, chenNishimori2025,hauserInformation2025,
putzFlow2025,zhuNishimoris2023,ecksteinLearning2025}:
The probability of logical errors, given a code and noise model, can be directly
expressed through partition functions of statistical mechanical (SM) Ising models
\cite{dennisTopological2002}. The error threshold---a critical physical error
rate, below which increasing the number of physical qubits leads to a lower
logical error rate
\cite{kitaevQuantum1997,shorFaulttolerant1997,bravyiQuantum1998,knillResilient1998,dennisTopological2002,
	kitaevFaulttolerant2003, aharonovFaultTolerant2008}---then emerges as a phase transition between ordered (error correcting) and
disordered (non-correcting) phases. Identifying these phase boundaries allows
one to assess the theoretical, decoder-agnostic, memory limits.

An important emerging frontier concerns how concepts of phases of matter can
capture inherently dynamical aspects of fault tolerance, e.g.,
syndrome-extraction circuits, logical operations for quantum computations, or
indeed intrinsically dynamic
codes~\cite{hastingsDynamically2021,haahBoundaries2022,tangPhases2025,claesDynamic2025,aasenMeasurement2023,ellisonFloquet2023,davydovaQuantum2024,vuStable2024,aitchisonCompeting2025,vodolaFundamental2022,xuError2025,shawLowering2025,sommersDynamically2025,negariSpacetime2024,
ecksteinRobust2024, zhuQubit2023,
cainFast2025,martonLattice2025,chirameStabilizing2025,chirameStable2025,takouEstimating2025,
zhuStructured2024,
gottesmanOpportunities2022,
derksDynamical2025, sommersDynamically2025, davydovaUniversal2025,
setiawanTailoring2025, derksDesigning2024, yadavalliNoisy2025, bombinLogical2023, williamsonDynamical2025,xuFaulttolerant2025,
tanggaraStrategic2024,tanggaraSimple2024, fuError2025, 
kishonyIncreasing2025}.
In this paper, we address this question by showing how to construct SM models for 
stabilizer circuits---with elementary Clifford gates, resets, and measurements,
furnishing the most general fabric behind fault-tolerant constructions.  

Leveraging stabilizer dynamics is of increasing focus in practical
QEC~\cite{bombinUnifying2024, derksDynamical2025,
davydovaUniversal2025,setiawanTailoring2025,
geherErrorcorrected2024,mcewenRelaxing2023,
eickbuschDemonstration2025,debroyLUCI2024, strikisQuantum2023,
ankerOptimized2025,
wolanskiAutomated2025,martielLowoverhead2025,gidneyLess2023,sahayError2025,
serra-peraltaDecoding2025, 
beverlandFault2024,delfosseSpacetime2023, derksDesigning2024, yadavalliNoisy2025, bombinLogical2023};
examples include designing circuits to overcome hardware limitations
\cite{mcewenRelaxing2023, eickbuschDemonstration2025}, avoid qubit or coupler
dropouts \cite{debroyLUCI2024, strikisQuantum2023, ankerOptimized2025,
wolanskiAutomated2025}, reduce overheads \cite{martielLowoverhead2025}, or turn
a static threshold-less code into a dynamical code with thresholds
\cite{gidneyLess2023}.
Despite these advancements, except for approaches specific to certain codes or
circuits~\cite{vodolaFundamental2022,tangPhases2025,xuError2025,sommersDynamically2025},
or cases where time describes repeated syndrome measurement in phenomenological
models~\cite{dennisTopological2002, chubbStatistical2021}, SM mappings have
largely focused on static codes. Our SM models for stabilizer circuits fill this
important gap.

In the dynamic setting, QEC and quantum computations correspond to $(d+1)$-dimensional [$(d+1)$D] circuits. Our SM models first represent these as
related stabilizer subsystem codes in $(d+1)$-spatial dimensions, known as 
spacetime codes \cite{baconSparse2017, gottesmanOpportunities2022,
delfosseSpacetime2023,
derksDesigning2024,pesahFaulttolerant2025,fuSubsystem2025}.
Considering generic uncorrelated Pauli noise channels, we then construct a
classical Hamiltonian whose partition function---by summing over related error
configurations--- accounts for fundamental error thresholds and the lowest
achievable logical error rates; our SM models are thus of the same status for
the dynamical settings as traditional SM models have been for quantum memories.

Mapping stabilizer circuits onto SM models opens up new avenues for analysis
beyond those seen in the static case. This framework allows us to compare
different compilations of the syndrome-measurement circuits of a QEC code, for
example. It also enables statistical mechanical analysis of the logical
circuits, such as through transversal gates, lattice surgery, or code switching.
A circuit-based analysis of decodability introduces spatially-correlated noise
processes---such as ``hook errors'' from \texttt{CNOT}s between ancilla and data
qubits \cite{beverlandFault2024, sahayError2025}---absent from abstracted
phenomenological representations of QEC syndrome measurements. These processes
are naturally captured by our SM models.
Moving SM mappings beyond static codes thus provides useful practical features,
beyond its key role in a holistic understanding of fault tolerance and how this defines noise-resilient dynamic phases of matter. 

A central technical ingredient in our approach is the introduction of modular
spin diagrams, a graphical language of elementary building blocks for
constructing  SM models for stabilizer circuits. While usable with general noise
models, this approach is particularly simple for independent circuit-level
$X$-$Z$ noise (where independent $X$ and $Z$ errors can occur before or after
any stabilizer operation in the circuit), for which the SM Hamiltonians can be
systematically simplified by integrating out certain spins. The spin diagrams we
introduce have similarities with those used in ZX-calculus
\cite{bridgemanHandwaving2017, weteringZXcalculus2020} with Pauli flows
\cite{xuFaulttolerant2025, bombinUnifying2024}; a core difference is that our
spin diagrams encode the noise channel, ensuring that simplifying the diagrams
preserves the SM partition functions. (See also
Ref.~\cite{ruschCompleteness2025} for a related construction that uses
ZX-diagrams to manipulate circuits while preserving the effects of faults, and
Ref.~\cite{pesahFaulttolerant2025} for an approach with similar simplification
processes but in the context of chain complexes.)

The rest of the paper is structured as follows: Section~\ref{sec:background}
presents a brief background on QEC and spacetime codes.
Section~\ref{sec:spacetimespins} outlines the spin models and diagrams for
stabilizer circuits. We then apply this to several demonstrative examples.
First, in Sec.~\ref{sec:examples}, we study the 1D repetition code, including
different syndrome extraction circuits and logical CNOT operations. These
examples will help convey the workings of our approach in the simplistic
settings of 2D random bond Ising models. In Sec.~\ref{sec:toric_code} we then
study the toric code and show how our approach can describe and make qualitative
predictions about the performance of recently introduced dynamic syndrome
extraction schemes~\cite{mcewenRelaxing2023,eickbuschDemonstration2025,claesDynamic2025,williamsonDynamical2025}
compared with standard syndrome extraction circuits. We conclude and provide
future directions in Sec.~\ref{sec:conclusion}. The appendices contain
derivations and details of the numerical methods.

\section{Preliminaries}
\label{sec:background}

In this section, we present a brief introduction to the theory of quantum
error-correcting codes, and how decoding of circuits can be mapped onto
spacetime codes.

\subsection{Error-correcting codes}

Error-correcting codes are designed to protect encoded logical quantum
information. In this paper we use subsystem codes \cite{bombinTopological2010},
which are generalizations of stabilizer codes~\cite{kitaevQuantum1997,
	kitaevFaulttolerant2003, bravyiQuantum1998, gottesmanStabilizer1997,
	steaneError1996, preskillFaultTolerant1998, shorFaulttolerant1997,
poulinStabilizer2005} and arise naturally in the spacetime code formalism
\cite{baconSparse2017, gottesmanOpportunities2022, delfosseSpacetime2023,
	fuSubsystem2025, derksDesigning2024,pesahFaulttolerant2025}. Information is
	encoded in a codespace $\mathfrak C$: the
simultaneous $+1$-eigenspace of a group $\mathcal S$ of commuting Pauli operators,
``stabilizers'', 
\begin{equation}
	\mathcal S = \braket{S_1,S_2,\ldots,S_k,\ldots,S_\kappa},\quad -I \notin
	\mathcal S.
\end{equation}
$\mathcal S$ is the center of a larger gauge group, $\mathcal G$, which may
contain non-commuting elements. The members of $\mathcal G$ are called gauge
operators.
Logical operators commute with all stabilizers,
but are not themselves in $\mathcal G$. For Pauli operators $A,B$, define the
scalar commutator $\com{A}{B} \in \{-1,1\}$ by the relation
\begin{equation}
	AB = \com{A}{B}BA.
\end{equation}
If $\mathfrak C$ is $2^l$-dimensional, then there
exist $l$ pairs of logical $\bar X_i, \bar Z_i$ operators that act as Pauli $X$
and $Z$ on $\mathfrak C$, with anticommutation relation 
\begin{equation}
	\com{\bar X_i}{\bar Z_j} = (-1)^{\delta_{ij}}.
	\label{eq:anticommutation}
\end{equation}
Each $\bar X_i, \bar Z_i$ pair defines a logical qubit that encodes logical
information. If a logical operator commutes with all the gauge
operators, it is known as a bare logical operator; otherwise, it is a dressed
logical operator. The anticommutation relation, Eq.~\eqref{eq:anticommutation},
must hold for bare logicals, but may not hold for dressed logicals.

These codes protect the logical state by signalling when (and information on where) an error
occurs. Specifically, for $N$-qubit density matrix $\rho$, we consider single-qubit Pauli error channels, 
\begin{equation}
		\mathcal E =\bigotimes_{i=1}^N\mathcal E_i,\quad 
		\mathcal E_i[\rho] = \sum_{\alpha \in \mathcal P_i} \mathbb
	P(\alpha) \, \alpha\rho\alpha^\dagger,
	\label{eq:error_channel}
\end{equation}
where $\mathcal P_i = \{I,X,Y,Z\}$ are the single-qubit Pauli operators (modulo
phase) for qubit $i$. $\mathbb P(\alpha)$ is the probability of an
$\alpha$-error on a qubit. We shall use the same error channel during each timestep of a circuit, for simplicity.\footnote{\label{fn:noise}Our results are readily generalizable to
	location or time-dependent error models, but for
notational simplicity we restrict attention to translation-invariant channels in
this work.} In the absence of errors, a stabilizer measurement returns $+1$.
Errors anticommute with some $S_k$,  and thus produce a syndrome,
$\textsf s \in \{-1, 1\}^{\kappa}$ for the stabilizer measurement outcomes. Given a
syndrome, a decoder is an algorithm that outputs a correction Pauli operator
$C_\textsf{s}$. An error $E$ with syndrome $\textsf s$ is successfully corrected
if $EC_\textsf{s} \in \mathcal G$, i.e., $E\in C_\textsf{s} \mathcal G$;
otherwise, $EC_\textsf{s} \in L \mathcal G$ (i.e., $E\in C_\textsf{s} L \mathcal
G$) where $L \in \{\bar X, \bar Z,\ldots\} $ is a nontrivial logical operator,
and a logical error occurs. These $L\mathcal G$ are logical equivalence
classes or error cosets.

The optimal decoder---known as the maximum-likelihood (ML) decoder---outputs the
$C_\textsf{s}$ whose error coset has the highest probability. The
success probability of a decoder is
\begin{equation}
	\mathbb P(\text{success}) = \sum_{\textsf s} \mathbb P(\textsf s)\mathbb
	P(\text{success} | \textsf s),
\end{equation}
where $\mathbb P(\textsf s)$ is the probability of observing a syndrome $\textsf
s$. The conditional probability of success is
\begin{equation}
	\mathbb P(\text{success}|\textsf s) = \frac{\mathbb P(\overline{C_\textsf
	s})}{\sum_{L \in \{I, \bar X, \bar Z,\ldots\}} \mathbb P(\overline{C_\textsf s
	L})},
	\label{eq:p_success}
\end{equation}
where the probability of an error coset $\overline E=E\mathcal G$ is
\begin{equation}
	\mathbb P(\overline E) = \sum_{\epsilon \in E\mathcal G} \mathbb P(\epsilon),
\end{equation}
and $L$ are representative logical errors.
The ML decoder chooses a $C_\textsf{s}$ such that its coset probability is
greater than all other cosets with the same syndrome,
\begin{equation}
	\mathbb P(\overline{C_\textsf s}) > \max \left\{\mathbb
		P(\overline{C_\textsf s \bar X}), \, \mathbb P(\overline{C_\textsf s \bar
	Z}), \, \ldots\right\}.
	\label{eq:ML-decoding}
\end{equation}
This maximizes $\mathbb P(\text{success}|\textsf s)$. Below the error threshold,
the success probability of an ML decoder approaches $1$ as the code size (i.e.,
the code distance) increases.

\subsection{Spacetime codes}
The codes described above protect information in a static quantum memory.
Quantum computation, however, requires a quantum state to evolve via the
unitaries and measurements of a quantum circuit. The spacetime code formalism
allows us to interpret this dynamical state as a static subsystem code---the
spacetime code---in one higher spatial dimension
\cite{baconSparse2017, gottesmanOpportunities2022, delfosseSpacetime2023,
fuSubsystem2025,derksDesigning2024,pesahFaulttolerant2025}.\footnote{Although the main results of
		\textcite{delfosseSpacetime2023} are for spacetime stabilizer codes, their
	construction also applies to spacetime subsystem codes, as we shall use here.}
A decoder of the spacetime code can be used to construct an
equivalent decoder for the corresponding circuit. We present here a
brief overview of spacetime codes; for more detail we refer the reader to
\textcite{delfosseSpacetime2023}.

We assume a Clifford circuit of $N$ physical qubits: at every timestep
$t=0,1,\ldots,T$ each qubit is either measured with a Pauli operator or is
acted on by a Clifford unitary (which may be the identity). This enforces that
all measurements and unitaries at the same timestep have disjoint support and
thus commute with each other.
In-between each timestep, an error channel, cf.
Eq.~\eqref{eq:error_channel}, acts on the physical qubits; we say that errors
(sometimes called faults in the context of noisy circuits
\cite{beverlandFault2024}) occur at half-integer times $\tau =
0.5,1.5,\ldots,T-0.5$. Although this error model does not explicitly include
measurement errors, they occur implicitly whenever an error at time $\tau$
anticommutes with a measurement at $\tau + 0.5$; this captures the essential
features of circuit-level noise without a separate measurement noise parameter.
Independent measurement errors may be incorporated by having an additional
spacetime qubit attached to each measurement and subjected to bit-flip errors
with probability $p_\text{meas}$.\footnote{This approach was used in
\textcite{gidneyBenchmarking2022}, for example.}

To construct the spacetime code, we model a Hilbert space of $NT$
so-called spacetime qubits, each labeled by a coordinate $(i,\tau)$ where
$i=1,\ldots,N$ corresponds to the location of physical qubit and $\tau$ is the
half-integer time. If an error in the circuit occurs on qubit $i=3$ between
times $t=2$ and $t=3$, for example, it occurs on the spacetime qubit at
$(i,\tau) = (3,2.5)$. 

Let $\mathcal P_{\text{ph}}$ and $\mathcal C_\text{ph}$ be the group of Pauli
and Clifford operators acting on the $N$ physical qubits of the circuit, and
$\mathcal P_{\text{st}}$ and $\mathcal C_\text{st}$ those acting on the $NT$
spacetime qubits. For $U\in \mathcal C_{\text{ph}}$, the notation $[U]_{\mathcal
I,\tau} \in \mathcal C_{\text{st}}$ denotes the spacetime operator that acts on
qubits $i \in \mathcal I$ at half-integer time $\tau$, and the identity
elsewhere. For example, $[XZ]_{\{i,j\},\tau}$ is a Pauli $X$ on qubit $i$ and a
$Z$ on qubit $j$ at half-integer time $\tau$. When the support is clear from the
operator itself, we write simply $[U]_{,\tau}$.

The spacetime code is a subsystem code; the gauge group defines equivalences
between errors in spacetime that have the same effects on the circuit (syndromes
and logical operator outcomes). The gauge group is typically generated by two
types of spacetime operators \cite{delfosseSpacetime2023}:
\begin{enumerate}
	\item $[M_i]_{, t_i-0.5}$ and $[M_i]_{,t_i+0.5}$, for each measurement $M_i
		\in \mathcal P_\text{ph}$ at integer time $t_i$. These encode that
		the qubits will be in an eigenstate of $M_i$ at time $t_i$.
	\item $[Q]_{,t_i-0.5}[\mathcal U_{t_i} Q \mathcal
		U_{t_i}^\dagger]_{,t_i+0.5}$, for each operator $Q \in \mathcal P_\text{ph}$
		that commutes with all $M_i$ at integer time $t_i$. Here, $\mathcal U_{t_i}
		\in \mathcal C_{\text{ph}}$ is the product of all unitaries that act at time
		$t_i$. These generators encode how errors propagate through the circuit.
\end{enumerate}
If qubit $i$ is not measured at time $t$, any single-qubit Pauli on qubit $i$
commutes with the measurements at $t$. The second type of generator is then just
\begin{equation}
	[X]_{i,t-0.5}[\mathcal U_t X\mathcal U_t^\dagger]_{i,t+0.5},\quad
	[Z]_{i,t-0.5}[\mathcal U_t Z \mathcal U_t^\dagger]_{i,t+0.5}.
	\label{eq:gauge_propagators}
\end{equation}

We explain in more details the spacetime stabilizers and logical operators in
Appendix~\ref{app:spacetimecodes}; their precise form is not needed for the
construction of the spin models. In brief, the stabilizers correspond to
detector cells: products of measurement outcomes that are deterministic in the
absence of errors, and thus reveal where errors occur. The bare logical operators
are concatenations of an initial logical operator's images under the
	circuit at each timestep;
for a $1$D string logical operator on a topological code, for
example, a fault-tolerant Clifford circuit extends these to $2$D membranes in
spacetime. These bare logical operators obey the anticommutation relation,
	Eq.~\eqref{eq:anticommutation}, whereas dressed logicals (such as the image of
a code's $\bar X$ at a particular timestep) may not. (Operators at different
timesteps, for example, always commute).

\section{Spacetime spin models}
\label{sec:spacetimespins}

A powerful property of quantum codes is that the probabilities of error cosets
can be related to statistical mechanics; models that delineate the limits up to
which QEC can work and encapsulate fundamental notions of noise-resilient phases
of matter. Initially introduced by \textcite{dennisTopological2002} for
uncorrelated Pauli errors in the toric code, such SM mappings have since
encompassed, for example, more stabilizer codes \cite{bombinStrong2012,
chubbStatistical2021, kubicaThreeDimensional2018}, subsystem codes
\cite{bombinTopological2010}, coherent errors
\cite{vennCoherentError2023,behrendsStatistical2025,behrendsSurface2025,leeMixedState2025,
baoPhases2024}, and
post-selection~\cite{englishThresholds2024,englishIsing2025}. We now describe
this mapping for general Clifford circuits---extending beyond static
codes---subject to noise channels of the form in Eq.~\eqref{eq:error_channel}.
In Sec.~\ref{sec:spin-diagrams}, we show how to represent these SM models as
spin diagrams that can be constructed modularly for each circuit element. These
diagrams are particularly simple when restricted to independent $X$-$Z$ noise
channels. Section~\ref{sec:gauge-equivalent} formalizes the process of
simplifying diagrams by removing low-connectivity spins. For a more detailed
derivation of this formulation, see Appendix~\ref{app:SM_mapping}.

We wish to express the probability of an error coset, $\mathbb P(\overline E)$,
as a partition function, $\mathcal Z_E$, where $E\in\mathcal P_\text{st}$.
To enumerate all errors in the coset $E\mathcal G$, we multiply the
representative error $E$ by all possible gauge group elements $g \in \mathcal
G$. We relate this to the SM model by assigning an Ising spin $\sigma_k \in
\{-1,1\}$ to each gauge generator $g_k$; we include $g_k$ in $g$ if $\sigma_k =
-1$. The spin configuration $\bm\sigma=[\sigma_1,\sigma_2,\ldots,\sigma_m]$ thus
specifies a group element 
\begin{equation}
	g(\bm\sigma) = \prod_{k=1}^{m} g_k^{(1-\sigma_k)/2} \in \mathcal G.
\end{equation}
Interaction signs 
\begin{equation}
	\eta_{i,\tau}(\alpha) \equiv \com{[\alpha]_{i,\tau}}{E}
\end{equation}
encode $E$ and create quenched disorder. An important
property is that for any error $\epsilon = Eg(\bm\sigma)$ in the coset, the
product $\eta_{i,\tau}(\alpha) \prod_k \sigma_k$, over all $k$ where
$\com{[\alpha]_{i,\tau}}{g_k} = -1$, equals $+1$ if
$\epsilon$ commutes with $[\alpha]_{(i,\tau)}$, and $-1$ if it
anticommutes. We use this product to assign $\epsilon$ the correct probability
via dimensionless interaction strengths $K_{i,\tau}(\alpha)$ that weight
each product such that $e^{-H_E}$ equals $\mathbb P(\epsilon)$ (with $H_E$ a
dimensionless Hamiltonian).\footnote{For simplicity,
we work with dimensionless quantities. One can equivalently define
$J_{i,\tau}(\alpha) = K_{i,\tau}(\alpha)/k_\text{B}T$, where $T$ is a
temperature to instead use a Hamiltonian in units of energy.}
Their value is specified by the noise channel, cf.~Eq.~\eqref{eq:error_channel},
via a Nishimori condition (see Appendix~\ref{app:SM_mapping}),
\begin{equation}
	K_{i,\tau}(\alpha) = \frac14 \sum_{Q \in \mathcal P}
	\ln[\mathbb P(Q)] \com{[\alpha]_{i,\tau}}{[Q]_{i,\tau}},
\end{equation}
where $\mathcal P = \{I, X, Y, Z\}$.\footnote{When some $\mathbb P(\epsilon) =
	0$, the expression for $K_{i,\tau}(\alpha)$ is not defined; however, by taking
	appropriate limits and using the convention $\exp(\ln 0) = 0$, the formalism
	remains well-defined \cite{chubbStatistical2021}.} Combining these components
	together, we have
\begin{equation}
	\mathbb P(\overline E) = \sum_{\bm\sigma} \mathbb
	P\left[Eg(\bm\sigma)\right] \equiv \sum_{\bm\sigma} e^{-H_E} =
	\mathcal Z_E
	\label{eq:partition_function_ES}
\end{equation}
with Hamiltonian
\begin{equation}
	H_{E} = -\sum_{i=1}^{N}\sum_{\tau=0.5}^{T-0.5}\sum_{\alpha \in \mathcal
	P} K_{i,\tau}(\alpha) \eta_{i,\tau}(\alpha)
	\prod_{k:\com{[\alpha]_{i,\tau}}{g_k}=-1}\sigma_k.
	\label{eq:H_ES_gauge}
\end{equation}

To illustrate the interactions in $H_E$, consider two examples: 
\textbf{Idling:} If qubit $i$ idles at time $t$, then by
Eq.~\eqref{eq:gauge_propagators} there are two relevant gauge generators $g_1 =
[X]_{i,t-0.5}[X]_{i,t+0.5}$ and $g_2=[Z]_{i,t-0.5}[Z]_{i,t+0.5}$. The spin
$\sigma_1$ associated with $g_1$ interacts via
$K_{i,t\pm0.5}(\alpha)\eta_{i,t\pm0.5}(\alpha)$ for $\alpha=Y, Z$, and similarly
for $\sigma_2$ for $\alpha=X,Y$. 
\textbf{Measurements:} If qubit $i$ is instead measured in the $Z$ basis at time
$t$, then we have gauge generators $g_3 = [Z]_{i,t-0.5}$ and $g_4 =
[Z]_{i,t+0.5}$. It is possible to construct a basis such that these are the only
two $Z$-flavored generators acting on qubits $(i, t\pm0.5)$ and the
interactions via $K_{i,t\pm0.5}(X)\eta_{i,t\pm0.5}(X)$ are field terms.

\subsection{Decoding}
\label{sec:decoding}
Since $\mathcal Z_E = \mathbb P(\overline E)$ for each logical equivalence class,
maximum-likelihood decoding, Eq.~\eqref{eq:ML-decoding}, reduces to finding
the coset with maximum partition function---equivalently, minimum free energy
$F_E = -\ln \mathcal Z_E$---consistent with the syndrome
\cite{bravyiEfficient2014}. The success
probability of a decoder, Eq.~\eqref{eq:p_success}, can be expressed as a
(softmax) function of these free energies,\footnote{There is not a unique basis
	for $\mathcal G$ and so the Hamiltonian is basis-dependent; the free energy
	and probability ratios, however, are independent of the basis.}
\begin{equation}
	\mathbb P(\text{success}|\textsf s) = \frac{e^{- F_{C_\textsf
	s}}}{\sum_{L} e^{- F_{C_\textsf s L}}}.
	\label{eq:free-energy}
\end{equation}
Below threshold, the free energy diverges in the thermodynamic limit for
all $C_\textsf{s}L$ ($L\neq I$, and $C_\textsf{s}$ the ML Pauli correction) as
the ML success probability goes to $1$~\cite{chubbStatistical2021}. Ordered
phases of matter are typically characterized by long-range correlations and
extensive free-energy costs associated with introducing macroscopic domain
walls; the error threshold generally coincides with ordered-disordered phase
transitions in the SM model~\cite{dennisTopological2002,chubbStatistical2021}.

The logical classes that are summed over in Eq.~\eqref{eq:free-energy} depend on
the spacetime code and its underlying circuit. A static code that supports $l$
logical qubits has $2^{2l}$ logical equivalence
classes for each syndrome ($C_{\textsf s}\mathcal G, \bar X_1C_{\textsf
s}\mathcal G, \bar Z_1C_{\textsf s}\mathcal G,\ldots$) with coset
representatives given by combinations of the logical operators of the code. When
a code is compiled into a syndrome-extraction circuit and mapped onto a
spacetime code, additional logical equivalence classes may arise. 

For example, in a stability experiment \cite{gidneyStability2022}, one
constructs a syndrome-extraction circuit and tests the ability for a decoder to
predict the parity of a product of stabilizers after being subjected to errors.
A decoder will certainly fail if an error occurs that flips this parity but does
not alter the syndrome; one such error is a timelike string of errors that
extends across the whole duration of the experiment (cf.
Fig.~\ref{fig:rep_stability}). These undetected timelike errors are coset
representatives of some of the spacetime code's logical equivalence classes. In
stability experiments, one typically uses spatial boundary conditions such that
no logical qubit is encoded: the code's $\bar X$ and $\bar Z$ anticommute with
stabilizers on the boundary and do not form logical equivalence classes. The
decoder now chooses between only the trivial logical equivalence class
$\mathcal G$ and those with undetected timelike errors. In contrast, in a memory
experiment, one tests the ability for a decoder to preserve the state of an
encoded logical qubit by asking it to predict the final eigenvalue of a
logical operator after the circuit and comparing it with the true value.
Undetected timelike errors add logical equivalence classes and increase the
complexity of decoding; one therefore typically uses temporal boundary
conditions with spacetime stabilizers that anticommute with these spanning
timelike errors to focus on the effect of the static code's logical equivalence
classes. Memory and stability experiments are discussed further in
Sec.~\ref{sec:memory-stability}, using the repetition code as a guiding
example.

\begin{figure*}
	\begin{center}
		\includegraphics[width=\linewidth]{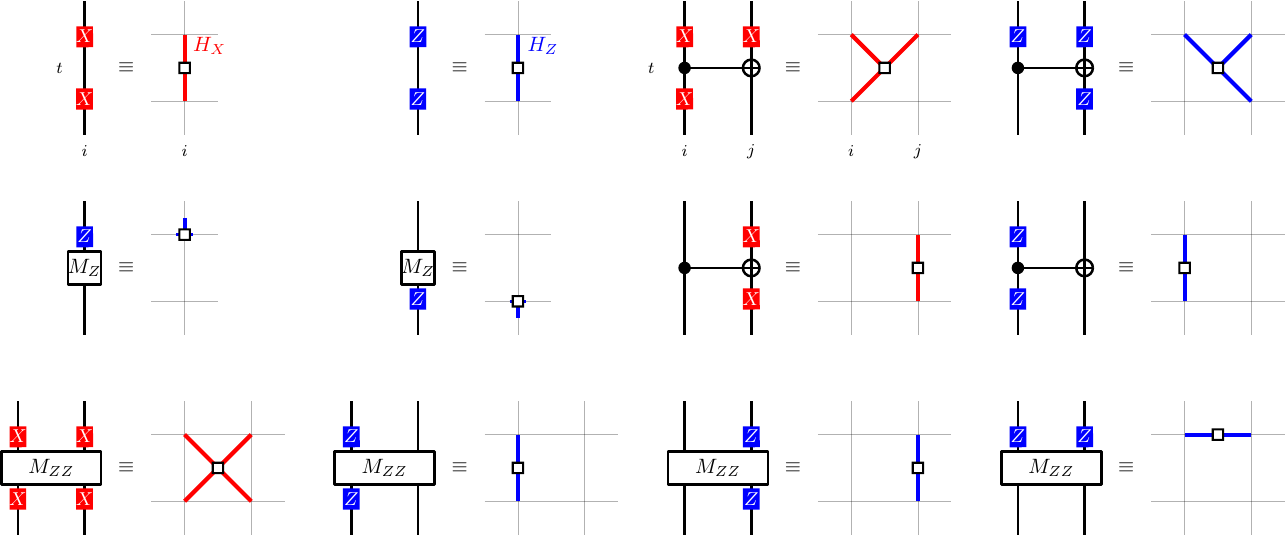}
	\end{center}
	\caption{The gauge generators (red and blue labels) from Clifford gates and
		their corresponding spin model ingredients.
		Each generator is associated with a spin (white square); interactions
		(colored lines) connect spins according to Eq.~\eqref{eq:H_ES_gauge_ind}.
		Red and blue lines are interactions on $H_X$ and $H_Z$ respectively.
		Gridlines show the spacetime locations $(i,\tau)$, with time going up the
		page. A qubit $i$ idling at integer time $t$ introduces two gauge
		generators: $[X]_{i,t-0.5}[X]_{i,t+0.5}$ and $[Z]_{i,t-0.5}[Z]_{i,t+0.5}$. A
		\texttt{CNOT} has four gauge generators; $g_1 =
		[X]_{i,t-0.5}[XX]_{\{i,j\},t+0.5}$ acts on three spacetime qubits and hence
		its spin has three legs (interactions). A single-qubit measurement $M_Z$ (a
		single-qubit reset, $R_Z$, produces the same spins and interactions) has
		gauge generators with only one interaction in $H_Z$ so we draw its spins
		atop the corresponding spacetime qubit. $M_Z$ introduces no spins to $H_X$
		(and similarly for $M_X$ and $H_Z$). Two-qubit measurements $M_{ZZ}$
		introduce a spin with four interactions and three spins with
		two-interactions. Shown here is one basis choice; $[ZZ]_{\{i,j\},t-0.5}$
		could also have been used instead of $[ZZ]_{\{i,j\},t+0.5}$.}
\label{fig:spins_intro}
\end{figure*}

\begin{figure}
	\begin{center}
		\includegraphics[width=0.95\linewidth]{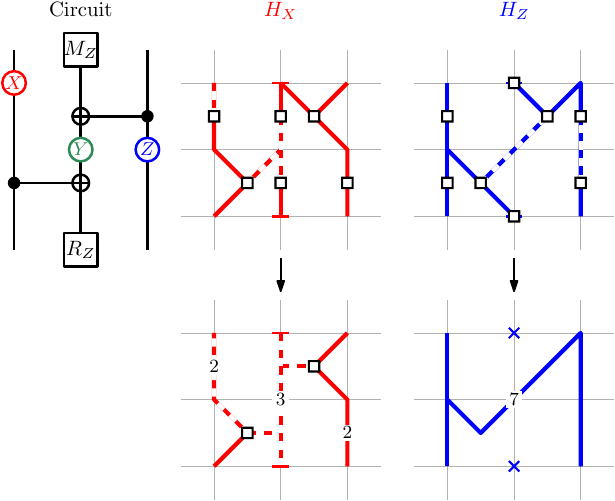}
	\end{center}
	\caption{(Top) An example
		circuit for a two-qubit $ZZ$ measurement via an ancilla qubit, with
		associated spin diagrams on the right, using the
		conventions in Fig.~\ref{fig:spins_intro}. Endcaps mark the corresponding
		termination of the interaction lines (open lines may join up with
		interactions at earlier/later times.) Errors occur in-between circuit
		operations; an example error realization is shown. These flip the sign of
		bonds on the spin model (dashed lines). $Y$ errors flip signs on both $H_X$
		and $H_Z$. (Bottom) The simplified spin diagram derived by applying the two
		rules in Fig.~\ref{fig:spins_rules}.}
	\label{fig:spins_examples}
\end{figure}

\begin{figure}
	\begin{center}
		\includegraphics[width=0.9\linewidth]{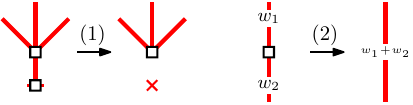}
	\end{center}
	\caption{Two rules used to simplify spin diagrams (and their associated
		Hamiltonians). (1) Spins with one interaction are integrated out
and the interaction removed from $H_\alpha$ (marked by crosses). (2) Spins with
two interactions are integrated out and their interactions grouped together. The
combined interaction is weighted by the number of spacetime locations grouped
together (unlabeled lines are weight one).}
	\label{fig:spins_rules}
\end{figure}

\subsection{Independent X-Z noise}
The above prescription holds for any noise channel. Under an independent $X$-$Z$
noise channel,
\begin{equation}
	\begin{gathered}
		\mathbb P(I) = (1-p_X)(1-p_Z),\, \mathbb P(X) = p_X(1-p_Z), \\\mathbb
		P(Y)=p_Xp_Z,\,
		\mathbb P(Z) = (1-p_X)p_Z
	\end{gathered}
	\label{eq:independentnoise}
\end{equation}
with $p_X, p_Z \in [0,1]$,\footnote{If a $p_\alpha=0$, care needs to be taken to
ensure interactions involving $K_\alpha\rightarrow\infty$ are satisfied. In CSS
codes, this usually involves removing all $\alpha$-flavored generators, and
ignoring $H_{\alpha, E}$.} the Nishimori conditions become 
\begin{align}
		K(I) & = \frac12\ln\left(p_Xp_Z(1-p_X)(1-p_Z)\right),         \\
		K(X) & = \frac12\ln\left(\frac{1-p_Z}{p_Z}\right) \equiv K_Z, \\
		K(Y) & = 0,                                                   \\
		K(Z) & = \frac12\ln\left(\frac{1-p_X}{p_X}\right) \equiv K_X.
\end{align}
We use the swapped subscripts on $K_X$ and $K_Z$ to increase interpretability:
$K_\alpha$ interaction strengths account for the probability of $\alpha$-errors
for $\alpha=X,Z$. $Y$ errors manifest as $X$ and $Z$ errors on the same qubit
and are not accounted for separately in the Hamiltonian.
Interaction signs are sampled from independent Bernoulli distributions, where
\begin{equation}
	\eta_{\alpha,(i, \tau)} \sim 1-2\text{Bernoulli}(p_\alpha) \equiv \begin{cases}
		\mathbb P(\eta=1) = p_\alpha \\ 
		\mathbb P(\eta=-1) = 1-p_\alpha\end{cases}
	\label{eq:eta_bernoulli}
\end{equation}
for every qubit $(i,\tau)$ and $\alpha = X,Z$. Ignoring the constant
contribution to the Hamiltonian from the $\alpha = I$ terms,
Eq.~\eqref{eq:H_ES_gauge} is then written as
\begin{equation}
	\begin{gathered}
		H_{E} = H_{X,E} + H_{Z,E},\\
		H_{\alpha,E} = - K_\alpha \sum_{i=1}^N \sum_{\tau=0.5}^{T-0.5} \eta_{\alpha,
		(i,\tau)} \prod_{k:\com{[\tilde\alpha]_{i,\tau}}{g_k} = -1}\sigma_k,
	\end{gathered}
	\label{eq:H_ES_gauge_ind}
\end{equation}
where $\tilde\alpha$ denotes the opposite Pauli: $\tilde X = Z$ and
$\tilde Z = X$; recall $K(X)\equiv K_Z$ and $K(Z)\equiv K_X$.
For CSS spacetime codes---where every gauge generator has pure $X$ or pure
$Z$ support---each
spin appears in only one of $H_X$ or $H_Z$. The partition
function, Eq.~\eqref{eq:partition_function_ES}, then factors into $\mathcal Z_{E}
= \mathcal Z_{X,E} \mathcal Z_{Z,E}$ where
\begin{equation}
	\mathcal Z_{\alpha,E} = \sum_{\bm\sigma} e^{-H_{\alpha,E}}.
\end{equation}

\subsection{Spin diagrams}
\label{sec:spin-diagrams}

One of our central results is to provide a framework to  systematically
construct these Hamiltonians  using spin diagrams: a graphical language where
each circuit element---idle wires, \texttt{CNOT}s, measurements,
etc.---contributes standard ``building blocks'' of spins and interactions. These
diagrams offer concise representations of all gauge-equivalent error
configurations and their associated probabilities. Spin diagrams work
for general noise channels; here we demonstrate them for independent
circuit-level $X$-$Z$ noise, where they are particularly transparent. For
general noise channels, each spin has more interactions, which are harder to
visualize.

We construct a diagram by drawing a square for each spin $\sigma_k$, and
connect red (blue) lines for each field term, bond, or many-body interaction
in $H_{X,E}$ ($H_{Z,E}$). The number of lines attached to a spin are the
number of interactions it participates in through the
Hamiltonian---we refer to this as the degree of the spin.
Figure~\ref{fig:spins_intro} illustrates some of these building blocks. An
idling qubit, for example, contributes two generators and therefore two spins,
one for $H_X$ (red) and one for $H_Z$ (blue). A \texttt{CNOT} gate couples
together spins on adjacent  qubits.
Composing these building blocks together, we can represent a circuit as a spin
diagram, such as in Fig.~\ref{fig:spins_examples}.

\begin{figure*}
	\begin{center}
		\includegraphics[width=0.98\linewidth]{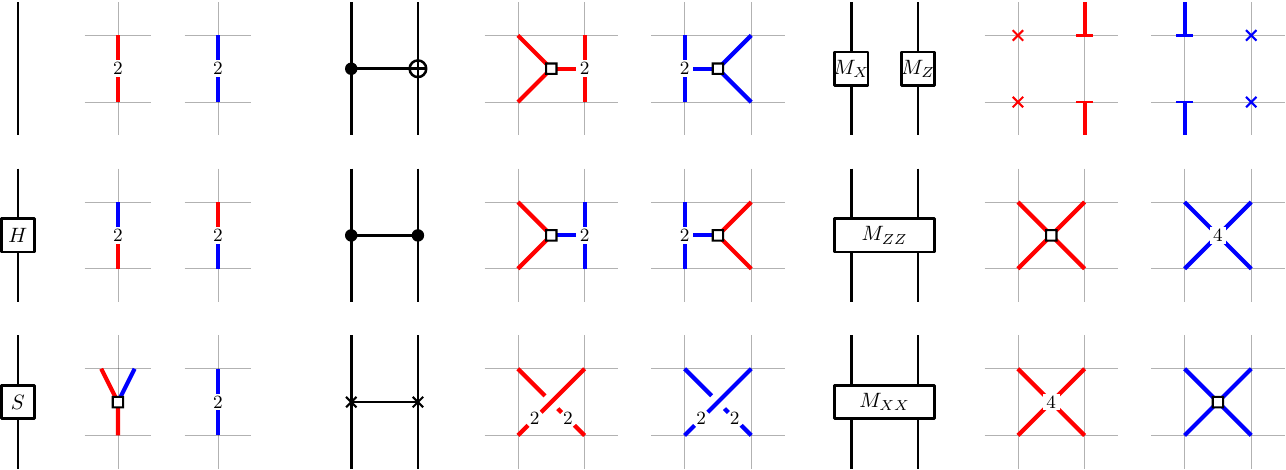}
	\end{center}
	\caption{Common circuit elements and their spin diagrams after integrating out
		spins with one or two interactions. $M_\alpha$ are measurements of Pauli
		$\alpha$. The \texttt{SWAP} gate has two weight-$2$ interactions for each
		$H_\alpha$ that cross over each other, swapping physical qubits. The
		two-qubit measurements each introduce three two-qubit gauge generators (cf.
		Fig.~\ref{fig:spins_intro}); when all three are integrated out, they create
		a weight-$4$ grouped interaction on all four spacetime qubits adjacent to
	the measurement. }
	\label{fig:spins}
\end{figure*}

Many of these spins can be eliminated, producing a simpler---but
equivalent---Hamiltonian: Spins with a single interaction, arising around
single-qubit resets and measurements, correspond to gauge operators that
identify single-qubit errors with no effect on the syndrome or logical
operators (called ``absolutely trivial fault configurations'' in
Ref.~\cite{beverlandFault2024}). Spins with exactly two interactions, such as
those found on idling qubits, encode gauge-equivalent pairs of single-qubit
errors at different spacetime locations that can be grouped together. Both cases
can be integrated out, simplifying the model while preserving the partition
function up to an $E$-independent multiplicative factor; relevant physical
quantities like probability ratios $\mathbb P(\overline E)/\mathbb P(\overline
E')$, free-energy differences $\Delta F = - \ln \mathcal Z_E + \ln \mathcal
Z_{E'}$, and thus ML decoding decisions remain unchanged. To preserve the
partition function, each interaction must now be assigned a weight equal to the
number of single-qubit errors grouped together as gauge-equivalent. This weight
determines the interaction's effective coupling strength $K$ and the probability
distribution of its sign $\eta$, see Eqs.~\eqref{eq:Keff} and \eqref{eq:etaeff}
in Sec.~\ref{sec:gauge-equivalent}, with additional details in
Appendix~\ref{app:walsh-hadamard}. These rules are illustrated diagrammatically
in Fig.~\ref{fig:spins_rules} (an additional rule is discussed in
Appendix~\ref{app:spin_diagrams}), which we then apply to simplify the example
$ZZ$ measurement circuit from Fig.~\ref{fig:spins_examples}. This simplification
reveals properties of the circuit; for example, $H_Z$ contains only one
interaction as a $Z$ error on any of the spacetime locations is equivalent (the
circuit only detects $X$ errors).

Using this prescription, we can identify common circuits elements with their
spin diagrams, as in Fig.~\ref{fig:spins}. Each diagram can be composed with
others to form a full circuit spin diagram and thus the Hamiltonian, bypassing
the computation of gauge generators and anticommuations. When composing circuit
elements, interactions of weight $w_1$ and $w_2$ that merge form a new
interaction of weight $w_1+w_2-1$ (since one spacetime location is shared). In
Sections~\ref{sec:examples} and \ref{sec:toric_code}, we use these building
blocks to construct SM models for a variety of examples.

\subsection{Gauge-equivalent components}
\label{sec:gauge-equivalent}

Having illustrated spin integration graphically, we now describe the precise
form of the simplified Hamiltonian. Appendix~\ref{app:walsh-hadamard} formalizes
the mathematical process behind integrating and re-weighting interactions, which
is applicable to generic noise channels (Hamiltonians of the form in
Eq.~\eqref{eq:H_ES_gauge}). However, under independent $X$-$Z$ noise the
integration process allows the Hamiltonian to take on a particularly simple and
interpretable form. 

To write down the Hamiltonian, we first identify grouped interactions terms in
the integrated Hamiltonian with connected components in a related vertex-labeled
graph. Specifically, we construct the graph $\Gamma = (V,E,L)$ where
$V=\{1,2,\ldots,2NT\}$ are vertices (representing interactions in the
unintegrated Hamiltonian), $E$ are edges (defined below), and $L:
	V\xrightarrow{\text{bijective}}
\{X,Z\}\times\{1,2,\ldots,N\}\times\{0.5,\ldots,T-0.5\}$
labels each vertex by a unique Pauli flavor and spacetime qubit. Then:
\begin{enumerate}
	\item For every single-qubit generator $g_k = [\alpha]_{i,\tau}$, remove
		from $\Gamma$ the vertex $v$ with $L(v) = (\alpha, i, \tau)$.
	\item For every generator $g_k = [\alpha]_{i,\tau}[\alpha']_{i',\tau'}$, where
		$\alpha,\alpha'=X$ or $Z$, add an edge $(v,v')$ to $E$ that joins vertices
		$v=L^{-1}(\alpha,i,\tau)$ and $v'=L^{-1}(\alpha',i',\tau')$.
	\item Form connected components $c\in\Gamma$ of vertices connected by
		edges.
\end{enumerate}
Steps 1 and 2 respectively carry out the process of removing degree-$1$ spins
(corresponding to single-qubit generators; we can remove these spins as
	this preserves the partition function up to a constant multiplicative factor)
	and grouping together interactions for degree-$2$ spins (cf.
Fig.~\ref{fig:spins_rules}). In this way, each connected component includes
interactions from the unintegrated Hamiltonian that are combined together into
one weighted interaction in the integrated Hamiltonian. Some connected
components may still consist of only one vertex: these are interactions from the
Hamiltonian that persist after integration.

Prior to integration, each interaction represented the probability of an error
occurring at a spacetime location and with a Pauli flavor $\alpha=X,Z$.
To preserve the partition function, the new interactions must take into account
the probabilities of errors occurring at any of the spacetime locations in each
of the connected components. Only the parity of the number of errors is
important: by construction, an even number of errors on a connected component is
gauge-equivalent to the identity. If a connected component includes $x$
vertices labeled $\alpha=X$, and $z$ for $\alpha=Z$, the probability that an
odd total number of errors
occurs across that component is
\begin{equation}
	p_{X,Z}^{(x,z)} = \frac12\left[1-(1-2p_X)^x(1-2p_Z)^z\right].
	\label{eq:peff}
\end{equation}
From this, we have effective coupling strength associated with that component
(cf. Appendix~\ref{app:walsh-hadamard})
\begin{equation}
	K_{XZ}^{(x,z)} = \frac12 \ln
	\left(\frac{1-p_{XZ}^{(x,z)}}{p_{XZ}^{(x,z)}}\right),
	\label{eq:Keff}
\end{equation}
and effective quenched disorder distribution
\begin{equation}
	\eta_{XZ}^{(x,z)} \sim 1-2\text{Bernoulli}\left(p_{XZ}^{(x,z)}\right).
	\label{eq:etaeff}
\end{equation}

Let the weight $w_{\alpha,c}$ be the number of Pauli-flavor $\alpha$ vertices in
component $c$. Using these, we define the effective Hamiltonian 
\begin{equation}
	H_E = - \sum_{c\in\Gamma}K_{XZ}^{(w_{X,c},\, w_{Z,c})}
	\eta_{XZ,c}^{(w_{X,c},\,w_{Z,c})} \prod_{k\in \Xi_c}\sigma_k,
	\label{eq:Heff}
\end{equation}
where 
\begin{equation}
	\Xi_c = \left\{k : \com{\prod_{(\alpha,i,\tau)\in c}
	[\tilde{\alpha}]_{i,\tau}}{g_k} = -1\right\}
	\label{eq:K_c}
\end{equation}
selects spins whose generators $g_k$ anticommute with an odd number of errors in
the connected component $c$.

If the gauge generators are all pure-$X$ or pure-$Z$, each connected component
contains only one Pauli flavor: $\Gamma$ splits into two disjoint graphs,
$\Gamma_X$ and $\Gamma_Z$. The Hamiltonian also separates, $H_E =
H_{X,E} + H_{Z,E}$, with $H_{\alpha,E}$ summing over only the connected
components in $\Gamma_\alpha$.
In this case, we simplify notation accordingly, writing
\begin{equation}
	p_{X}^{(x)} \equiv p_{XZ}^{(x,0)}, \, p_Z^{(z)} \equiv p_{XZ}^{(0,z)},
\end{equation}
and similarly for $K_X^{(x)}$ and $\eta_X^{(x)}$. 

For single-vertex connected
components, $p_\alpha^{(1)} = p_\alpha$ and $K_\alpha^{(1)} = K_\alpha$. As the
size of a connected component increases, $p_X^{(w_{X,c})} \rightarrow \frac12$ and
$K_X^{(w_{X,c})}\rightarrow 0$: physically, a net odd-parity error is more likely
to occur on larger components, and consequently the Hamiltonian places a smaller
energy penalty on that interaction being violated. If the size of a component
scales extensively with the number of spacetime qubits, in the thermodynamic
limit its interaction becomes irrelevant to the free energy.

\section{Repetition code}
\label{sec:examples}

We now illustrate our framework through three simple examples, each using the 1D
repetition code. First, we derive the spin models for a memory experiment and a
stability experiment on the code, using a simple measurement-only brickwork
circuit that elucidates the spacetime duality of these two diagnostic tools.
Secondly, we compare different syndrome-extraction circuits, each mapping onto
random-bond Ising models with distinct bond anisotropies. Thirdly, we examine
transversal \texttt{CNOT} gates between repetition code patches, showing how
logical operations introduce defects into these models. In these examples, we
estimate maximum-likelihood decoding thresholds using Monte Carlo simulations
and compare to minimum-weight perfect-matching (MWPM) decoders implemented with
\verb|Stim| \cite{gidneyStim2021} and \verb|pymatching|
\cite{higgottSparse2025}. Numerical methods are detailed in
Appendix~\ref{app:methods}.

The repetition code stores one logical qubit in $N$ physical qubits, using $N-1$
stabilizer generators of $ZZ$ parity checks on adjacent qubits. The $\bar X$
logical operator is $X_1\otimes X_2\otimes\cdots\otimes X_N$ over all physical
qubits, while the $\bar Z$ logical operator is simply $Z_1$ (on any one physical
qubit). Since there are no $X$ stabilizers, it protects against $X$ noise only
and is essentially a classical code. We use it here as a demonstrative example
for our framework, which can be readily generalized to more complicated and
useful codes (cf. Sec.~\ref{sec:toric_code}).

\subsection{Memory and stability experiments}
\label{sec:memory-stability}

\begin{figure*}
	\centering 
	\subfloat[Memory experiment with $N=7$ physical qubits and $T=5$ timesteps.]{%
		\begin{minipage}[c][8.5cm]{0.3\textwidth}%
    \centering
	\includegraphics[width=\textwidth]{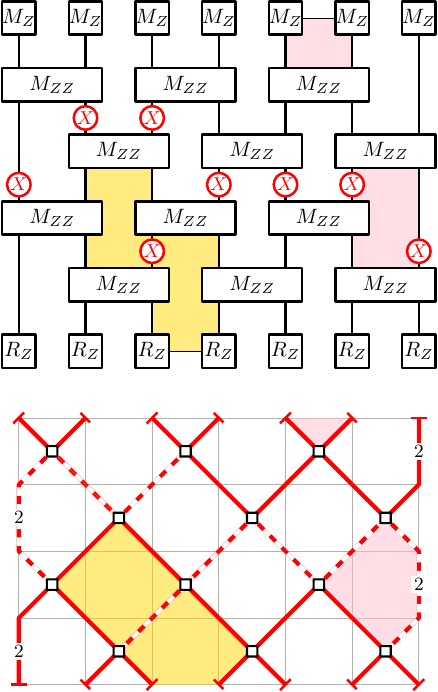}
	\label{fig:rep_memory}
  \end{minipage}
}
	\hfill
	\subfloat[Stability experiment with $N=7$ physical qubits and $T=7$ timesteps.]{%
		\begin{minipage}[c][8.5cm]{0.65\textwidth}%
    \centering
		\includegraphics[width=\textwidth]{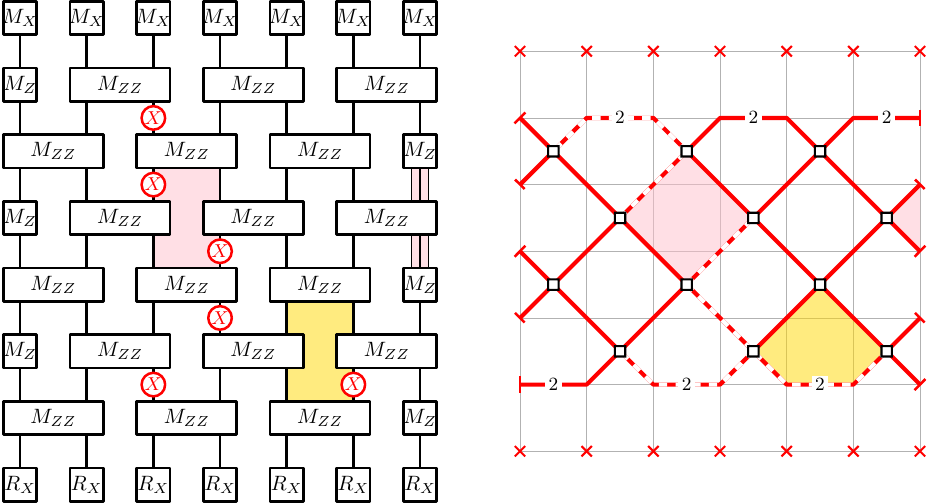}
		\label{fig:rep_stability}
		\vfill
  \end{minipage}
	}
	\caption{Brickwork measurement circuits for repetition code
		experiments, where errors occur in-between each circuit operation. We show
		representatives of error cosets that lead to decoding failures. The detector
		cells triggered by these errors are highlighted in yellow; some untriggered
		detector cells are shown in pink. Both circuits map onto a random-bond Ising
		model on a rotated square lattice. The errors flip the sign of bonds,
		indicated by dashed lines, and form a chain on the dual lattice. Detector
		cells form plaquettes of the lattice, and triggered detector cells host
		Ising vortices at the endpoints of the chains. The spatial and temporal
		(open) boundary conditions are interchanged between the two experiments.}
    \label{fig:memstab}
\end{figure*}

Syndrome-extraction circuits can be used for different experimental tasks; two
common diagnostic tools are the memory and stability experiments (cf.
Sec.~\ref{sec:decoding}). Notably, the two experiments can be interpreted as
spacetime duals of each other \cite{gidneyStability2022, liPerturbative2025}. As
we explain in this section, this is clearly seen using their spacetime spin
diagrams. The structure of these diagrams gives insight to the differences (or
lack thereof) between memory and stability experimental thresholds.

We use two-qubit $ZZ$ measurements as the elementary building blocks. Taking
into account that each qubit may undergo only one measurement at a time, the
circuit takes on a brickwork spatiotemporal layout, shown in
Fig.~\ref{fig:memstab}. This already describes a circuit-level model in systems
with native pairwise measurements~\cite{grans-samuelssonImproved2024,
grans-samuelssonFaulttolerant2025}, and otherwise presents a phenomenological
representation of a measurement circuit.

\paragraph{Memory experiments.}

Figure~\ref{fig:rep_memory} shows a memory experiment circuit.  It begins and
ends with single-qubit $Z$ resets and measurements, respectively, creating
partial detector cells along the temporal boundaries that prevent undetected
timelike errors from spanning the duration of the experiment, cf.
Sec.~\ref{sec:decoding}. Spacelike strings of errors that are gauge
equivalent to a logical $\bar X$ do not trigger the detectors along the spatial
boundaries; these contribute towards logical equivalence classes $\bar
XC_\textsf{s}\mathcal G$ that result in decoding failures.
We construct the spin diagram by tiling components from Fig.~\ref{fig:spins}.
The bulk of $H_X$ is a random bond Ising model (RBIM) on a rotated square
lattice:
\begin{equation}
	H_{X,\text{ bulk}} = - K_X^{(1)} \sum_{\braket{i,j}}
	\eta_{X,\braket{i,j}}^{(1)} \sigma_i \sigma_j,
	\label{eq:rbim}
\end{equation}
with $\braket{i, j}$ the bonds of the lattice. 
The open spatial boundaries produce
weight-$2$ timelike bonds 
\begin{equation}
	-K_X^{(2)} \eta_{X,\braket{i,j}}^{(2)} \sigma_i\sigma_j.
\end{equation}
Due to the qubit initialization and final readouts, on the temporal boundaries
we have field terms
\begin{equation}
	-2K_X^{(1)} \eta_{X,i}^{(1)} \sigma_i,
    \  -K_X^{(2)}\eta_{X,j}^{(2)} \sigma_j.
\end{equation}
As there are no $X$-measurements, all $Z$ errors are undetected and are
gauge-equivalent to anywhere in the circuit. Consequently, all
vertices in $\Gamma_Z$ are connected, forming a single extensive component.
$H_Z$ is a constant, independent of the spin configuration: all $\mathbb
P(\overline E) = \mathbb P(\overline{E'})$ where $E, E'$ are pure-$Z$.

It is well-established that the repetition code subjected to measurement errors
under a phenomenological noise model maps onto the square lattice RBIM
\cite{dennisTopological2002}. Here, spin diagrams have allowed us to
easily derive the associated spin model for our circuit: an RBIM, but on a
rotated square lattice. Unlike previous works, we have no explicit measurement
errors; the non-simultaneous layering of gates (i.e., brickwork) and the
repeated syndrome measurements bring about the two-dimensional structure.

\paragraph{Stability experiments.} Figure~\ref{fig:rep_stability} shows a
stability experiment for the repetition code, again using only
measurements. The initialization and final measurements are now in the $X$ basis
to prevent detector cells from forming on the temporal boundaries,
enabling logical equivalence classes with undetected spanning timelike errors,
while partial stabilizers\footnote{These are logical $Z$ operators for the
repetition code, hence they remove the logical degree of freedom; equivalently,
they define boundary stabilizers for the spacetime code on which $X$-strings
cannot terminate.} are measured on the spatial boundaries to detect
spanning spacelike errors and remove these unwanted equivalence classes from the
experiment \cite{gidneyStability2022}. 
The bulk of the Hamiltonian is again a rotated RBIM. On the spatial boundaries,
however, we now have field terms due to the single-qubit $Z$ measurements
truncating the interactions from each $M_{ZZ}$ spin.
On the temporal boundaries, weight-$2$ spatial bonds are created after
removing degree-$1$
spins from each $R_X$ and $M_X$, leaving the adjacent $M_{ZZ}$ spins as
degree-$2$.

\begin{figure}
	\begin{center}
		\includegraphics[width=\linewidth]{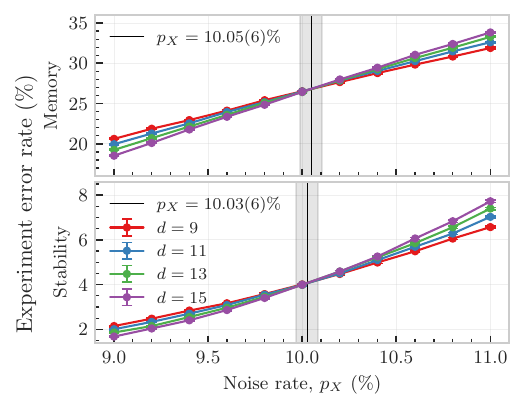}
	\end{center}
	\caption{Error rates for memory and stability experiments corresponding to the
		repetition code measurement circuits in Figs.~\ref{fig:rep_memory} and
		\ref{fig:rep_stability} respectively, using MWPM decoders with circuits of
		size $N=d$, $T=2d+1$. Both exhibit the same threshold, the noise rate $p_X$
	below which increasing $d$ decreases the experimental error rate:
approximately $10.0\%$.}
	\label{fig:rep_m_experiment}
\end{figure}

From the spin diagrams, the spacetime duality between the memory experiment and
stability experiments is apparent. The error thresholds of the repetition code
in a memory experiment (determined by the free energy cost of a spacelike
logical operator) and in a stability experiment (from the free energy cost of an
undetected spanning timelike error)
should be identical. To test this, we perform
numerical simulations of the circuit using the MWPM decoder. Each run, a
syndrome is passed to the decoder, which estimates the underlying error. The
memory experiment is a success if the decoder predicts an eigenvalue of $\bar Z
= Z_1$ (based on anticommutation with its estimated error) that is consistent
with the true eigenvalue at the end of the circuit. Similarly, the stability
experiment is a success if the decoder predicts the correct value for the
product of stabilizer measurement outcomes
$Z_1,\,Z_1Z_2,\,Z_2Z_3,\,\ldots,\,Z_{N-1}Z_N,\,Z_N$.\footnote{Qubit $1$ is on
	the left of Fig.~\ref{fig:rep_stability}; $Z_1$ and $Z_N$ are the single-qubit
$M_Z$ on the spatial boundaries.} In the absence of errors, this product is
always trivial (each $Z_i$ is included twice).
In Fig.~\ref{fig:rep_stability}, for example, the
true outcome would be $-1$ due to the $X$ error before the final round of
$M_{ZZ}$.
Figure~\ref{fig:rep_m_experiment} shows the failure rates of
these experiments at different bit-flip noise rates $p_X$.
Vertical lines indicate the approximate location of the threshold
computed via least-squares estimation in the vicinity of the crossing (cf.
Appendix~\ref{app:additional_details}).
Both experiments, as predicted, exhibit an equal threshold of approximately
$10.0\%$. 

\begin{figure}
	\begin{center}
		\includegraphics[width=\linewidth]{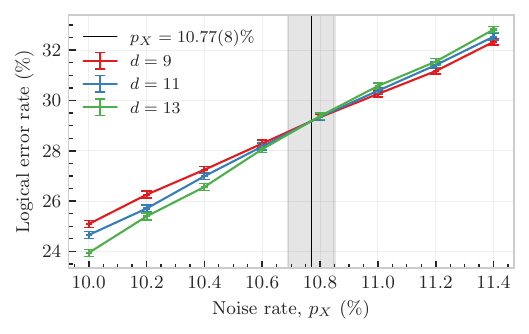}
	\end{center}
	\caption{Monte Carlo simulations of the logical error rate from free-energy
		decoding of the repetition code memory experiment from
		Fig.~\ref{fig:rep_memory} with $N=d$, $T=2d+1$. A crossing occurs at $p_X
		=10.77(8)\%$, higher than the $10.05(6)\%$ from MWPM decoding
        in Fig.~\ref{fig:rep_m_experiment}.}
\label{fig:rep_m_SM}
\end{figure}

The ordered-disordered phase transition of a square lattice RBIM along the
Nishimori line is known to occur
at $p=10.94(2)\%$ \cite{honeckerUniversality2001, merzTwodimensional2002}.
Accounting for boundaries---contributing field
terms and enhanced finite-size effects---it is reasonable to expect a threshold
to deviate from this value. Indeed, running Monte Carlo simulations (cf.
Appendix~\ref{app:methods}) to compute the free energy of the spin model for the
memory experiment, Fig.~\ref{fig:rep_m_SM} highlights a crossing point in the
logical error rate at $p=10.77(8)\%$. MWPM decoders 
consider only the
most-likely error rather than the error cosets and are therefore not optimal;
the lower threshold of approximately $p_X = 10\%$
and consistently higher logical error rates (at $p_X =
10\%$, the MWPM exhibits a logical error rate of approximately $26\%$ compared
to the SM
Monte Carlo $24-25\%$, for example) reflect this. This discrepancy in MWPM
threshold from the $10.5(2)\%$ zero-temperature critical value found in Ref.~\cite{kawashimaZeroTemperature1999} may also be accounted for by finite size
effects.

\subsection{Circuit compilations}
\label{sec:compilations}

The stabilizers of a QEC code can compiled into syndrome-extraction circuits in
a variety of ways. The choice of implementation may be influenced by factors
such as the native hardware-efficient gate set or a need to tolerate fabrication
defects \cite{debroyLUCI2024, wolanskiAutomated2025, mcewenRelaxing2023,
eickbuschDemonstration2025}. In this section, we demonstrate how different
syndrome-extraction circuits for the repetition code can be mapped onto spin
diagrams with distinct bond anisotropies. We compare these models to offer
insight into the differing performance of these circuits. 

\paragraph{Standard ancilla circuit.}

\begin{figure*}
	\centering 
	\subfloat[Standard circuit for the repetition code.]{\includegraphics[width=0.25\linewidth]{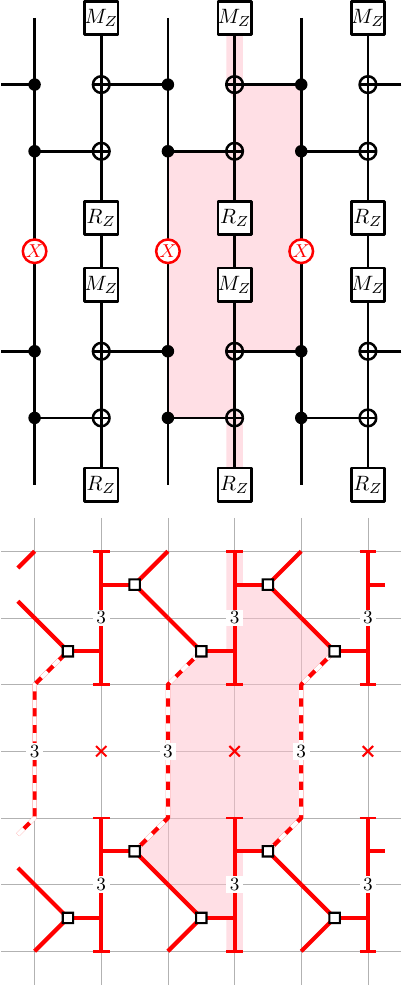}\label{fig:rep_standard}}
	\hfill
	\subfloat[``Wiggling'' circuit that swaps data and ancilla qubits.]{\includegraphics[width=0.25\linewidth]{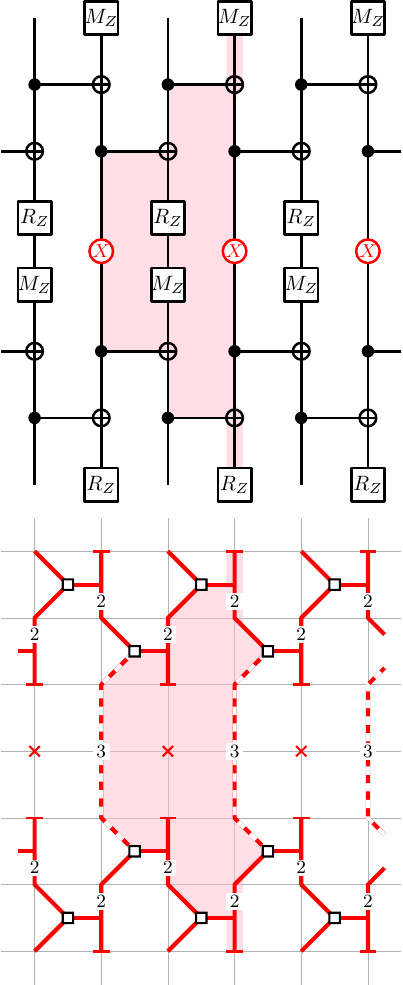}\label{fig:rep_wiggle}}
	\hfill
	\begin{minipage}[b]{0.33\linewidth}
		\centering
	\subfloat[Simplified spin model for the standard circuit is an anisotropic RBIM on a sheared honeycomb lattice.]{\includegraphics[width=\linewidth]{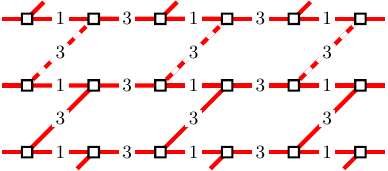}\label{fig:rep_spins}} \\
	\vspace{24pt}
	\subfloat[Simplified spin model for the wiggling circuit is an anisotropic RBIM on a regular honeycomb lattice.]{\includegraphics[width=\linewidth]{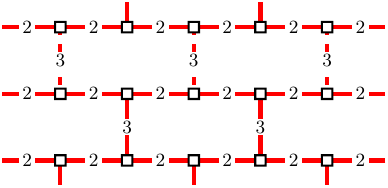}\label{fig:rep_spins_wiggle}}
	\vspace{24pt}
\end{minipage}
\caption{Different syndrome-extraction circuits for the repetition code.
		Spin diagrams have degree-$2$ spins integrated out and replaced with
		weighted interactions (labeled; $1$ if omitted). A cross indicates a removed
		interaction from a degree-$1$ spin.
		Detector cells (one highlighted in pink) map onto plaquettes of the
		spin lattice; triggered detectors (none shown) again host Ising
		vortices.
			Part of a logical $\bar X$ error is shown on the circuit; dashed bonds of
			the spin model show those flipped by the error. Re-arranging bonds to
			simplify reveals an equivalence to
		anisotropic random-bond Ising models on honeycomb lattices.}
	\label{fig:spins_rep}
\end{figure*}

Figure~\ref{fig:spins_rep} shows detector cells for a compiled syndrome
extraction circuit for the repetition code, performing $ZZ$ measurements using
ancilla qubits and \texttt{CNOT} operations (cf.~Fig.~\ref{fig:spins_examples}).
Unlike the memory and stability experiment circuits, measurements of all
stabilizers are performed in parallel, while
the \texttt{CNOT}s introduce complexities to the spread of errors (e.g.,
hook errors).
Applying the spins for \texttt{CNOT}s from
Fig.~\ref{fig:spins}, each weight-$2$ interaction combines with an additional
weight-$2$ interaction to form a single one of weight-$3$,
and we get the spin diagram in Fig.~\ref{fig:rep_standard}. Re-arranging the
bonds (cf. Fig.~\ref{fig:rep_spins}) reveals a random-bond Ising model on a
sheared honeycomb lattice, with anisotropic bond strengths (with weights
$1,3,3$). The bulk Hamiltonian is
\begin{equation}
	H_X = -\sum_{\braket{i,j}} K_X^{(w_{\braket{i,j}})}
	\eta_{X,\braket{i,j}}^{(w_{\braket{i,j}})} \sigma_i \sigma_j,
\end{equation}
where $w_{\braket{i,j}} \in \{1,3,3\}$ depends on the bond.
The shearing is necessary to preserve the timelike direction of the lattice; far
from any boundaries, however, the behavior should be
equivalent to an un-sheared honeycomb lattice. The anisotropy and boundary
conditions place this SM model in a different universality class from the
typical honeycomb RBIM; its critical values will be different and we will  determine them numerically.

\paragraph{Wiggling ancilla circuit.}

\textcite{mcewenRelaxing2023} introduced an alternate, dynamic
syndrome-extraction circuit for the repetition code that exchanges the roles of
data and ancilla qubits, producing a ``wiggling'' pattern in the detector cells.
Figure~\ref{fig:rep_wiggle} shows this circuit and the corresponding spin
diagrams. As with the regular circuit, the simplified model
(cf. Fig.~\ref{fig:rep_spins_wiggle}) is a random bond Ising model,
this time on an un-sheared honeycomb lattice with anisotropic weights
$w_{\braket{i,j}}\in\{2,2,3\}$.
In the bulk of the model, the shear does not affect local interactions---the
difference in memory-experiment threshold between these models should be solely
due to the differing anisotropy.

\begin{figure}
	\begin{center}
		\includegraphics[width=\linewidth]{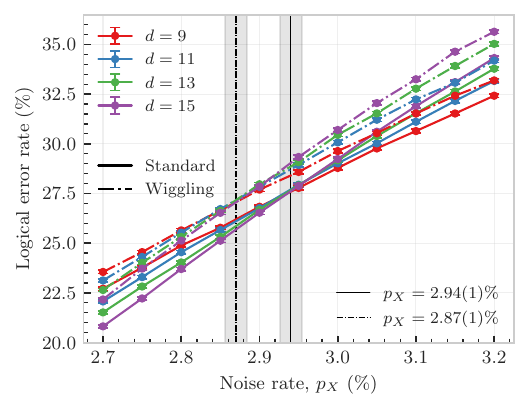}
	\end{center}
	\caption{MWPM error rates for memory experiments of the standard (solid line)
	and wiggling (dot-dashed line) repetition code circuits with $N=2d-1$,
	$T=8d-1$. The wiggling circuit has a lower threshold at $2.87(1)\%$ compared
	to $2.94(1)\%$ for the standard circuit.} \label{fig:rep_compilations}
\end{figure}

To understand the difference in decoding performance of these two circuits,
consider the free energy cost of inserting a spatial logical operator. 
One representative of this logical operator is the product of $X$ over all
qubits in-between each measurement and reset, indicated by the circular errors
and dashed bonds in Fig.~\ref{fig:spins_rep}.
Starting from a ground state with $\eta_{X,\braket{i,j}} = 1$ and $\sigma_i = 1$
for both models, applying the logical error flips vertical weight-$3$ bonds.
The energy cost is identical for both circuits. However, we need to also account
for the contributions to the free energy from gauge-equivalent errors.
Flipping a spin attached to the affected bonds (i.e., including a gauge
operator to deform the logical error)
changes energy by $\Delta \mathcal E_\text{standard} = 2K_X^{(1)}$
in the standard circuit, and $\Delta \mathcal E_\text{wiggling} = 4K_X^{(2)} -
2K_X^{(3)}$
in the wiggling circuit. Taking the low-error limit of Eq.~\eqref{eq:Keff},
one can show that 
\begin{equation}
	\lim_{p_X\rightarrow 0} \left(K_{X}^{(x)} - K_{X}^{(x')}\right) =
	\frac1{2} \ln \left(\frac{x'}{x}\right),
	\label{eq:K_limit}
\end{equation}
and thus 
\begin{equation}
	\lim_{p_X\rightarrow 0} \left(\Delta \mathcal E_\text{standard} - \Delta
	\mathcal E_\text{wiggling}\right)
	= \ln \left(\frac{4}{3}\right).
	\label{eq:delta_E_wiggle}
\end{equation}
From this, we can infer that domain walls separating regions of flipped spins
cost more energy to move in the standard circuit, leading to larger free-energy
barriers. An ML decoder of the standard circuit will more accurately discern the
correct logical class. This suggests lower sub-threshold ML logical error
rates for the standard circuit;
the larger energetic barriers also suggest a higher error threshold
than for the wiggling circuit.

The suppressed fluctuation of errors strings for the standard circuit also
suggest that its MWPM decoding (picking the minimum-energy string) will be more
accurate at low $p_X$ than for the wiggling circuit (resulting in lower logical
error rate). We also expect the MWPM thresholds to reflect the ML threshold
differences: a higher threshold for the standard circuit.

\begin{figure}
	\begin{center}
		\includegraphics[width=\linewidth]{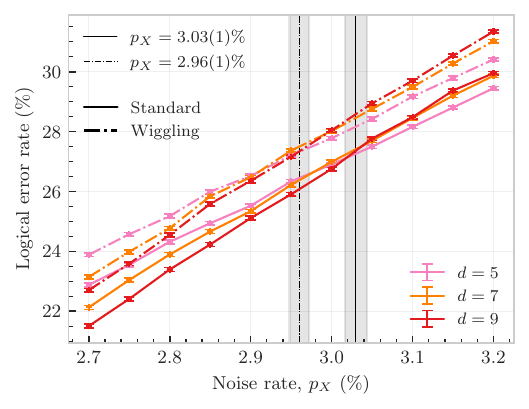}
	\end{center}
	\caption{Monte Carlo simulations of the ML logical error rate for two repetition
		code circuits with $N=2d-1$, $T=8d-1$. The standard circuit again exhibits a
		higher threshold and lower logical error rate compared to the wiggling
		circuit. For the standard circuit, a threshold of $3.03(1)\%$ exceeds the
		MWPM threshold of $2.94(1)\%$ from Fig.~\ref{fig:rep_compilations}.}
	\label{fig:rep_ancilla_SM}
\end{figure}

We performed simulations of MWPM decoding for a memory experiment with both
circuits.
We simulated a circuit depth of $T=8d-1$; this equates to $2d$ rounds of
syndrome measurements, and initialization and final readouts. We plot the
results of the memory experiment in Fig.~\ref{fig:rep_compilations}:
the standard circuit consistently achieves lower logical error rates, as we
expected based on our spin models. The standard circuit has a MWPM threshold of
approximately $2.94(1)\%$, while the wiggling circuit has a lower threshold
of $2.87(1)\%$. This is also consistent with our expectations. Running Monte
Carlo simulations of the spin models and estimating the ML decoding error rates,
Fig.~\ref{fig:rep_ancilla_SM} shows a threshold of $3.03(1)\%$ for the standard
circuit and $2.96(1)\%$ for the wiggling circuit, both slightly exceeding their
MWPM thresholds. The logical error rates themselves are also lower for the ML
decoder, achieving a rate of $21.5\%$ at $p_X = 2.7\%$ with the $d=9$ standard
circuit, compared to $23\%$ for the MWPM decoder, for example. 

As this circuit compilation comparison illustrates, our framework serves both as
a source of intuition for performance differences, as well as a tool for
quantitative comparisons to ML decoders.

\subsection{Logical circuits}
\label{sec:logical_circuits}

\begin{figure}
	\begin{center}
		\includegraphics[width=\linewidth]{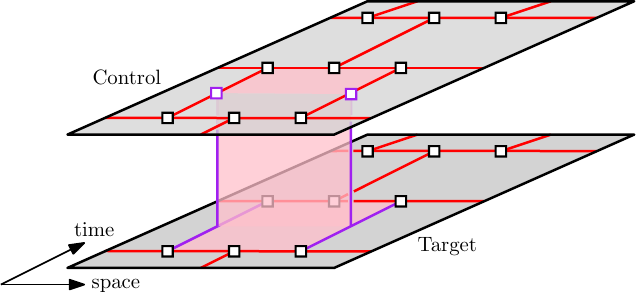}
	\end{center}
	\caption{A transversal \texttt{CNOT} between two repetition codes modifies the
		$H_X$ spin diagram of the system (here shown for syndrome extraction
		circuits as in Fig.~\ref{fig:spins_rep}) by introducing a row of additional
		spins on the control logical qubit that couple with pairs of spins on the
	target (highlighted in purple). Detector cells at the inter-code couplings now
span both code patches (such a cell shown in pink). }
	\label{fig:cnot_3D}
\end{figure}

\begin{figure}
	\begin{center}
		\includegraphics[width=.9\linewidth]{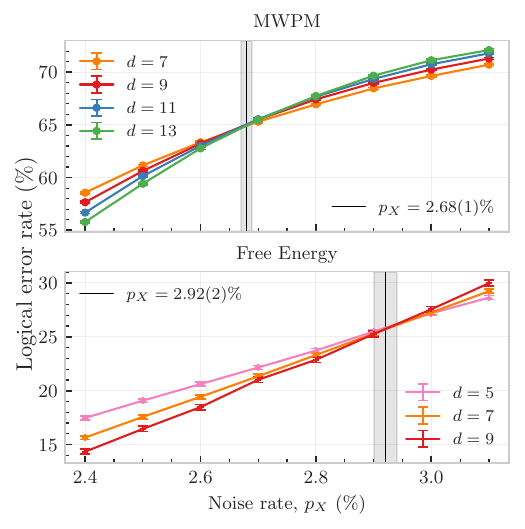}
	\end{center}
	\caption{Logical error rates for a pair of repetition codes with
		standard syndrome extraction circuits
		subjected to one transversal \texttt{CNOT} gate, using a MWPM decoder
		and free energy calculations, with $N=4d-2$,
		$T=16d-1$ for MWPM and $T=4d-1$ for the free energy (ML)
		decoder. (The MWPM decoder required a larger circuit depth to
		observe a threshold, cf. Fig.~\ref{fig:rep_cnot_appendix}; the error rates
		between these two decoders are therefore not directly comparable.) 
		The ML threshold is significantly higher than the MWPM threshold.
	}
	\label{fig:rep_cnot}
\end{figure}

\begin{figure}
	\begin{center}
		\includegraphics[width=.9\linewidth]{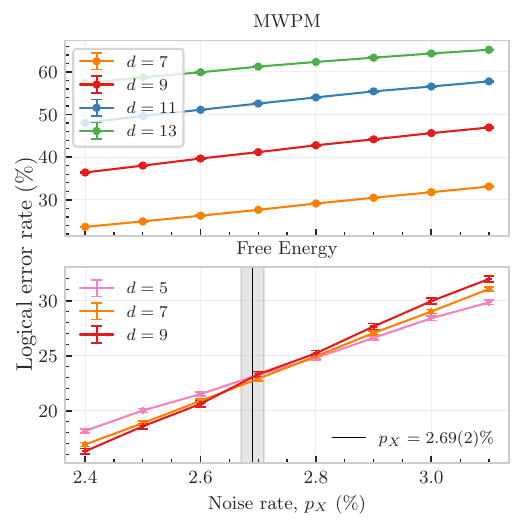}
	\end{center}
	\caption{Logical error rates for a pair of repetition codes with
		standard syndrome extraction circuits
		subjected to $d$ transversal \texttt{CNOT} gates, using both a MWPM decoder
		and free energy calculations, with $N=4d-2$,
		$T=4d-1$ for both. A threshold persists for only the free energy (ML)
		decoder.
	}
	\label{fig:rep_cnot_d}
\end{figure}

Our dynamical spin model framework also naturally encompasses logical
circuits---the QEC of quantum computations. For example, a logical \texttt{CNOT}
gate between two repetition codes running syndrome extraction circuits can be
implemented transversally via \texttt{CNOT} gates on each corresponding data
qubit in the two code patches. Compared to a circuit with idling logical qubits
(cf. Fig.~\ref{fig:spins_rep}), each introduced \texttt{CNOT} modifies the local
structure in two ways (cf. Fig.~\ref{fig:cnot_3D}): on each control qubit, $H_X$
is modified to include an additional spin; and this is coupled to $H_X$ of the
target logical qubit through a three-spin interaction with a pair of target
spins. Acting across all data qubits, a transversal \texttt{CNOT} introduces a
defect line, coupling spins between the two code patches. If a spin on the
control patch involved in this coupling flips to $\sigma_\text{control}=-1$,
this effectively inverts the sign of the interaction on the target, converting
ferromagnetic to antiferromagnetic bonds (and vice-versa). Increasing the error
rate (i.e., weakening the interaction strengths) in the control patch promotes
flipping the spins associated with the transversal \texttt{CNOT}, and this
increases the disorder on the target patch. We thus expect the overall logical
error rate, at a given $p_X$, to be larger than for two decoupled patches---and
the error threshold to be lower. This spin-coupled behavior between the two code
patches encodes the effect of hook errors on the circuit---a single-qubit error
on the control qubit is gauge-equivalent to a two-qubit error on both qubits
after the \texttt{CNOT}.

We expect the threshold of the model to decrease when subjected to transversal
\texttt{CNOT} gates. Figure~\ref{fig:rep_cnot} shows the estimated logical error
rates for the standard syndrome extraction circuit, using both MWPM decoding and
Monte Carlo simulations of the free energy. A logical error occurs if the
decoder fails to predict the correct error coset for either of the two code
patches. Simulations of the MWPM decoder were unable to produce thresholds for
circuit depths smaller than $T=16d-1$, and it was computationally infeasible to
simulate spin models for these larger system sizes---for this reason, the
logical error rates in Fig.~\ref{fig:rep_cnot} are not directly comparable.
Simulations of the MWPM decoder at $T=4d-1$ (cf.
Fig.~\ref{fig:rep_cnot_appendix}) consistent with the ML simulations produced a
logical error rate around $36\%$ for $d=9$ and $p_X=2.8\%$, which is notably
larger than the $23\%$ of the ML decoder. Thresholds, on the other hand, are
properties of the circuit and noise model, and (up to finite-size effects)
remain comparable at different circuit depths. In Fig.~\ref{fig:rep_cnot}, for
one transversal \texttt{CNOT}, we observe an ML threshold at $2.92(2)\%$---lower
than the $3.03(1)\%$ without the gate. For MWPM, we observe a threshold at
$2.68(1)\%$, well below the $2.94(1)\%$ threshold without the gate.

Instead of one transversal \texttt{CNOT} occurring at the midpoint of the
repetition code circuit, we may also consider \texttt{CNOT}s occurring within
each detector cell of a circuit. Performing numerical simulations (cf.
Fig.~\ref{fig:rep_cnot_d}), the MWPM decoder is now unable to produce a
threshold, even at extended circuit depths (cf.
Fig.~\ref{fig:rep_cnot_appendix}). The ML decoder, on the other hand, has a
threshold at $p_X = 2.69(2)\%$. Despite the presence of defects introduced by
the transversal \texttt{CNOT}s to every unit cell of the spin model, there
nevertheless persists an error-correcting phase at non-zero $p_X$.

Whereas the difference between ML and MWPM thresholds for the circuits in
Sec.~\ref{sec:compilations} was around only $0.09$ percentage points, the larger
discrepancy seen here may be accounted for by the MWPM algorithm, which requires
that each single-qubit error flips at most two detectors (a ``graphlike''
detector-error model). An $X$ error on a control qubit after the \texttt{CNOT},
however, flips four detectors: two from its own code patch (as usual) and two
additional from the target code patch (that now extend across both patches due
to the \texttt{CNOT}); an example detector that spans both code patches is shown
in Fig.~\ref{fig:cnot_3D}. While the MWPM decoder can overcome this by
decomposing such errors into products of graphlike errors, doing so introduces
suboptimality \cite{higgottSparse2025}. The same considerations apply for the
notable gap between the with-\texttt{CNOT} and without-\texttt{CNOT} MWPM
thresholds.

The discussions in this section naturally extend to other codes with transversal
\texttt{CNOT}s, different transversal logic gates, lattice surgery, or
code-switching protocols. Spin models of logical circuits thus allow us to probe
the limits of decoding techniques, considering the error-correcting behavior of
quantum circuits, rather than merely quantum memories.

\begin{figure}
	\begin{center}
		\includegraphics[width=\linewidth]{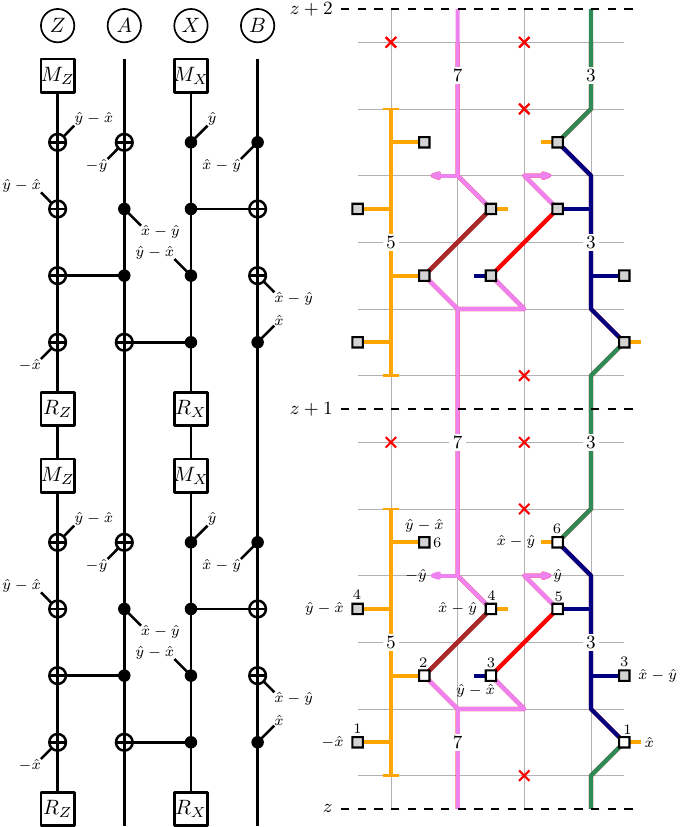}
	\end{center}
	\vskip -.3cm
	\caption{Part of a standard syndrome-extraction circuit for the toric code,
		creating both an $X$-error and $Z$-error detector cell. The code's unit cell
		has four physical qubits $Z$, $A$, $X$ and $B$, cf.
		Fig.~\ref{fig:spins_toric}, and is indexed by lattice directions $(x,y,z)$
		(with $z$ timelike; separated by dashed lines in the spin
		diagram). Partially shown \texttt{CNOT}s couple to other unit cells in the
		directions indicated. The spin diagram for $H_X$ shows one unit cell with
		six spins (white squares and labeled $1$-$6$; spins from adjacent sites are
		gray) and six bonds [colored to reference Eq.~\eqref{eq:H_toric}]. Arrows
		connect bonds to other unit cells.	}
	\vskip -.5cm
	\label{fig:circuit_toric}
\end{figure}

\section{Toric code}
\label{sec:toric_code}

The repetition code, detecting only $X$ errors, is an instructive albeit
simplistic example of the usefulness of SM mappings. Moving to codes that detect
both $X$ and $Z$ errors, our mappings allow us to model and investigate
	different aspects of its error-correcting behavior, such as the effects of
	different circuit compilations or of logical operations.
We also observe richer structure in the form of local $\mathbb Z_2$ gauge
symmetries, which emerge from the spacetime geometry of detector cells. We
demonstrate this now using the toric code.

\subsection{Circuit compilations}

\begin{figure}
	\begin{center}
		\includegraphics[width=\linewidth]{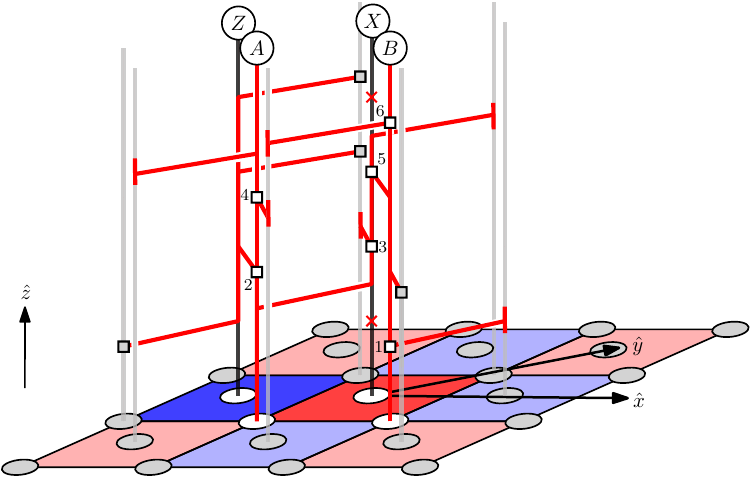}
	\end{center}
	\vskip -.2cm
	\caption{A $3$D unit cell for the toric code $H_X$, based on the syndrome
	extraction circuit in Fig.~\ref{fig:circuit_toric}. Each unit cell is half a
	detector cell, and has $4$ physical qubits, $6$ spins (labeled), and $6$ bonds
	(weights unlabeled for simplicity in the diagram). The lattice directions
	follow the checkerboard pattern of the plaquettes; to keep $\hat
	x$ and $\hat y$ orthogonal we would need to extend to a $12$-spin unit cell.
}
	\vskip -.2cm
	\label{fig:spins_toric}
\end{figure}

The toric code encodes two logical qubits on a periodic $2$D lattice using
weight-four $X$ and $Z$ stabilizers on alternating plaquettes
\cite{dennisTopological2002, bravyiQuantum1998, kitaevFaulttolerant2003}.
Figure~\ref{fig:circuit_toric} shows a standard syndrome-extraction circuit
using ancilla qubits \cite{mcewenRelaxing2023, dennisTopological2002}.
The corresponding SM model is
$3$D and has more complex spin geometries than our prior examples. A unit cell
for $H_X$ is sketched in Fig.~\ref{fig:circuit_toric} and
Fig.~\ref{fig:spins_toric} by applying the elementary spin components from
Fig.~\ref{fig:spins} to the circuit. (The spacetime code is CSS and $H_Z$
has an equivalent structure; we focus on $H_X$ for simplicity.)
Each unit cell at lattice coordinates $(x,y,z)$ has four physical qubits
(labeled $Z$, $A$, $X$, and $B$), six spins (labeled $\sigma_{i,x,y,z}$ for
$i=1,\ldots,6$), and six bonds (signs labeled as $\eta_{j,x,y,z}^{(w)}$ for
$j=1,\ldots,6$). Writing down the interactions from the model, we have a bulk
Hamiltonian
\begin{equation}
	\begin{aligned}
		H_X &= - 
		\sum_{x,y,z} \bigg(
		{\color{Brown} K^{(1)}\eta_{1,x,y,z}^{(1)} \sigma_{2,x,y,z}\sigma_{4,x,y,z}} \\ 
				&+ {\color{Red}K^{(1)}\eta_{2,x,y,z}^{(1)}\sigma_{3,x,y,z}\sigma_{5,x,y,z} }\\ 
				&+{\color{NavyBlue}K^{(3)}\eta_{3,x,y,z}^{(3)}\sigma_{1,x,y,z}\sigma_{3,x+1,y-1,z}\sigma_{5,x,y,z}\sigma_{6,x,y,z}} \\
				&+ {\color{PineGreen}K^{(3)}\eta_{4,x,y,z}^{(3)}\sigma_{1,x,y,z}\sigma_{6,x,y,z-1}} \\ 
		&+
		{\color{Orange}K^{(5)}\eta_{5,x,y,z}^{(5)}\sigma_{1,x-1,y,z}\sigma_{2,x,y,z}\sigma_{4,x-1,y+1,z}\sigma_{6,x-1,y+1,z}}
		\\ 
		&+ {\color{Rhodamine}K^{(7)} \eta_{6,x,y,z}^{(7)}
	\sigma_{2,x,y,z}\sigma_{3,x,y,z}\sigma_{4,x,y,z-1}\sigma_{5,x,y-1,z-1}} \bigg)
\end{aligned}
	\label{eq:H_toric}
\end{equation}
with colors matching Fig.~\ref{fig:circuit_toric} ($X$ subscripts
on $K$ and $\eta$ are omitted for clarity).
Compared to the repetition codes, this is a more complex Ising model, with two
or four spins per interaction, and weights up to $7$.

\begin{figure}
	\begin{center}
		\includegraphics[width=\linewidth]{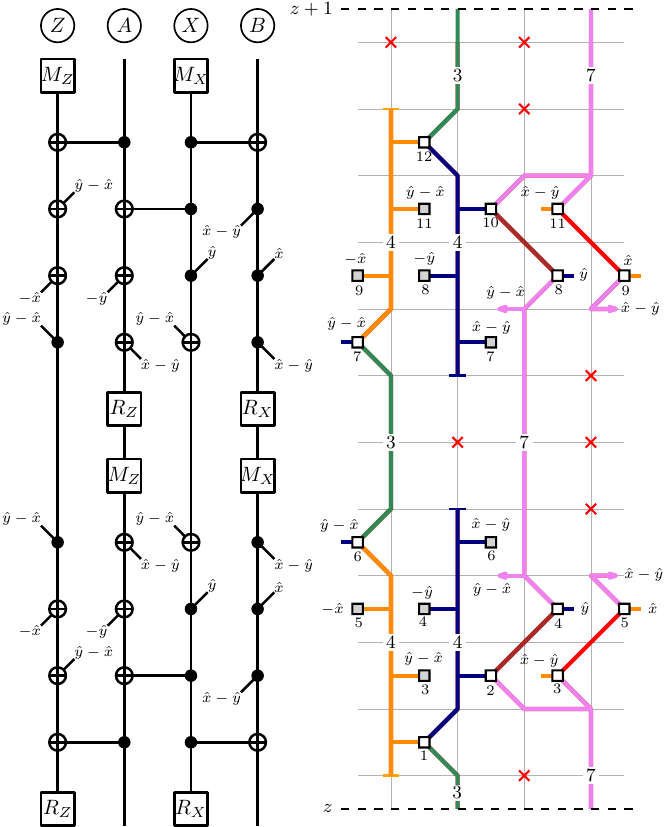}
	\end{center}
	\caption{Part of a wiggling syndrome-extraction circuit for the toric code,
		along with its spin diagram for $H_X$. Each unit cell again contains four
		physical qubits $Z$, $A$, $X$ and $B$ (cf. Fig.~\ref{fig:spins_toric}), but
		the timelike direction is twice as long to accommodate the wiggling
		structure. There are $12$ spins (white squares; spins from adjacent sites
		are gray) per unit cell and $12$ bonds [colored to reference
		Eq.~\eqref{eq:H_toric_wiggle}, and highlighting the symmetric structure of
		the unit cell].}
	\label{fig:circuit_toric_wiggle}
\end{figure}

As we did with the repetition code, we can also consider alternate
syndrome-extraction circuits. The wiggling
toric code circuit from
\textcite{mcewenRelaxing2023} is shown in Fig.~\ref{fig:circuit_toric_wiggle}.
Because ancilla and data qubits are interchanged in the circuit, the second half
of each detector cell circuit is the reverse of the first half. The spin model
unit cell encompasses both halves and now extends twice as long in time and
includes $12$ spins and $12$ bonds. These form the Hamiltonian
\begin{equation}
	\begin{aligned}
		H_X &= - 
		\sum_{x,y,z} \bigg(
			{\color{Brown} K^{(1)}\eta_{1,x,y,z}^{(1)} \sigma_{2,x,y,z}\sigma_{4,x,y,z}}
			\\
				&+ {\color{Brown}
				K^{(1)}\eta_{7,x,y,z}^{(1)}\sigma_{8,x,y,z}\sigma_{4,x,y,z}}
		\\ 
				&+ {\color{Red}K^{(1)}\eta_{2,x,y,z}^{(1)}\sigma_{3,x,y,z}\sigma_{5,x,y,z} }\\ 
				&+{\color{Red}K^{(1)}\eta_{8,x,y,z}^{(1)}\sigma_{9,x,y,z}\sigma_{11,x,y,z}}\\
				&+{\color{NavyBlue}K^{(4)}\eta_{3,x,y,z}^{(4)}\sigma_{1,x,y,z}\sigma_{2,x,y,z}\sigma_{4,x,y-1,z}\sigma_{6,x+1,y-1,z}}\\
				&+{\color{NavyBlue}K^{(4)}\eta_{9,x,y,z}^{(4)}\sigma_{7,x+1,y-1,z}\sigma_{8,x,y-1,z}\sigma_{10,x,y,z}\sigma_{12,x,y,z}}\\
				&+ {\color{PineGreen}K^{(3)}\eta_{4,x,y,z}^{(3)}\sigma_{1,x,y,z}\sigma_{12,x,y,z-12}} \\ 
				&+{\color{PineGreen}K^{(3)}\eta_{10,x,y,z}^{(3)}\sigma_{6,x,y,z}\sigma_{7,x,y,z}} \\
		&+
		{\color{Orange}K^{(4)}\eta_{5,x,y,z}^{(4)}\sigma_{1,x,y,z}\sigma_{3,x-1,y+1,z}\sigma_{5,x-1,y,z}\sigma_{6,x,y,z}}\\
		&+{\color{Orange}K^{(4)}\eta_{11,x,y,z}^{(4)}\sigma_{7,x,y,z}\sigma_{9,x-1,y,z}\sigma_{11,x-1,y+1,z}\sigma_{12,x,y,z}}\\
		&+ {\color{Rhodamine}K^{(7)} \eta_{6,x,y,z}^{(7)}
		\sigma_{2,x,y,z}\sigma_{3,x,y,z}\sigma_{10,x,y,z-1}\sigma_{11,x+1,y-1,z-1}}
		\\
		&+{\color{Rhodamine}K^{(7)}\eta_{12,x,y,z}^{(7)}\sigma_{4,x,y,z}\sigma_{5,x-1,y+1,z}\sigma_{8,x,y,z}\sigma_{9,x-1,y+1,z}}
\bigg)
\end{aligned}
	\label{eq:H_toric_wiggle}
\end{equation}

As with the repetition code, we can form a qualitative understanding of the
differences in ML decoding by examining the free energy cost of inserting a
spatial logical operator. One representation of the $\bar X$ logical is $X$
on qubits along a non-contractible cycle of the lattice, in-between each
measurement and reset. On both circuits' spin diagrams, this flips
${\color{PineGreen}\eta^{(3)}_{4,x,y,z}}$ and
${\color{Rhodamine}\eta^{(7)}_{6,x,y,z}}$ along all $x$ at some $y,z$. For
a ground state of all $\eta=1$ and $\sigma=1$, the energy cost is identical for
both circuits. We then consider gauge-equivalent errors. Flipping spin
$\sigma_{1,x,y,z}$ deforms the logical error by a gauge operator,\footnote{This
is the lowest-energy deformation, and therefore the most likely at low
temperature (low noise).} and changes energy by $\Delta\mathcal
E_{\text{standard}} = 2K^{(5)}$ in the standard circuit and $\Delta\mathcal
E_{\text{wiggling}} = 4K^{(4)}-2K^{(3)}$ in the wiggling circuit. Using
Eq.~\eqref{eq:K_limit}, in the low-noise limit we get 
\begin{equation}
	\lim_{p_X\rightarrow 0}\left(\Delta\mathcal E_\text{standard} - \Delta
		E_\text{wiggling} \right) = \ln \left(\frac{16}{15}\right).
\end{equation}
We obtain the same result as with the repetition code: the wiggling circuit
assigns a lower energy penalty to fluctuations of the logical error strings
compared to the standard circuit, and hence ML and MWPM
decoders for the standard circuit should
exhibit (slightly) lower logical error rates and higher thresholds.

\begin{figure}
	\begin{center}
		\includegraphics[width=\linewidth]{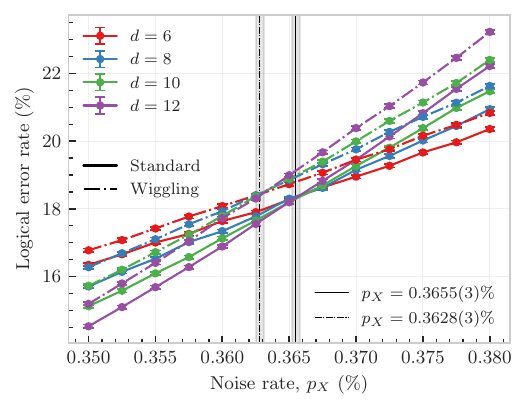}
	\end{center}
	\caption{MWPM logical error rates for the standard (solid line) and wiggling
	(dashed) toric code syndrome-extraction circuits, with $N=2d^2$, $T=12d$. As
	predicted from interaction-weight arguments, the wiggling circuit has a
	slightly lower threshold and higher error rate.}
	\label{fig:toric_compilations}
\end{figure}

To verify this, we  simulate MWPM decoding of toric code memory
experiments. Figure.~\ref{fig:toric_compilations} shows a threshold of
$0.3655(3)\%$ for the standard circuit, higher than the $0.3628(3)\%$ for the
wiggling circuit, in agreement with our energy analysis. Logical error rates for
the standard circuit are also consistently lower (for the same distance and noise rate) compared with the wiggling circuit.

These thresholds appear to be considerably lower than the $2.93(2)\%$ threshold
for MWPM decoding of a toric code with a phenomenological noise model of
bit-flip and measurement errors \cite{wangConfinementHiggs2003}. Our
	circuit-level noise models are not directly comparable to phenomenological
noise models, however, as every detector cell in our circuit supports $44$
spacetime locations where an error can trigger it, compared to six in the
phenomenological. A rough comparison can be drawn by considering the threshold
as an effective noise rate per detector cell (the probability of a detector
being triggered) using Eq.~\eqref{eq:peff}: $\frac12[1-(1-2\times0.003655)^{44}]
= 13.8\%$ for our results, compared to $\frac12[1-(1-2\times0.0293)^{6}]=
15.2\%$ for the phenomenological model. These results are more consistent, with
a lower threshold potentially due to the increased complexities of error
propagation through the circuit's \texttt{CNOT}s.
For further comparison, under circuit-level depolarizing noise the toric
	code has a MWPM threshold of $0.60\%$ \cite{fowlerHighthreshold2009}.
	Depolarizing noise often results in higher thresholds due to
	induced correlations; a toric code with a phenomenological bit-flip noise
	channel without measurement errors gives a $10.94(2)\%$ ML threshold
	\cite{honeckerUniversality2001, merzTwodimensional2002}, compared to
	$18.9(3)\%$ for depolarizing noise \cite{bombinStrong2012, ohzekiError2012}.

\subsection{Gauge symmetries}

Each $X$-flavored detector cell (corresponding to an $X$ stabilizer in the
spacetime code, and covering two unit cells in the standard circuit's $H_X$)
introduces a local constraint: the product of all $X$-flavored gauge generators
within a single detector cell equals the identity. In the spin language, this
means $H_X$ is invariant under local $\mathbb Z_2$ gauge symmetries. For the standard toric code [Eq.~\eqref{eq:H_toric}], these are
\begin{equation}
	\begin{aligned}
		\phi_{x,y,z} &= \text{flip all }
		\{\sigma_{2,x,y,z},\,\sigma_{3,x,yz}\,\sigma_{4,x,y,z},\,\sigma_{5,x,y,z},\\ 
							&\sigma_{6,x,y,z},\,\sigma_{6,x-1,y+1,z},\,\sigma_{1,x,y,z+1},\,\sigma_{1,x-1,y+1,z+1},\\
							&\sigma_{2,x,y+1,z+1},\,\sigma_{3,x,y,z+1},\,\sigma_{4,x,y+1,z+1},\,\sigma_{5,x,y,z+1}\}
	\end{aligned}
	\label{eq:toric_symmetries}
\end{equation}

These symmetries arise generically in any model with an overcomplete set
of chosen gauge generators
$g_k$. For example, around an $M_{ZZ}$
measurement (cf. Fig.~\ref{fig:spins_intro}), including all four generators $g_1
= [Z]_{i,t-0.5}[Z]_{i,t+0.5}$, $g_2 = [Z]_{j,t-0.5}[Z]_{j,t+0.5}$,
$g_3=[ZZ]_{\{i,j\},t-0.5}$ and $g_4=[ZZ]_{\{i,j\},t+0.5}$ creates a redundancy
with $g_1g_2g_3g_4 = I$. This introduces the local gauge symmetry
$\phi
= \text{flip all } \{\sigma_1,\,\sigma_2,\,\sigma_3,\,\sigma_4\}$.
In this example, the redundancy can be easily avoided (the gauge symmetry fixed)
by including only three of these $g_k$ (as we did in
Fig.~\ref{fig:spins_intro}).
In detector cells, however, the involved spins may span several unit
	cells and be included in other overcomplete sets; the dependency
	is not generically fixable without careful examination of the model.
	Removing a spin may also obfuscate the relationship between gauge-equivalent
	error strings. For example, gauge-fixing $\phi_{x,y,z}$ by removing
	$\sigma_{2,x,y,z}$ hides that errors affecting interactions
	${\color{Brown}1}$, ${\color{Orange}5}$, and ${\color{Rhodamine}6}$ are
related via the gauge group.

Gauge symmetries have practical implications for the treatment of the spin
models.
Gauge-invariant quantities---such as Wilson loops formed by products of spins
around closed paths---correspond to observables relevant for determining phases
of matter. In the toric code and other topological codes, appropriate Wilson
loops may be formed by considering the bonds affected when logical operators are
translated in space and time; the complicated bond interactions in our SM
model,
however, obfuscate the geometric significance of such objects
[Eq.~\eqref{eq:H_toric} and Eq.~\eqref{eq:H_toric_wiggle} are not the
simple random plaquette gauge theories of phenomenological noise models~\cite{dennisTopological2002}].

These gauge symmetries also constrain which Monte Carlo algorithms are efficient.
Standard single-spin Metropolis updates respect the gauge symmetry but can
suffer from critical slowing down near the phase transition, as they cannot
efficiently sample between gauge-equivalent configurations
separated by large energy barriers---an important aspect for accurately
determining thermodynamic expectations that require averages over the entire
configuration space.
 Gauge fixing removes the local gauge-equivalent spin configurations from the
 simulation landscape, but does not eliminate these energy barriers.
 Simulations of the spin model that perform single-spin updates would
	 encounter artificial inefficiencies when attempting to propagate errors
 around this removed spin.
Cluster algorithms that respect the gauge structure,
or algorithms suited for exploring complex energy landscapes such as population
annealing, may offer improved sampling performances.

\begin{figure}
	\begin{center}
		\includegraphics[width=\linewidth]{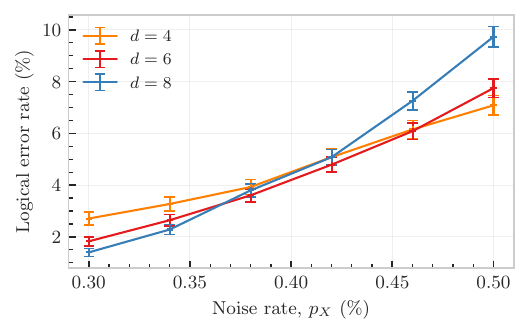}
	\end{center}
	\caption{Estimates of the ML logical error rates for the toric code memory
		experiment with $N=2d^2$, $T=3d$ using Monte Carlo SM simulations. A rough
		threshold is observed in the regime of $0.35\%$-$0.45\%$.
    }
	\label{fig:toric}
\end{figure}

Using Monte Carlo simulations, we estimate the error rates of an ML decoder for
the standard toric code syndrome-extraction circuit; compared to the repetition
code simulations, the higher dimensionality and gauge symmetries create a
rougher energy landscape and precise determination of the threshold is
computationally difficult. Increased compute power, as well as cluster update
algorithms tailored to the local symmetries of Eq.~\eqref{eq:toric_symmetries},
for example, may allow one to observe higher-precision results. Nevertheless,
Fig.~\ref{fig:toric} shows an approximate threshold consistent with the MWPM
result. Unlike the logical circuit from Sec.~\ref{sec:logical_circuits}, the
error model produces only graphlike
errors in this circuit and it is expected that MWPM performs
near-optimally. The system sizes able to be simulated require that we use
a circuit depth $T=3d$, compared to the $T=12d$ used in
Fig.~\ref{fig:toric_compilations}. Running MWPM decoding using $T=3d$ (not
plotted) produces logical error rates consistent (within error bars) with the ML
decoder.

\subsection{Logical circuits}
We can also consider the effect that logical operations have on the spin
models for the toric code.
As with the repetition code, a transversal \texttt{CNOT} on all the data qubits
of the toric code is a logical \texttt{CNOT} operation. The effect on the spin
diagram is qualitatively equivalent to that observed with the toric code: 
Applied to the standard or wiggling circuits, the \texttt{CNOT} modifies the
{\color{PineGreen}weight-$3$} and {\color{Rhodamine}weight-$7$} interactions
that connect two unit cells in the $z$ direction. On the control logical qubit,
these interactions would be replaced by two interactions, joined by an
additional spin; on the code patch for the target logical qubit, these
interactions would now involve {\color{PineGreen}three}
({\color{Rhodamine}five}) spins, instead of their usual {\color{PineGreen}two}
({\color{Rhodamine}four}) for an idling logical qubit. The additional spin and
modified interaction couple together the spin models for the two logical qubits
and promote increased disorder at a given noise rate $p_X$. The detector cells
are also modified, with the $X$-error-detecting cell on the target code patch
now including support on the control code patch, and the $Z$-error-detecting
cell on the control code patch now including support on the target code patch.
These modified detector cells result in non-graphlike errors, where single-qubit
errors on the data qubits after the logical \texttt{CNOT} trigger four detector
cells. The effect of the logical operation is therefore to decrease the
efficiency of decoding, expected to result in higher logical error rates and
lower thresholds, as was seen with the repetition code. 
We leave explicit Monte Carlo simulations of these models to future works.

\section{Conclusion}
\label{sec:conclusion}

In this work, we established a general framework for mapping stabilizer circuits
onto classical statistical mechanics models. Our approach leverages the
spacetime code representation of circuits, where the $(d+1)$D circuit
becomes a subsystem code in $(d+1)$ spatial dimensions. The probability of any
logical error class becomes the partition function of a classical spin
Hamiltonian, with gauge operators of the spacetime code corresponding to spin
degrees of freedom. This mapping applies to arbitrary noise channels and can be
visualized using spin diagrams, which we described in detail for independent
circuit-level $X$-$Z$ noise.

The spin diagram formalism is a modular graphical language that encodes
gauge-equivalent errors and their relative probabilities, and offers both
conceptual clarity and practical utility.
Each circuit element---idle wires, \texttt{CNOT}s, measurements,
resets---contributes a standard building block of spins and interactions that
combine to create the SM Hamiltonian without requiring explicit
computation of gauge generators or their commutation relations. These diagrams
can be simplified by integrating out low-degree spins, grouping gauge-equivalent
errors into weighted interactions while preserving the partition function ratios
that determine the phases of the system (error-correcting or non-correcting) and
hence maximum-likelihood decoding. The resulting models often take the form of
generalized random-bond Ising models, connecting the rich phenomenology of
disordered classical systems to the decodability of stabilizer circuits.

Our examples in Sections~\ref{sec:examples} and \ref{sec:toric_code} illustrate
the capabilities of this framework. For the repetition code, we showed that
memory and stability experiments map onto spacetime-dual random-bond Ising
models on rotated square
lattices, providing a clear geometric interpretation of why their thresholds
coincide. Different
syndrome-extraction circuits, such as a standard ancilla circuit and the
wiggling circuit of \textcite{mcewenRelaxing2023}, produce honeycomb lattice
models with distinct bond anisotropies. By analyzing the free-energy cost of
logical operators in these models, we explained why the standard circuit
achieves a slightly higher threshold and lower logical error rates, and
confirmed these with MWPM and Monte Carlo free energy simulations. These results
already illustrate how SM models can provide an analytical handle for comparing
circuit compilations.

Logical operations also fit naturally within our framework. For example, for the
repetition code, we showed how transversal \texttt{CNOT} gates between two code
patches introduce defect lines into the SM model, coupling spins across the two
lattices and creating higher-order interactions. Monte Carlo simulations
revealed that while a single transversal \texttt{CNOT} has minimal impact on the
threshold, repeated applications---one per detector cell---reduce the threshold
from $2.92(2)\%$ to $2.69(2)\%$. Crucially, an error-correcting phase persists
also with these defects---QEC can succeed even when considering evolving logical
circuits, and not just static memories---which we not only confirm but also
establish its ultimate (i.e., maximum-likelihood decoding) boundaries. 

We also showed how local $\mathbb{Z}_2$ gauge symmetries arise from the
spacetime structure of detector cells. Each $X$-flavor detector cell introduces
a constraint that renders the Hamiltonian invariant under simultaneous flips of
all associated spins. These gauge symmetries connect circuit-level decoding to
lattice gauge theory, suggesting that the spin models for topological codes are
not merely random-bond Ising models but classical lattice gauge theories with
gauge-invariant observables analogous to Wilson loops. This has practical
implications for Monte Carlo simulations, where gauge-fixing or gauge-respecting
algorithms may offer improved sampling efficiency near criticality or at low
temperatures (i.e., low physical error rates).

Our framework opens several avenues for future research. An immediate next
direction concerns dynamical codes~\cite{gottesmanOpportunities2022,
derksDynamical2025, sommersDynamically2025, davydovaUniversal2025,
setiawanTailoring2025, williamsonDynamical2025,xuFaulttolerant2025, tanggaraStrategic2024,tanggaraSimple2024, fuError2025, derksDesigning2024, kishonyIncreasing2025,shawLowering2025},
such as Floquet codes~\cite{hastingsDynamically2021,haahBoundaries2022, blackwellCode2025,
alamDynamical2024, tangPhases2025, kesselringAnyon2024, claesDynamic2025, rodatzFloquetifying2024, alamDynamical2024, tangPhases2025, sutcliffeDistributed2025, zhangXcube2023, aasenFaultTolerant2023, duaEngineering2024, mclauchlanAccommodating2024, davydovaFloquet2023, motamarriSymTFT2024,moylettLogical2025} or dynamic automorphism
codes~\cite{davydovaQuantum2024, aitchisonCompeting2025}, both in
platforms with native two-qubit measurements~
\cite{grans-samuelssonImproved2024, grans-samuelssonFaulttolerant2025}, and in
other systems where one can again compare distinct circuit compilations. 
It may be fruitful to also consider other circuits, such as different
compilations of the surface code syndrome-extraction circuit, surface code
transversal logic gates or lattice surgery, magic state factories, or
code-switching architectures. There has been recent progress in developing
syndrome-extraction circuits that tolerate qubit dropouts \cite{debroyLUCI2024,
wolanskiAutomated2025, strikisQuantum2023, ankerOptimized2025}; it would be
interesting to see how these modified circuits affect the spin models---if lost
qubits manifest as defects, and if so, what are their properties and effect on
the phases of the SM models? Through our framework, these dynamical perspectives
can now all be readily understood through the perspective of statistical
mechanics.

For these and further problems, one can leverage the systematic and modular
nature of spin diagrams, which makes them amenable to automation: given a
circuit description, one could algorithmically generate the corresponding
Hamiltonian and estimate thresholds via Monte Carlo methods. This would
complement existing approaches for comparisons of circuit compilations, gate
orderings, or scheduling strategies for a given code. Spin diagrams
fundamentally encode gauge-equivalent error configurations, and by considering
dominant contributions to the partition function, it may also be possible to
form analytic bounds on the entropic contributions to logical failure rates of
circuits \cite{beverlandRole2019, beverlandFault2024}.

Our framework also opens conceptual directions for exploration. One of these
concerns mixed-state phases of matter~\cite{lyonsUnderstanding2024,
baoMixedstate2023, liReplica2024, fanDiagnostics2024, sohalNoisy2025,
sangMixedState2024, rakovszkyDefining2024, luoTopological2025,
chenSeparability2024, lessaStrongtoWeak2025, leeMixedState2025},
providing an alternative viewpoint on noise-resilient phases. This approach can
also relate the decodability of a quantum code to classical SM systems, but
instead of spins enumerating equivalent errors consistent with a syndrome, they
encode the syndrome itself when a codestate is subjected to a noise channel.
These mixed-state models are Kramers-Wannier duals to the SM models discussed in
this paper, and the mapping from spacetime codes discussed here would follow
analogously. The resulting models will describe a static mixed state in $(d+1)$
spatial dimensions---a counterpart of the time-evolving error-corrupted physical
system in $d$ spatial dimensions. (See also Ref.~\cite{negariSpacetime2024} for
a mixed-state approach to fault tolerance via static $(d+1)$D states.) It would
be interesting to study how the corresponding information-theoretic quantities
can capture features accounted for by our SM models.

From a broader perspective, our work contributes to the growing understanding
that QEC is intimately connected to noise-resilient phases
of matter. Static codes exhibit phase transitions between correctable and
uncorrectable regimes; we have shown that this picture extends naturally to the
full dynamical setting of circuits in spacetime. 
By investigating the spin models, we may reveal connections and dualities
between initially-unrelated circuits and codes. As QEC moves from theoretical
constructs to experimental implementations, tools that bridge circuit-level
dynamics and statistical mechanics will prove increasingly valuable for
understanding and optimizing fault-tolerant quantum computation.

\acknowledgments{CA thanks Alaric Sanders for helpful discussions on Monte Carlo
	simulations. This research was supported by the Gates Cambridge Trust and by
EPSRC grant EP/V062654/1.}

\bibliography{ref}

@misc{aasenFaultTolerant2023,
  title = {Fault-{{Tolerant Hastings-Haah Codes}} in the {{Presence}} of {{Dead Qubits}}},
  author = {Aasen, David and Haah, Jeongwan and Bonderson, Parsa and Wang, Zhenghan and Hastings, Matthew},
  year = 2023,
  month = jul,
  number = {arXiv:2307.03715},
  eprint = {2307.03715},
  primaryclass = {quant-ph},
  publisher = {arXiv},
  urldate = {2024-07-29},
  archiveprefix = {arXiv}
}

@misc{aasenMeasurement2023,
  title = {Measurement {{Quantum Cellular Automata}} and {{Anomalies}} in {{Floquet Codes}}},
  author = {Aasen, David and Haah, Jeongwan and Li, Zhi and Mong, Roger S. K.},
  year = 2023,
  month = aug,
  number = {arXiv:2304.01277},
  eprint = {2304.01277},
  primaryclass = {cond-mat, physics:math-ph, physics:quant-ph},
  publisher = {arXiv},
  urldate = {2023-10-12},
  archiveprefix = {arXiv}
}

@article{aharonovFaultTolerant2008,
  title = {Fault-{{Tolerant Quantum Computation}} with {{Constant Error Rate}}},
  author = {Aharonov, Dorit and {Ben-Or}, Michael},
  year = 2008,
  month = jan,
  journal = {SIAM Journal on Computing},
  volume = {38},
  number = {4},
  pages = {1207--1282},
  issn = {0097-5397, 1095-7111},
  doi = {10.1137/S0097539799359385},
  urldate = {2025-06-11},
  langid = {english}
}

@article{aitchisonCompeting2025,
  title = {Competing Automorphisms and Disordered {{Floquet}} Codes},
  author = {Aitchison, Cory T. and B{\'e}ri, Benjamin},
  year = 2025,
  month = jun,
  journal = {Physical Review B},
  volume = {111},
  number = {23},
  pages = {235112},
  issn = {2469-9950, 2469-9969},
  doi = {10.1103/PhysRevB.111.235112},
  urldate = {2025-06-06},
  langid = {english}
}

@misc{alamDynamical2024,
  title = {Dynamical {{Logical Qubits}} in the {{Bacon-Shor Code}}},
  author = {Alam, M. Sohaib and Rieffel, Eleanor},
  year = 2024,
  month = dec,
  number = {arXiv:2403.03291},
  eprint = {2403.03291},
  primaryclass = {quant-ph},
  publisher = {arXiv},
  doi = {10.48550/arXiv.2403.03291},
  urldate = {2025-04-30},
  archiveprefix = {arXiv}
}

@misc{ankerOptimized2025,
  title = {Optimized {{Measurement Schedules}} for the {{Surface Code}} with {{Dropout}}},
  author = {Anker, Benjamin and Debroy, Dripto M.},
  year = 2025,
  month = dec,
  number = {arXiv:2512.10871},
  eprint = {2512.10871},
  primaryclass = {quant-ph},
  publisher = {arXiv},
  doi = {10.48550/arXiv.2512.10871},
  urldate = {2025-12-15},
  archiveprefix = {arXiv}
}

@inproceedings{baconSparse2017,
  title = {Sparse {{Quantum Codes}} from {{Quantum Circuits}}},
  booktitle = {{{IEEE Transactions}} on {{Information Theory}}},
  author = {Bacon, Dave and Flammia, Steven T. and Harrow, Aram W. and Shi, Jonathan},
  year = 2017,
  month = apr,
  series = {4},
  volume = {63},
  eprint = {1411.3334},
  primaryclass = {quant-ph},
  pages = {2464--2479},
  doi = {10.1109/TIT.2017.2663199},
  urldate = {2025-02-28},
  archiveprefix = {arXiv}
}

@misc{baoMixedstate2023,
  title = {Mixed-State Topological Order and the Errorfield Double Formulation of Decoherence-Induced Transitions},
  author = {Bao, Yimu and Fan, Ruihua and Vishwanath, Ashvin and Altman, Ehud},
  year = 2023,
  month = jan,
  number = {arXiv:2301.05687},
  eprint = {2301.05687},
  primaryclass = {quant-ph},
  publisher = {arXiv},
  doi = {10.48550/arXiv.2301.05687},
  urldate = {2025-03-31},
  archiveprefix = {arXiv}
}

@misc{baoPhases2024,
  title = {Phases of Decodability in the Surface Code with Unitary Errors},
  author = {Bao, Yimu and Anand, Sajant},
  year = 2024,
  month = nov,
  number = {arXiv:2411.05785},
  eprint = {2411.05785},
  primaryclass = {quant-ph},
  publisher = {arXiv},
  doi = {10.48550/arXiv.2411.05785},
  urldate = {2025-12-12},
  archiveprefix = {arXiv}
}

@article{behrendsStatistical2025,
  title = {Statistical {{Mechanical Mapping}} and {{Maximum-Likelihood Thresholds}} for the {{Surface Code}} under {{Generic Single-Qubit Coherent Errors}}},
  author = {Behrends, Jan and B{\'e}ri, Benjamin},
  year = 2025,
  month = oct,
  journal = {PRX Quantum},
  volume = {6},
  number = {4},
  pages = {040305},
  issn = {2691-3399},
  doi = {10.1103/gskb-t5ql},
  urldate = {2025-12-12},
  langid = {english}
}

@article{behrendsSurface2025,
  title = {The {{Surface Code}} beyond {{Pauli Channels}}: {{Logical Noise Coherence}}, {{Information-Theoretic Measures}}, and {{Errorfield-Double Phenomenology}}},
  shorttitle = {The {{Surface Code}} beyond {{Pauli Channels}}},
  author = {Behrends, Jan and B{\'e}ri, Benjamin},
  year = 2025,
  month = dec,
  journal = {PRX Quantum},
  volume = {6},
  number = {4},
  pages = {040350},
  issn = {2691-3399},
  doi = {10.1103/psf5-b6j2},
  urldate = {2025-12-12},
  langid = {english}
}

@article{bennettEfficient1976,
  title = {Efficient Estimation of Free Energy Differences from {{Monte Carlo}} Data},
  author = {Bennett, Charles H},
  year = 1976,
  month = oct,
  journal = {Journal of Computational Physics},
  volume = {22},
  number = {2},
  pages = {245--268},
  issn = {00219991},
  doi = {10.1016/0021-9991(76)90078-4},
  urldate = {2025-10-03},
  copyright = {https://www.elsevier.com/tdm/userlicense/1.0/},
  langid = {english}
}

@misc{berloffExact2025,
  title = {Exact {{Spin Elimination}} in {{Ising Hamiltonians}} and {{Energy-Based Machine Learning}}},
  author = {Berloff, Natalia G.},
  year = 2025,
  month = may,
  number = {arXiv:2505.07163},
  eprint = {2505.07163},
  primaryclass = {quant-ph},
  publisher = {arXiv},
  doi = {10.48550/arXiv.2505.07163},
  urldate = {2025-08-23},
  archiveprefix = {arXiv}
}

@misc{beverlandFault2024,
  title = {Fault Tolerance of Stabilizer Channels},
  author = {Beverland, Michael E. and Huang, Shilin and Kliuchnikov, Vadym},
  year = 2024,
  month = feb,
  number = {arXiv:2401.12017},
  eprint = {2401.12017},
  primaryclass = {quant-ph},
  publisher = {arXiv},
  urldate = {2024-10-11},
  archiveprefix = {arXiv}
}

@article{beverlandRole2019,
  title = {The Role of Entropy in Topological Quantum Error Correction},
  author = {Beverland, Michael E and Brown, Benjamin J and Kastoryano, Michael J and Marolleau, Quentin},
  year = 2019,
  month = jul,
  journal = {Journal of Statistical Mechanics: Theory and Experiment},
  volume = {2019},
  number = {7},
  pages = {073404},
  issn = {1742-5468},
  doi = {10.1088/1742-5468/ab25de},
  urldate = {2026-02-04}
}

@misc{blackwellCode2025,
  title = {The Code Distance of {{Floquet}} Codes},
  author = {Blackwell, Keller and Haah, Jeongwan},
  year = 2025,
  month = oct,
  number = {arXiv:2510.05549},
  eprint = {2510.05549},
  primaryclass = {quant-ph},
  publisher = {arXiv},
  doi = {10.48550/arXiv.2510.05549},
  urldate = {2025-10-08},
  archiveprefix = {arXiv}
}

@article{bombinLogical2023,
  title = {Logical {{Blocks}} for {{Fault-Tolerant Topological Quantum Computation}}},
  author = {Bomb{\'i}n, H{\'e}ctor and Dawson, Chris and Mishmash, Ryan V. and Nickerson, Naomi and Pastawski, Fernando and Roberts, Sam},
  year = 2023,
  month = apr,
  journal = {PRX Quantum},
  volume = {4},
  number = {2},
  pages = {020303},
  issn = {2691-3399},
  doi = {10.1103/PRXQuantum.4.020303},
  urldate = {2026-02-04},
  langid = {english}
}

@article{bombinStrong2012,
  title = {Strong {{Resilience}} of {{Topological Codes}} to {{Depolarization}}},
  author = {Bombin, H. and Andrist, Ruben S. and Ohzeki, Masayuki and Katzgraber, Helmut G. and {Martin-Delgado}, M. A.},
  year = 2012,
  month = apr,
  journal = {Physical Review X},
  volume = {2},
  number = {2},
  pages = {021004},
  issn = {2160-3308},
  doi = {10.1103/PhysRevX.2.021004},
  urldate = {2025-03-25},
  copyright = {http://creativecommons.org/licenses/by/3.0/},
  langid = {english}
}

@article{bombinTopological2010,
  title = {Topological Subsystem Codes},
  author = {Bombin, H.},
  year = 2010,
  month = mar,
  journal = {Physical Review A},
  volume = {81},
  number = {3},
  pages = {032301},
  issn = {1050-2947, 1094-1622},
  doi = {10.1103/PhysRevA.81.032301},
  urldate = {2022-08-09},
  langid = {english}
}

@article{bombinUnifying2024,
  title = {Unifying Flavors of Fault Tolerance with the {{ZX}} Calculus},
  author = {Bombin, Hector and Litinski, Daniel and Nickerson, Naomi and Pastawski, Fernando and Roberts, Sam},
  year = 2024,
  month = jun,
  journal = {Quantum},
  volume = {8},
  pages = {1379},
  issn = {2521-327X},
  doi = {10.22331/q-2024-06-18-1379},
  urldate = {2024-08-15},
  langid = {english}
}

@article{bravyiEfficient2014,
  title = {Efficient {{Algorithms}} for {{Maximum Likelihood Decoding}} in the {{Surface Code}}},
  author = {Bravyi, Sergey and Suchara, Martin and Vargo, Alexander},
  year = 2014,
  month = sep,
  journal = {Physical Review A},
  volume = {90},
  number = {3},
  eprint = {1405.4883},
  primaryclass = {quant-ph},
  pages = {032326},
  issn = {1050-2947, 1094-1622},
  doi = {10.1103/PhysRevA.90.032326},
  urldate = {2023-02-15},
  archiveprefix = {arXiv}
}

@article{bravyiQuantum1998,
  title = {Quantum Codes on a Lattice with Boundary},
  author = {Bravyi, S. B. and Kitaev, A. Yu},
  year = 1998,
  month = nov,
  journal = {arXiv:quant-ph/9811052},
  eprint = {quant-ph/9811052},
  urldate = {2022-02-10},
  archiveprefix = {arXiv}
}

@article{bridgemanHandwaving2017,
  title = {Hand-Waving and {{Interpretive Dance}}: {{An Introductory Course}} on {{Tensor Networks}}},
  shorttitle = {Hand-Waving and {{Interpretive Dance}}},
  author = {Bridgeman, Jacob C. and Chubb, Christopher T.},
  year = 2017,
  month = jun,
  journal = {Journal of Physics A: Mathematical and Theoretical},
  volume = {50},
  number = {22},
  eprint = {1603.03039},
  primaryclass = {cond-mat, physics:hep-th, physics:quant-ph},
  pages = {223001},
  issn = {1751-8113, 1751-8121},
  doi = {10.1088/1751-8121/aa6dc3},
  urldate = {2023-03-15},
  archiveprefix = {arXiv}
}

@misc{cainFast2025,
  title = {Fast Correlated Decoding of Transversal Logical Algorithms},
  author = {Cain, Madelyn and Bluvstein, Dolev and Zhao, Chen and Gu, Shouzhen and Maskara, Nishad and Kalinowski, Marcin and Geim, Alexandra A. and Kubica, Aleksander and Lukin, Mikhail D. and Zhou, Hengyun},
  year = 2025,
  month = jun,
  number = {arXiv:2505.13587},
  eprint = {2505.13587},
  primaryclass = {quant-ph},
  publisher = {arXiv},
  doi = {10.48550/arXiv.2505.13587},
  urldate = {2025-12-21},
  archiveprefix = {arXiv}
}

@article{chengEmergent2025,
  title = {Emergent {{Unitary Designs}} for {{Encoded Qubits}} from {{Coherent Errors}} and {{Syndrome Measurements}}},
  author = {Cheng, Zihan and Huang, Eric and Khemani, Vedika and Gullans, Michael J. and Ippoliti, Matteo},
  year = 2025,
  month = aug,
  journal = {PRX Quantum},
  volume = {6},
  number = {3},
  pages = {030333},
  issn = {2691-3399},
  doi = {10.1103/bnld-2chd},
  urldate = {2025-12-25},
  langid = {english}
}

@article{chenNishimori2025,
  title = {Nishimori Transition across the Error Threshold for Constant-Depth Quantum Circuits},
  author = {Chen, Edward H. and Zhu, Guo-Yi and Verresen, Ruben and Seif, Alireza and B{\"a}umer, Elisa and Layden, David and Tantivasadakarn, Nathanan and Zhu, Guanyu and Sheldon, Sarah and Vishwanath, Ashvin and Trebst, Simon and Kandala, Abhinav},
  year = 2025,
  month = jan,
  journal = {Nature Physics},
  volume = {21},
  number = {1},
  pages = {161--167},
  issn = {1745-2473, 1745-2481},
  doi = {10.1038/s41567-024-02696-6},
  urldate = {2025-12-25},
  langid = {english}
}

@article{chenSeparability2024,
  title = {Separability {{Transitions}} in {{Topological States Induced}} by {{Local Decoherence}}},
  author = {Chen, Yu-Hsueh and Grover, Tarun},
  year = 2024,
  month = apr,
  journal = {Physical Review Letters},
  volume = {132},
  number = {17},
  pages = {170602},
  issn = {0031-9007, 1079-7114},
  doi = {10.1103/PhysRevLett.132.170602},
  urldate = {2025-03-31},
  langid = {english}
}

@article{chirameStabilizing2025,
  title = {Stabilizing {{Non-Abelian Topological Order Against Heralded Noise}} via {{Local Lindbladian Dynamics}}},
  author = {Chirame, Sanket and Prem, Abhinav and Gopalakrishnan, Sarang and Burnell, Fiona J.},
  year = 2025,
  month = sep,
  journal = {PRX Quantum},
  volume = {6},
  number = {3},
  pages = {030363},
  issn = {2691-3399},
  doi = {10.1103/zf7y-hxtq},
  urldate = {2025-12-25},
  langid = {english}
}

@article{chirameStable2025,
  title = {Stable {{Symmetry-Protected Topological Phases}} in {{Systems}} with {{Heralded Noise}}},
  author = {Chirame, Sanket and Burnell, Fiona J. and Gopalakrishnan, Sarang and Prem, Abhinav},
  year = 2025,
  month = jan,
  journal = {Physical Review Letters},
  volume = {134},
  number = {1},
  pages = {010403},
  issn = {0031-9007, 1079-7114},
  doi = {10.1103/PhysRevLett.134.010403},
  urldate = {2025-12-25},
  langid = {english}
}

@article{chubbStatistical2021,
  title = {Statistical Mechanical Models for Quantum Codes with Correlated Noise},
  author = {Chubb, Christopher T. and Flammia, Steven T.},
  year = 2021,
  month = may,
  journal = {Annales de l'Institut Henri Poincar\'e D, Combinatorics, Physics and their Interactions},
  volume = {8},
  number = {2},
  pages = {269--321},
  issn = {2308-5827, 2308-5835},
  doi = {10.4171/aihpd/105},
  urldate = {2024-10-11}
}

@article{claesDynamic2025,
  title = {Dynamic Circuit for the Honeycomb {{Floquet}} Code},
  author = {Claes, Jahan},
  year = 2025,
  month = dec,
  journal = {Physical Review A},
  volume = {112},
  number = {6},
  pages = {062406},
  issn = {2469-9926, 2469-9934},
  doi = {10.1103/qklt-4jnj},
  urldate = {2025-12-23},
  langid = {english}
}

@article{davydovaFloquet2023,
  title = {Floquet {{Codes}} without {{Parent Subsystem Codes}}},
  author = {Davydova, Margarita and Tantivasadakarn, Nathanan and Balasubramanian, Shankar},
  year = 2023,
  month = jun,
  journal = {PRX Quantum},
  volume = {4},
  number = {2},
  pages = {020341},
  issn = {2691-3399},
  doi = {10.1103/PRXQuantum.4.020341},
  urldate = {2024-05-14},
  langid = {english}
}

@article{davydovaQuantum2024,
  title = {Quantum Computation from Dynamic Automorphism Codes},
  author = {Davydova, Margarita and Tantivasadakarn, Nathanan and Balasubramanian, Shankar and Aasen, David},
  year = 2024,
  month = aug,
  journal = {Quantum},
  volume = {8},
  pages = {1448},
  issn = {2521-327X},
  doi = {10.22331/q-2024-08-27-1448},
  urldate = {2024-08-28},
  langid = {english}
}

@misc{davydovaUniversal2025,
  title = {Universal Fault Tolerant Quantum Computation in {{2D}} without Getting Tied in Knots},
  author = {Davydova, Margarita and Bauer, Andreas and de la Fuente, Julio C. Magdalena and Webster, Mark and Williamson, Dominic J. and Brown, Benjamin J.},
  year = 2025,
  month = mar,
  number = {arXiv:2503.15751},
  eprint = {2503.15751},
  primaryclass = {quant-ph},
  publisher = {arXiv},
  doi = {10.48550/arXiv.2503.15751},
  urldate = {2025-03-21},
  archiveprefix = {arXiv}
}

@misc{debroyLUCI2024,
  title = {{{LUCI}} in the {{Surface Code}} with {{Dropouts}}},
  author = {Debroy, Dripto M. and McEwen, Matt and Gidney, Craig and Shutty, Noah and Zalcman, Adam},
  year = 2024,
  month = oct,
  number = {arXiv:2410.14891},
  eprint = {2410.14891},
  primaryclass = {quant-ph},
  publisher = {arXiv},
  doi = {10.48550/arXiv.2410.14891},
  urldate = {2025-07-23},
  archiveprefix = {arXiv}
}

@misc{delfosseSpacetime2023,
  title = {Spacetime Codes of {{Clifford}} Circuits},
  author = {Delfosse, Nicolas and Paetznick, Adam},
  year = 2023,
  month = may,
  number = {arXiv:2304.05943},
  eprint = {2304.05943},
  primaryclass = {quant-ph},
  publisher = {arXiv},
  urldate = {2024-10-15},
  archiveprefix = {arXiv}
}

@article{dennisTopological2002,
  title = {Topological Quantum Memory},
  author = {Dennis, Eric and Kitaev, Alexei and Landahl, Andrew and Preskill, John},
  year = 2002,
  month = sep,
  journal = {Journal of Mathematical Physics},
  volume = {43},
  number = {9},
  eprint = {quant-ph/0110143},
  pages = {4452--4505},
  issn = {0022-2488, 1089-7658},
  doi = {10.1063/1.1499754},
  urldate = {2022-02-10},
  archiveprefix = {arXiv}
}

@misc{derksDesigning2024,
  title = {Designing Fault-Tolerant Circuits Using Detector Error Models},
  author = {Derks, Peter-Jan H. S. and {Townsend-Teague}, Alex and Burchards, Ansgar G. and Eisert, Jens},
  year = 2024,
  month = dec,
  number = {arXiv:2407.13826},
  eprint = {2407.13826},
  primaryclass = {quant-ph},
  doi = {10.48550/arXiv.2407.13826},
  urldate = {2025-02-28},
  archiveprefix = {arXiv}
}

@misc{derksDynamical2025,
  title = {Dynamical Codes for Hardware with Noisy Readouts},
  author = {Derks, Peter-Jan H. S. and {Townsend-Teague}, Alex and Eisert, Jens and Kesselring, Markus S. and Higgott, Oscar and Brown, Benjamin J.},
  year = 2025,
  month = may,
  number = {arXiv:2505.07658},
  eprint = {2505.07658},
  primaryclass = {quant-ph},
  publisher = {arXiv},
  doi = {10.48550/arXiv.2505.07658},
  urldate = {2025-05-14},
  archiveprefix = {arXiv}
}

@article{duaEngineering2024,
  title = {Engineering {{3D Floquet Codes}} by {{Rewinding}}},
  author = {Dua, Arpit and Tantivasadakarn, Nathanan and Sullivan, Joseph and Ellison, Tyler D.},
  year = 2024,
  month = apr,
  journal = {PRX Quantum},
  volume = {5},
  number = {2},
  pages = {020305},
  issn = {2691-3399},
  doi = {10.1103/PRXQuantum.5.020305},
  urldate = {2024-07-29},
  langid = {english}
}

@misc{ecksteinLearning2025,
  title = {Learning Transitions of Topological Surface Codes},
  author = {Eckstein, Finn and Han, Bo and Trebst, Simon and Zhu, Guo-Yi},
  year = 2025,
  month = dec,
  number = {arXiv:2512.19786},
  eprint = {2512.19786},
  primaryclass = {quant-ph},
  publisher = {arXiv},
  doi = {10.48550/arXiv.2512.19786},
  urldate = {2025-12-25},
  archiveprefix = {arXiv}
}

@article{ecksteinRobust2024,
  title = {Robust {{Teleportation}} of a {{Surface Code}} and {{Cascade}} of {{Topological Quantum Phase Transitions}}},
  author = {Eckstein, Finn and Han, Bo and Trebst, Simon and Zhu, Guo-Yi},
  year = 2024,
  month = oct,
  journal = {PRX Quantum},
  volume = {5},
  number = {4},
  pages = {040313},
  issn = {2691-3399},
  doi = {10.1103/PRXQuantum.5.040313},
  urldate = {2025-12-21},
  langid = {english}
}

@article{eickbuschDemonstration2025,
  title = {Demonstration of Dynamic Surface Codes},
  author = {Eickbusch, Alec and McEwen, Matt and Sivak, Volodymyr and Bourassa, Alexandre and Atalaya, Juan and Claes, Jahan and Kafri, Dvir and Gidney, Craig and Warren, Christopher W. and Gross, Jonathan and Opremcak, Alex and Zobrist, Nicholas and Miao, Kevin C. and Roberts, Gabrielle and Satzinger, Kevin J. and Bengtsson, Andreas and Neeley, Matthew and Livingston, William P. and Greene, Alex and Acharya, Rajeev and Aghababaie Beni, Laleh and Aigeldinger, Georg and Alcaraz, Ross and Andersen, Trond I. and Ansmann, Markus and Arute, Frank and Arya, Kunal and Asfaw, Abraham and Babbush, Ryan and Ballard, Brian and Bardin, Joseph C. and Bilmes, Alexander and Bovaird, Jenna and Bowers, Dylan and Brill, Leon and Broughton, Michael and Browne, David A. and Buchea, Brett and Buckley, Bob B. and Burger, Tim and Burkett, Brian and Bushnell, Nicholas and Cabrera, Anthony and Campero, Juan and Chang, Hung-Shen and Chiaro, Ben and Chih, Liang-Ying and Cleland, Agnetta Y. and Cogan, Josh and Collins, Roberto and Conner, Paul and Courtney, William and Crook, Alexander L. and Curtin, Ben and Das, Sayan and Del Toro Barba, Alexander and Demura, Sean and De Lorenzo, Laura and Di Paolo, Agustin and Donohoe, Paul and Drozdov, Ilya K. and Dunsworth, Andrew and Elbag, Aviv Moshe and Elzouka, Mahmoud and Erickson, Catherine and Ferreira, Vinicius S. and Flores Burgos, Leslie and Forati, Ebrahim and Fowler, Austin G. and Foxen, Brooks and Ganjam, Suhas and Garcia, Gonzalo and Gasca, Robert and Genois, {\'E}lie and Giang, William and Gilboa, Dar and Gosula, Raja and Grajales Dau, Alejandro and Graumann, Dietrich and Ha, Tan and Habegger, Steve and Hamilton, Michael C. and Hansen, Monica and Harrigan, Matthew P. and Harrington, Sean D. and Heslin, Stephen and Heu, Paula and Higgott, Oscar and Hiltermann, Reno and Hilton, Jeremy and Huang, Hsin-Yuan and Huff, Ashley and Huggins, William J. and Jeffrey, Evan and Jiang, Zhang and Jin, Xiaoxuan and Jones, Cody and Joshi, Chaitali and Juhas, Pavol and Kabel, Andreas and Kang, Hui and Karamlou, Amir H. and Kechedzhi, Kostyantyn and Khaire, Trupti and Khattar, Tanuj and Khezri, Mostafa and Kim, Seon and Kobrin, Bryce and Korotkov, Alexander N. and Kostritsa, Fedor and Kreikebaum, John Mark and Kurilovich, Vladislav D. and Landhuis, David and {Lange-Dei}, Tiano and Langley, Brandon W. and Lau, Kim-Ming and Ledford, Justin and Lee, Kenny and Lester, Brian J. and Le Guevel, Lo{\"i}ck and Li, Wing Yan and Lill, Alexander T. and Locharla, Aditya and Lucero, Erik and Lundahl, Daniel and Lunt, Aaron and Madhuk, Sid and Maloney, Ashley and Mandr{\`a}, Salvatore and Martin, Leigh S. and Martin, Orion and Maxfield, Cameron and McClean, Jarrod R. and Meeks, Seneca and Megrant, Anthony and Molavi, Reza and Molina, Sebastian and Montazeri, Shirin and Movassagh, Ramis and Newman, Michael and Nguyen, Anthony and Nguyen, Murray and Ni, Chia-Hung and Oas, Logan and Orosco, Raymond and Ottosson, Kristoffer and Pizzuto, Alex and Potter, Rebecca and Pritchard, Orion and Quintana, Chris and Ramachandran, Ganesh and Reagor, Matthew J. and Rhodes, David M. and Rosenberg, Eliott and Rossi, Elizabeth and Sankaragomathi, Kannan and Schurkus, Henry F. and Shearn, Michael J. and Shorter, Aaron and Shutty, Noah and Shvarts, Vladimir and Small, Spencer and Smith, W. Clarke and Springer, Sofia and Sterling, George and Suchard, Jordan and Szasz, Aaron and Sztein, Alex and Thor, Douglas and Tomita, Eifu and Torres, Alfredo and Torunbalci, M. Mert and Vaishnav, Abeer and Vargas, Justin and Vdovichev, Sergey and Vidal, Guifre and Vollgraff Heidweiller, Catherine and Waltman, Steven and Waltz, Jonathan and Wang, Shannon X. and Ware, Brayden and Weidel, Travis and White, Theodore and Wong, Kristi and Woo, Bryan W. K. and Woodson, Maddy and Xing, Cheng and Yao, Z. Jamie and Yeh, Ping and Ying, Bicheng and Yoo, Juhwan and Yosri, Noureldin and Young, Grayson and Zalcman, Adam and Zhang, Yaxing and Zhu, Ningfeng and Boixo, Sergio and Kelly, Julian and Smelyanskiy, Vadim and Neven, Hartmut and Bacon, Dave and Chen, Zijun and Klimov, Paul V. and Roushan, Pedram and Neill, Charles and Chen, Yu and Morvan, Alexis},
  year = 2025,
  month = dec,
  journal = {Nature Physics},
  volume = {21},
  number = {12},
  pages = {1994--2001},
  issn = {1745-2473, 1745-2481},
  doi = {10.1038/s41567-025-03070-w},
  urldate = {2025-12-16},
  langid = {english}
}

@misc{ellisonFloquet2023,
  title = {Floquet Codes with a Twist},
  author = {Ellison, Tyler D. and Sullivan, Joseph and Dua, Arpit},
  year = 2023,
  month = sep,
  number = {arXiv:2306.08027},
  eprint = {2306.08027},
  primaryclass = {quant-ph},
  publisher = {arXiv},
  urldate = {2023-10-11},
  archiveprefix = {arXiv}
}

@misc{englishIsing2025,
  title = {Ising on the Donut: {{Regimes}} of Topological Quantum Error Correction from Statistical Mechanics},
  shorttitle = {Ising on the Donut},
  author = {English, Lucas H. and Roberts, Sam and Bartlett, Stephen D. and Doherty, Andrew C. and Williamson, Dominic J.},
  year = 2025,
  month = dec,
  number = {arXiv:2512.10399},
  eprint = {2512.10399},
  primaryclass = {quant-ph},
  publisher = {arXiv},
  doi = {10.48550/arXiv.2512.10399},
  urldate = {2025-12-12},
  archiveprefix = {arXiv}
}

@misc{englishThresholds2024,
  title = {Thresholds for Post-Selected Quantum Error Correction from Statistical Mechanics},
  author = {English, Lucas H. and Williamson, Dominic J. and Bartlett, Stephen D.},
  year = 2024,
  month = oct,
  number = {arXiv:2410.07598},
  eprint = {2410.07598},
  primaryclass = {quant-ph},
  publisher = {arXiv},
  urldate = {2024-10-11},
  archiveprefix = {arXiv}
}

@article{fanDiagnostics2024,
  title = {Diagnostics of {{Mixed-State Topological Order}} and {{Breakdown}} of {{Quantum Memory}}},
  author = {Fan, Ruihua and Bao, Yimu and Altman, Ehud and Vishwanath, Ashvin},
  year = 2024,
  month = may,
  journal = {PRX Quantum},
  volume = {5},
  number = {2},
  pages = {020343},
  issn = {2691-3399},
  doi = {10.1103/PRXQuantum.5.020343},
  urldate = {2025-03-31},
  langid = {english}
}

@article{fowlerHighthreshold2009,
  title = {High-Threshold Universal Quantum Computation on the Surface Code},
  author = {Fowler, Austin G. and Stephens, Ashley M. and Groszkowski, Peter},
  year = 2009,
  month = nov,
  journal = {Physical Review A},
  volume = {80},
  number = {5},
  pages = {052312},
  issn = {1050-2947, 1094-1622},
  doi = {10.1103/PhysRevA.80.052312},
  urldate = {2025-12-14},
  copyright = {http://link.aps.org/licenses/aps-default-license},
  langid = {english}
}

@article{fuError2025,
  title = {Error {{Correction}} in {{Dynamical Codes}}},
  author = {Fu, Esther Xiaozhen and Gottesman, Daniel},
  year = 2025,
  month = oct,
  journal = {Quantum},
  volume = {9},
  pages = {1886},
  issn = {2521-327X},
  doi = {10.22331/q-2025-10-20-1886},
  urldate = {2025-12-23},
  langid = {english}
}

@misc{fuSubsystem2025,
  type = {Poster},
  title = {Subsystem {{Spacetime Code}}},
  author = {Fu, Xiaozhen and Gottesman, Daniel},
  year = 2025,
  month = feb
}

@article{geherErrorcorrected2024,
  title = {Error-Corrected {{Hadamard}} Gate Simulated at the Circuit Level},
  author = {Geh{\'e}r, Gy{\"o}rgy P. and McLauchlan, Campbell and Campbell, Earl T. and Moylett, Alexandra E. and Crawford, Ophelia},
  year = 2024,
  month = jul,
  journal = {Quantum},
  volume = {8},
  pages = {1394},
  issn = {2521-327X},
  doi = {10.22331/q-2024-07-02-1394},
  urldate = {2025-10-01},
  langid = {english}
}

@article{gidneyBenchmarking2022,
  title = {Benchmarking the {{Planar Honeycomb Code}}},
  author = {Gidney, Craig and Newman, Michael and McEwen, Matt},
  year = 2022,
  month = sep,
  journal = {Quantum},
  volume = {6},
  pages = {813},
  issn = {2521-327X},
  doi = {10.22331/q-2022-09-21-813},
  urldate = {2025-12-22},
  langid = {english}
}

@misc{gidneyLess2023,
  title = {Less {{Bacon More Threshold}}},
  author = {Gidney, Craig and Bacon, Dave},
  year = 2023,
  month = may,
  number = {arXiv:2305.12046},
  eprint = {2305.12046},
  primaryclass = {quant-ph},
  publisher = {arXiv},
  doi = {10.48550/arXiv.2305.12046},
  urldate = {2025-05-08},
  archiveprefix = {arXiv}
}

@article{gidneyStability2022,
  title = {Stability {{Experiments}}: {{The Overlooked Dual}} of {{Memory Experiments}}},
  shorttitle = {Stability {{Experiments}}},
  author = {Gidney, Craig},
  year = 2022,
  month = aug,
  journal = {Quantum},
  volume = {6},
  pages = {786},
  publisher = {Verein zur Forderung des Open Access Publizierens in den Quantenwissenschaften},
  issn = {2521-327X},
  doi = {10.22331/q-2022-08-24-786},
  urldate = {2025-07-15},
  copyright = {https://creativecommons.org/licenses/by/4.0/},
  langid = {english}
}

@article{gidneyStim2021,
  title = {Stim: A Fast Stabilizer Circuit Simulator},
  shorttitle = {Stim},
  author = {Gidney, Craig},
  year = 2021,
  month = jul,
  journal = {Quantum},
  volume = {5},
  pages = {497},
  issn = {2521-327X},
  doi = {10.22331/q-2021-07-06-497},
  urldate = {2024-11-07},
  langid = {english}
}

@misc{gottesmanOpportunities2022,
  title = {Opportunities and {{Challenges}} in {{Fault-Tolerant Quantum Computation}}},
  author = {Gottesman, Daniel},
  year = 2022,
  month = oct,
  number = {arXiv:2210.15844},
  eprint = {2210.15844},
  primaryclass = {quant-ph},
  publisher = {arXiv},
  urldate = {2024-04-22},
  archiveprefix = {arXiv}
}

@phdthesis{gottesmanStabilizer1997,
  title = {Stabilizer {{Codes}} and {{Quantum Error Correction}}},
  author = {Gottesman, Daniel Eric},
  year = 1997,
  month = may,
  doi = {10.7907/RZR7-DT72},
  urldate = {2023-07-24},
  copyright = {No commercial reproduction, distribution, display or performance rights in this work are provided.},
  langid = {english},
  school = {California Institute of Technology}
}

@article{grans-samuelssonFaulttolerant2025,
  title = {Fault-Tolerant Pairwise Measurement-Based Code on Eight Qubits},
  author = {{Grans-Samuelsson}, Linnea and Aasen, David and Bonderson, Parsa},
  year = 2025,
  month = oct,
  journal = {Physical Review A},
  volume = {112},
  number = {4},
  pages = {042413},
  issn = {2469-9926, 2469-9934},
  doi = {10.1103/c54n-y9vw},
  urldate = {2025-12-17},
  langid = {english}
}

@article{grans-samuelssonImproved2024,
  title = {Improved {{Pairwise Measurement-Based Surface Code}}},
  author = {{Grans-Samuelsson}, Linnea and Mishmash, Ryan V. and Aasen, David and Knapp, Christina and Bauer, Bela and Lackey, Brad and Silva, Marcus P. Da and Bonderson, Parsa},
  year = 2024,
  month = aug,
  journal = {Quantum},
  volume = {8},
  pages = {1429},
  issn = {2521-327X},
  doi = {10.22331/q-2024-08-02-1429},
  urldate = {2025-12-17},
  langid = {english}
}

@article{haahBoundaries2022,
  title = {Boundaries for the {{Honeycomb Code}}},
  author = {Haah, Jeongwan and Hastings, Matthew B.},
  year = 2022,
  month = apr,
  journal = {Quantum},
  volume = {6},
  eprint = {2110.09545},
  primaryclass = {quant-ph},
  pages = {693},
  issn = {2521-327X},
  doi = {10.22331/q-2022-04-21-693},
  urldate = {2023-11-28},
  archiveprefix = {arXiv}
}

@article{hastingsDynamically2021,
  title = {Dynamically {{Generated Logical Qubits}}},
  author = {Hastings, Matthew B. and Haah, Jeongwan},
  year = 2021,
  month = oct,
  journal = {Quantum},
  volume = {5},
  pages = {564},
  issn = {2521-327X},
  doi = {10.22331/q-2021-10-19-564},
  urldate = {2023-10-11},
  langid = {english}
}

@misc{hauserInformation2025,
  title = {Information Dynamics and Symmetry Breaking in Generic Monitored \$\textbackslash mathbb\textbraceleft{{Z}}\textbraceright\_2\$-Symmetric Open Quantum Systems},
  author = {Hauser, Jacob and Lavasani, Ali and Vijay, Sagar and Fisher, Matthew P. A.},
  year = 2025,
  month = dec,
  number = {arXiv:2512.03031},
  eprint = {2512.03031},
  primaryclass = {quant-ph},
  publisher = {arXiv},
  doi = {10.48550/arXiv.2512.03031},
  urldate = {2025-12-25},
  archiveprefix = {arXiv}
}

@article{higgottSparse2025,
  title = {Sparse {{Blossom}}: Correcting a Million Errors per Core Second with Minimum-Weight Matching},
  shorttitle = {Sparse {{Blossom}}},
  author = {Higgott, Oscar and Gidney, Craig},
  year = 2025,
  month = jan,
  journal = {Quantum},
  volume = {9},
  pages = {1600},
  issn = {2521-327X},
  doi = {10.22331/q-2025-01-20-1600},
  urldate = {2025-09-21},
  langid = {english}
}

@article{honeckerUniversality2001,
  title = {Universality {{Class}} of the {{Nishimori Point}} in the {{2D}} \textpm{} {{J Random-Bond Ising Model}}},
  author = {Honecker, A. and Picco, M. and Pujol, P.},
  year = 2001,
  month = jul,
  journal = {Physical Review Letters},
  volume = {87},
  number = {4},
  pages = {047201},
  issn = {0031-9007, 1079-7114},
  doi = {10.1103/PhysRevLett.87.047201},
  urldate = {2025-12-17},
  copyright = {http://link.aps.org/licenses/aps-default-license},
  langid = {english}
}

@misc{huangRobust2025,
  title = {A Robust Phase of Continuous Transversal Gates in Quantum Stabilizer Codes},
  author = {Huang, Eric and Rozon, Pierre-Gabriel and Dua, Arpit and Gopalakrishnan, Sarang and Gullans, Michael J.},
  year = 2025,
  month = oct,
  number = {arXiv:2510.01319},
  eprint = {2510.01319},
  primaryclass = {quant-ph},
  publisher = {arXiv},
  doi = {10.48550/arXiv.2510.01319},
  urldate = {2025-12-25},
  archiveprefix = {arXiv}
}

@misc{jingIntrinsic2025,
  title = {Intrinsic {{Heralding}} and {{Optimal Decoders}} for {{Non-Abelian Topological Order}}},
  author = {Jing, Dian and Sala, Pablo and Jiang, Liang and Verresen, Ruben},
  year = 2025,
  month = oct,
  number = {arXiv:2507.23765},
  eprint = {2507.23765},
  primaryclass = {quant-ph},
  publisher = {arXiv},
  doi = {10.48550/arXiv.2507.23765},
  urldate = {2025-12-25},
  archiveprefix = {arXiv}
}

@misc{kawashimaZeroTemperature1999,
  title = {Zero-{{Temperature Critical Phenomena}} in {{Two-Dimensional Spin Glasses}}},
  author = {Kawashima, Naoki and Aoki, Takayuki},
  year = 1999,
  month = nov,
  number = {arXiv:cond-mat/9911120},
  eprint = {cond-mat/9911120},
  publisher = {arXiv},
  doi = {10.48550/arXiv.cond-mat/9911120},
  urldate = {2025-12-17},
  archiveprefix = {arXiv}
}

@article{kesselringAnyon2024,
  title = {Anyon {{Condensation}} and the {{Color Code}}},
  author = {Kesselring, Markus S. and Magdalena De La Fuente, Julio C. and Thomsen, Felix and Eisert, Jens and Bartlett, Stephen D. and Brown, Benjamin J.},
  year = 2024,
  month = mar,
  journal = {PRX Quantum},
  volume = {5},
  number = {1},
  eprint = {2212.00042},
  primaryclass = {cond-mat, physics:quant-ph},
  pages = {010342},
  issn = {2691-3399},
  doi = {10.1103/PRXQuantum.5.010342},
  urldate = {2024-05-13},
  archiveprefix = {arXiv},
  langid = {english}
}

@misc{kishonyIncreasing2025,
  title = {Increasing the Distance of Topological Codes with Time Vortex Defects},
  author = {Kishony, Gilad and Berg, Erez},
  year = 2025,
  month = feb,
  number = {arXiv:2502.12236},
  eprint = {2502.12236},
  primaryclass = {quant-ph},
  publisher = {arXiv},
  doi = {10.48550/arXiv.2502.12236},
  urldate = {2025-02-19},
  archiveprefix = {arXiv}
}

@article{kitaevFaulttolerant2003,
  title = {Fault-Tolerant Quantum Computation by Anyons},
  author = {Kitaev, A. Yu},
  year = 2003,
  month = jan,
  journal = {Annals of Physics},
  volume = {303},
  number = {1},
  eprint = {quant-ph/9707021},
  pages = {2--30},
  issn = {00034916},
  doi = {10.1016/S0003-4916(02)00018-0},
  urldate = {2022-02-10},
  archiveprefix = {arXiv}
}

@incollection{kitaevQuantum1997,
  title = {Quantum {{Error Correction}} with {{Imperfect Gates}}},
  booktitle = {Quantum {{Communication}}, {{Computing}}, and {{Measurement}}},
  author = {Kitaev, A. {\relax Yu}.},
  editor = {Hirota, O. and Holevo, A. S. and Caves, C. M.},
  year = 1997,
  pages = {181--188},
  publisher = {Springer US},
  address = {Boston, MA},
  doi = {10.1007/978-1-4615-5923-8_19},
  urldate = {2022-08-01},
  isbn = {978-1-4613-7716-0},
  langid = {english}
}

@article{knillResilient1998,
  title = {Resilient {{Quantum Computation}}},
  author = {Knill, Emanuel and Laflamme, Raymond and Zurek, Wojciech H.},
  year = 1998,
  month = jan,
  journal = {Science},
  volume = {279},
  number = {5349},
  pages = {342--345},
  issn = {0036-8075, 1095-9203},
  doi = {10.1126/science.279.5349.342},
  urldate = {2025-06-11},
  langid = {english}
}

@article{kubicaThreeDimensional2018,
  title = {Three-{{Dimensional Color Code Thresholds}} via {{Statistical-Mechanical Mapping}}},
  author = {Kubica, Aleksander and Beverland, Michael E. and Brand{\~a}o, Fernando and Preskill, John and Svore, Krysta M.},
  year = 2018,
  month = may,
  journal = {Physical Review Letters},
  volume = {120},
  number = {18},
  pages = {180501},
  issn = {0031-9007, 1079-7114},
  doi = {10.1103/PhysRevLett.120.180501},
  urldate = {2025-03-20},
  langid = {english}
}

@article{lavasaniStability2025,
  title = {Stability of Gapped Quantum Matter and Error-Correction with Adiabatic Noise},
  author = {Lavasani, Ali and Vijay, Sagar},
  year = 2025,
  month = may,
  journal = {Physical Review Research},
  volume = {7},
  number = {2},
  pages = {023166},
  issn = {2643-1564},
  doi = {10.1103/PhysRevResearch.7.023166},
  urldate = {2025-12-25},
  langid = {english}
}

@article{leeMixedState2025,
  title = {Mixed-{{State Topological Order}} under {{Coherent Noise}}},
  author = {Lee, Seunghun and Moon, Eun-Gook},
  year = 2025,
  month = sep,
  journal = {PRX Quantum},
  volume = {6},
  number = {3},
  pages = {030355},
  issn = {2691-3399},
  doi = {10.1103/hchr-rqq9},
  urldate = {2025-12-12},
  langid = {english}
}

@article{lessaStrongtoWeak2025,
  title = {Strong-to-{{Weak Spontaneous Symmetry Breaking}} in {{Mixed Quantum States}}},
  author = {Lessa, Leonardo A. and Ma, Ruochen and Zhang, Jian-Hao and Bi, Zhen and Cheng, Meng and Wang, Chong},
  year = 2025,
  month = mar,
  journal = {PRX Quantum},
  volume = {6},
  number = {1},
  pages = {010344},
  issn = {2691-3399},
  doi = {10.1103/PRXQuantum.6.010344},
  urldate = {2025-12-23},
  langid = {english}
}

@article{liPerturbative2025,
  title = {Perturbative {{Stability}} and {{Error-Correction Thresholds}} of {{Quantum Codes}}},
  author = {Li, Yaodong and O'Dea, Nicholas and Khemani, Vedika},
  year = 2025,
  month = feb,
  journal = {PRX Quantum},
  volume = {6},
  number = {1},
  pages = {010327},
  issn = {2691-3399},
  doi = {10.1103/PRXQuantum.6.010327},
  urldate = {2025-08-07},
  langid = {english}
}

@misc{liReplica2024,
  title = {Replica Topological Order in Quantum Mixed States and Quantum Error Correction},
  author = {Li, Zhuan and Mong, Roger S. K.},
  year = 2024,
  month = feb,
  number = {arXiv:2402.09516},
  eprint = {2402.09516},
  primaryclass = {quant-ph},
  publisher = {arXiv},
  doi = {10.48550/arXiv.2402.09516},
  urldate = {2025-03-31},
  archiveprefix = {arXiv}
}

@misc{liuCoherent2025,
  title = {Coherent Error Induced Phase Transition},
  author = {Liu, Hanchen and Chen, Xiao},
  year = 2025,
  month = may,
  number = {arXiv:2506.00650},
  eprint = {2506.00650},
  primaryclass = {quant-ph},
  publisher = {arXiv},
  doi = {10.48550/arXiv.2506.00650},
  urldate = {2025-12-25},
  archiveprefix = {arXiv}
}

@article{luoTopological2025,
  title = {Topological {{Holography}} for {{Mixed-State Phases}} and {{Phase Transitions}}},
  author = {Luo, Ran and Wang, Yi-Nan and Bi, Zhen},
  year = 2025,
  month = dec,
  journal = {PRX Quantum},
  volume = {6},
  number = {4},
  pages = {040358},
  issn = {2691-3399},
  doi = {10.1103/9kmh-gjf8},
  urldate = {2025-12-23},
  langid = {english}
}

@misc{lyonsUnderstanding2024,
  title = {Understanding {{Stabilizer Codes Under Local Decoherence Through}} a {{General Statistical Mechanics Mapping}}},
  author = {Lyons, Anasuya},
  year = 2024,
  month = mar,
  number = {arXiv:2403.03955},
  eprint = {2403.03955},
  primaryclass = {quant-ph},
  doi = {10.48550/arXiv.2403.03955},
  urldate = {2025-02-26},
  archiveprefix = {arXiv}
}

@misc{martielLowoverhead2025,
  title = {Low-Overhead Error Detection with Spacetime Codes},
  author = {Martiel, Simon and {Javadi-Abhari}, Ali},
  year = 2025,
  month = apr,
  number = {arXiv:2504.15725},
  eprint = {2504.15725},
  primaryclass = {quant-ph},
  publisher = {arXiv},
  doi = {10.48550/arXiv.2504.15725},
  urldate = {2025-04-24},
  archiveprefix = {arXiv}
}

@article{martonLattice2025,
  title = {Lattice Surgery-Based Logical State Teleportation via Noisy Links},
  author = {M{\'a}rton, {\'A}ron and Colmenarez, Luis and B{\"o}deker, Lukas and M{\"u}ller, Markus},
  year = 2025,
  month = sep,
  journal = {Physical Review Research},
  volume = {7},
  number = {3},
  pages = {033238},
  issn = {2643-1564},
  doi = {10.1103/ppng-vbqj},
  urldate = {2025-10-07},
  langid = {english}
}

@article{mcewenRelaxing2023,
  title = {Relaxing {{Hardware Requirements}} for {{Surface Code Circuits}} Using {{Time-dynamics}}},
  author = {McEwen, Matt and Bacon, Dave and Gidney, Craig},
  year = 2023,
  month = nov,
  journal = {Quantum},
  volume = {7},
  pages = {1172},
  issn = {2521-327X},
  doi = {10.22331/q-2023-11-07-1172},
  urldate = {2024-12-20},
  langid = {english}
}

@misc{mclauchlanAccommodating2024,
  title = {Accommodating {{Fabrication Defects}} on {{Floquet Codes}} with {{Minimal Hardware Requirements}}},
  author = {McLauchlan, Campbell and Geh{\'e}r, Gy{\"o}rgy P. and Moylett, Alexandra E.},
  year = 2024,
  month = jun,
  number = {arXiv:2405.15854},
  eprint = {2405.15854},
  primaryclass = {quant-ph},
  publisher = {arXiv},
  urldate = {2024-07-29},
  archiveprefix = {arXiv}
}

@article{merzTwodimensional2002,
  title = {Two-Dimensional Random-Bond {{Ising}} Model, Free Fermions, and the Network Model},
  author = {Merz, F. and Chalker, J. T.},
  year = 2002,
  month = jan,
  journal = {Physical Review B},
  volume = {65},
  number = {5},
  pages = {054425},
  issn = {0163-1829, 1095-3795},
  doi = {10.1103/PhysRevB.65.054425},
  urldate = {2025-12-13},
  copyright = {http://link.aps.org/licenses/aps-default-license},
  langid = {english}
}

@misc{motamarriSymTFT2024,
  title = {{{SymTFT}} out of Equilibrium: From Time Crystals to Braided Drives and {{Floquet}} Codes},
  shorttitle = {{{SymTFT}} out of Equilibrium},
  author = {Motamarri, Vedant and McLauchlan, Campbell and B{\'e}ri, Benjamin},
  year = 2024,
  month = feb,
  number = {arXiv:2312.17176},
  eprint = {2312.17176},
  primaryclass = {cond-mat, physics:hep-th, physics:quant-ph},
  publisher = {arXiv},
  urldate = {2024-03-13},
  archiveprefix = {arXiv}
}

@misc{moylettLogical2025,
  title = {Logical Gates on {{Floquet}} Codes via Folds and Twists},
  author = {Moylett, Alexandra E. and Jonnadula, Bhargavi},
  year = 2025,
  month = dec,
  number = {arXiv:2512.17999},
  eprint = {2512.17999},
  primaryclass = {quant-ph},
  publisher = {arXiv},
  doi = {10.48550/arXiv.2512.17999},
  urldate = {2025-12-23},
  archiveprefix = {arXiv}
}

@misc{negariSpacetime2024,
  title = {Spacetime {{Markov}} Length: A Diagnostic for Fault Tolerance via Mixed-State Phases},
  shorttitle = {Spacetime {{Markov}} Length},
  author = {Negari, Amir-Reza and Ellison, Tyler D. and Hsieh, Timothy H.},
  year = 2024,
  month = nov,
  number = {arXiv:2412.00193},
  eprint = {2412.00193},
  primaryclass = {quant-ph},
  publisher = {arXiv},
  doi = {10.48550/arXiv.2412.00193},
  urldate = {2025-04-03},
  archiveprefix = {arXiv}
}

@article{ohzekiError2012,
  title = {Error Threshold Estimates for Surface Code with Loss of Qubits},
  author = {Ohzeki, Masayuki},
  year = 2012,
  month = jun,
  journal = {Physical Review A},
  volume = {85},
  number = {6},
  pages = {060301},
  issn = {1050-2947, 1094-1622},
  doi = {10.1103/PhysRevA.85.060301},
  urldate = {2025-12-17},
  copyright = {http://link.aps.org/licenses/aps-default-license},
  langid = {english}
}

@misc{pesahFaulttolerant2025,
  title = {Fault-Tolerant Transformations of Spacetime Codes},
  author = {Pesah, Arthur and Daniel, Austin K. and Tzitrin, Ilan and Vasmer, Michael},
  year = 2025,
  month = sep,
  number = {arXiv:2509.09603},
  eprint = {2509.09603},
  primaryclass = {quant-ph},
  publisher = {arXiv},
  doi = {10.48550/arXiv.2509.09603},
  urldate = {2025-12-23},
  archiveprefix = {arXiv}
}

@article{poulinStabilizer2005,
  title = {Stabilizer {{Formalism}} for {{Operator Quantum Error Correction}}},
  author = {Poulin, David},
  year = 2005,
  month = dec,
  journal = {Physical Review Letters},
  volume = {95},
  number = {23},
  eprint = {quant-ph/0508131},
  pages = {230504},
  issn = {0031-9007, 1079-7114},
  doi = {10.1103/PhysRevLett.95.230504},
  urldate = {2023-10-11},
  archiveprefix = {arXiv}
}

@inbook{preskillFaultTolerant1998,
  title = {Fault-{{Tolerant Quantum Computation}}},
  booktitle = {Introduction to {{Quantum Computation}} and {{Information}}},
  author = {Preskill, John},
  year = 1998,
  month = oct,
  pages = {213--269},
  publisher = {WORLD SCIENTIFIC},
  doi = {10.1142/9789812385253_0008},
  urldate = {2023-07-27},
  collaborator = {Lo, Hoi-Kwong and Spiller, Tim and Popescu, Sandu},
  isbn = {978-981-02-3399-0 978-981-238-525-3},
  langid = {english}
}

@misc{putzFlow2025,
  title = {Flow to {{Nishimori}} Universality in Weakly Monitored Quantum Circuits with Qubit Loss},
  author = {P{\"u}tz, Malte and Vasseur, Romain and Ludwig, Andreas W. W. and Trebst, Simon and Zhu, Guo-Yi},
  year = 2025,
  month = may,
  number = {arXiv:2505.22720},
  eprint = {2505.22720},
  primaryclass = {quant-ph},
  publisher = {arXiv},
  doi = {10.48550/arXiv.2505.22720},
  urldate = {2025-12-25},
  archiveprefix = {arXiv}
}

@article{rakovszkyDefining2024,
  title = {Defining {{Stable Phases}} of {{Open Quantum Systems}}},
  author = {Rakovszky, Tibor and Gopalakrishnan, Sarang and Von Keyserlingk, Curt},
  year = 2024,
  month = nov,
  journal = {Physical Review X},
  volume = {14},
  number = {4},
  pages = {041031},
  issn = {2160-3308},
  doi = {10.1103/PhysRevX.14.041031},
  urldate = {2025-12-23},
  langid = {english}
}

@misc{rodatzFloquetifying2024,
  title = {Floquetifying Stabiliser Codes with Distance-Preserving Rewrites},
  author = {Rodatz, Benjamin and Po{\'o}r, Boldizs{\'a}r and Kissinger, Aleks},
  year = 2024,
  month = dec,
  number = {arXiv:2410.17240},
  eprint = {2410.17240},
  primaryclass = {quant-ph},
  publisher = {arXiv},
  doi = {10.48550/arXiv.2410.17240},
  urldate = {2025-07-03},
  archiveprefix = {arXiv}
}

@misc{ruschCompleteness2025,
  title = {Completeness for {{Fault Equivalence}} of {{Clifford ZX Diagrams}}},
  author = {R{\"u}sch, Maximilian and Rodatz, Benjamin and Kissinger, Aleks},
  year = 2025,
  month = nov,
  number = {arXiv:2510.08477},
  eprint = {2510.08477},
  primaryclass = {quant-ph},
  publisher = {arXiv},
  doi = {10.48550/arXiv.2510.08477},
  urldate = {2025-12-22},
  archiveprefix = {arXiv}
}

@article{sahayError2025,
  title = {Error {{Correction}} of {{Transversal}} Cnot {{Gates}} for {{Scalable Surface-Code Computation}}},
  author = {Sahay, Kaavya and Lin, Yingjia and Huang, Shilin and Brown, Kenneth R. and Puri, Shruti},
  year = 2025,
  month = may,
  journal = {PRX Quantum},
  volume = {6},
  number = {2},
  pages = {020326},
  issn = {2691-3399},
  doi = {10.1103/PRXQuantum.6.020326},
  urldate = {2025-12-21},
  langid = {english}
}

@article{sangMixedState2024,
  title = {Mixed-{{State Quantum Phases}}: {{Renormalization}} and {{Quantum Error Correction}}},
  shorttitle = {Mixed-{{State Quantum Phases}}},
  author = {Sang, Shengqi and Zou, Yijian and Hsieh, Timothy H.},
  year = 2024,
  month = sep,
  journal = {Physical Review X},
  volume = {14},
  number = {3},
  pages = {031044},
  issn = {2160-3308},
  doi = {10.1103/PhysRevX.14.031044},
  urldate = {2025-03-31},
  langid = {english}
}

@article{sangStability2025,
  title = {Stability of {{Mixed-State Quantum Phases}} via {{Finite Markov Length}}},
  author = {Sang, Shengqi and Hsieh, Timothy H.},
  year = 2025,
  month = feb,
  journal = {Physical Review Letters},
  volume = {134},
  number = {7},
  pages = {070403},
  issn = {0031-9007, 1079-7114},
  doi = {10.1103/PhysRevLett.134.070403},
  urldate = {2025-12-21},
  langid = {english}
}

@misc{serra-peraltaDecoding2025,
  title = {Decoding across Transversal {{Clifford}} Gates in the Surface Code},
  author = {{Serra-Peralta}, Marc and Shaw, Mackenzie H. and Terhal, Barbara M.},
  year = 2025,
  month = oct,
  number = {arXiv:2505.13599},
  eprint = {2505.13599},
  primaryclass = {quant-ph},
  publisher = {arXiv},
  doi = {10.48550/arXiv.2505.13599},
  urldate = {2025-12-21},
  archiveprefix = {arXiv}
}

@article{setiawanTailoring2025,
  title = {Tailoring Dynamical Codes for Biased Noise: The {{X3Z3 Floquet}} Code},
  shorttitle = {Tailoring Dynamical Codes for Biased Noise},
  author = {Setiawan, F. and McLauchlan, Campbell},
  year = 2025,
  month = sep,
  journal = {npj Quantum Information},
  volume = {11},
  number = {1},
  pages = {149},
  issn = {2056-6387},
  doi = {10.1038/s41534-025-01074-1},
  urldate = {2025-12-21},
  langid = {english}
}

@article{shawLowering2025,
  title = {Lowering {{Connectivity Requirements}} for {{Bivariate Bicycle Codes Using Morphing Circuits}}},
  author = {Shaw, Mackenzie H. and Terhal, Barbara M.},
  year = 2025,
  month = mar,
  journal = {Physical Review Letters},
  volume = {134},
  number = {9},
  pages = {090602},
  issn = {0031-9007, 1079-7114},
  doi = {10.1103/PhysRevLett.134.090602},
  urldate = {2025-12-23},
  langid = {english}
}

@misc{shorFaulttolerant1997,
  title = {Fault-Tolerant Quantum Computation},
  author = {Shor, Peter W.},
  year = 1997,
  month = mar,
  number = {arXiv:quant-ph/9605011},
  eprint = {quant-ph/9605011},
  publisher = {arXiv},
  urldate = {2023-07-27},
  archiveprefix = {arXiv}
}

@article{sohalNoisy2025,
  title = {Noisy {{Approach}} to {{Intrinsically Mixed-State Topological Order}}},
  author = {Sohal, Ramanjit and Prem, Abhinav},
  year = 2025,
  month = jan,
  journal = {PRX Quantum},
  volume = {6},
  number = {1},
  pages = {010313},
  issn = {2691-3399},
  doi = {10.1103/PRXQuantum.6.010313},
  urldate = {2025-03-31},
  langid = {english}
}

@article{sommersDynamically2025,
  title = {Dynamically Generated Concatenated Codes and Their Phase Diagrams},
  author = {Sommers, Grace M. and Huse, David A. and Gullans, Michael J.},
  year = 2025,
  month = apr,
  journal = {Physical Review Research},
  volume = {7},
  number = {2},
  pages = {023086},
  issn = {2643-1564},
  doi = {10.1103/PhysRevResearch.7.023086},
  urldate = {2025-08-01},
  langid = {english}
}

@article{steaneError1996,
  title = {Error {{Correcting Codes}} in {{Quantum Theory}}},
  author = {Steane, A. M.},
  year = 1996,
  month = jul,
  journal = {Physical Review Letters},
  volume = {77},
  number = {5},
  pages = {793--797},
  issn = {0031-9007, 1079-7114},
  doi = {10.1103/PhysRevLett.77.793},
  urldate = {2022-09-13},
  langid = {english}
}

@article{strikisQuantum2023,
  title = {Quantum {{Computing}} Is {{Scalable}} on a {{Planar Array}} of {{Qubits}} with {{Fabrication Defects}}},
  author = {Strikis, Armands and Benjamin, Simon C. and Brown, Benjamin J.},
  year = 2023,
  month = jun,
  journal = {Physical Review Applied},
  volume = {19},
  number = {6},
  pages = {064081},
  issn = {2331-7019},
  doi = {10.1103/PhysRevApplied.19.064081},
  urldate = {2025-08-12},
  langid = {english}
}

@misc{sutcliffeDistributed2025,
  title = {Distributed Quantum Error Correction Based on Hyperbolic {{Floquet}} Codes},
  author = {Sutcliffe, Evan and Jonnadula, Bhargavi and Gall, Claire Le and Moylett, Alexandra E. and Westoby, Coral M.},
  year = 2025,
  month = jan,
  number = {arXiv:2501.14029},
  eprint = {2501.14029},
  primaryclass = {quant-ph},
  publisher = {arXiv},
  doi = {10.48550/arXiv.2501.14029},
  urldate = {2025-01-27},
  archiveprefix = {arXiv}
}

@misc{takouEstimating2025,
  title = {Estimating and Decoding Coherent Errors of {{QEC}} Experiments with Detector Error Models},
  author = {Takou, Evangelia and Brown, Kenneth R.},
  year = 2025,
  month = oct,
  number = {arXiv:2510.23797},
  eprint = {2510.23797},
  primaryclass = {quant-ph},
  publisher = {arXiv},
  doi = {10.48550/arXiv.2510.23797},
  urldate = {2025-12-25},
  archiveprefix = {arXiv}
}

@misc{tanggaraSimple2024,
  title = {Simple {{Construction}} of {{Qudit Floquet Codes}} on a {{Family}} of {{Lattices}}},
  author = {Tanggara, Andrew and Gu, Mile and Bharti, Kishor},
  year = 2024,
  month = oct,
  number = {arXiv:2410.02022},
  eprint = {2410.02022},
  primaryclass = {quant-ph},
  publisher = {arXiv},
  urldate = {2024-10-08},
  archiveprefix = {arXiv}
}

@misc{tanggaraStrategic2024,
  title = {Strategic {{Code}}: {{A Unified Spatio-Temporal Framework}} for {{Quantum Error-Correction}}},
  shorttitle = {Strategic {{Code}}},
  author = {Tanggara, Andrew and Gu, Mile and Bharti, Kishor},
  year = 2024,
  month = may,
  number = {arXiv:2405.17567},
  eprint = {2405.17567},
  primaryclass = {quant-ph},
  publisher = {arXiv},
  doi = {10.48550/arXiv.2405.17567},
  urldate = {2024-12-08},
  archiveprefix = {arXiv}
}

@misc{tangPhases2025,
  title = {Phases of {{Floquet}} Code under Local Decoherence},
  author = {Tang, Yuchen and Bao, Yimu},
  year = 2025,
  month = apr,
  number = {arXiv:2504.19041},
  eprint = {2504.19041},
  primaryclass = {quant-ph},
  publisher = {arXiv},
  doi = {10.48550/arXiv.2504.19041},
  urldate = {2025-04-29},
  archiveprefix = {arXiv}
}

@article{vennCoherentError2023,
  title = {Coherent-{{Error Threshold}} for {{Surface Codes}} from {{Majorana Delocalization}}},
  author = {Venn, Florian and Behrends, Jan and B{\'e}ri, Benjamin},
  year = 2023,
  month = aug,
  journal = {Physical Review Letters},
  volume = {131},
  number = {6},
  pages = {060603},
  issn = {0031-9007, 1079-7114},
  doi = {10.1103/PhysRevLett.131.060603},
  urldate = {2024-03-14},
  langid = {english}
}

@article{vodolaFundamental2022,
  title = {Fundamental Thresholds of Realistic Quantum Error Correction Circuits from Classical Spin Models},
  author = {Vodola, Davide and Rispler, Manuel and Kim, Seyong and M{\"u}ller, Markus},
  year = 2022,
  month = jan,
  journal = {Quantum},
  volume = {6},
  pages = {618},
  issn = {2521-327X},
  doi = {10.22331/q-2022-01-05-618},
  urldate = {2025-08-18},
  langid = {english}
}

@article{voterMonte1985,
  title = {A {{Monte Carlo}} Method for Determining Free-Energy Differences and Transition State Theory Rate Constants},
  author = {Voter, Arthur F.},
  year = 1985,
  month = feb,
  journal = {The Journal of Chemical Physics},
  volume = {82},
  number = {4},
  pages = {1890--1899},
  issn = {0021-9606, 1089-7690},
  doi = {10.1063/1.448373},
  urldate = {2025-10-02},
  langid = {english}
}

@article{vuStable2024,
  title = {Stable {{Measurement-Induced Floquet Enriched Topological Order}}},
  author = {Vu, DinhDuy and Lavasani, Ali and Lee, Jong Yeon and Fisher, Matthew P. A.},
  year = 2024,
  month = feb,
  journal = {Physical Review Letters},
  volume = {132},
  number = {7},
  eprint = {2303.01533},
  primaryclass = {cond-mat, physics:quant-ph},
  pages = {070401},
  issn = {0031-9007, 1079-7114},
  doi = {10.1103/PhysRevLett.132.070401},
  urldate = {2024-05-13},
  archiveprefix = {arXiv},
  langid = {english}
}

@article{wangConfinementHiggs2003,
  title = {Confinement-{{Higgs}} Transition in a Disordered Gauge Theory and the Accuracy Threshold for Quantum Memory},
  author = {Wang, Chenyang and Harrington, Jim and Preskill, John},
  year = 2003,
  month = jan,
  journal = {Annals of Physics},
  volume = {303},
  number = {1},
  pages = {31--58},
  issn = {00034916},
  doi = {10.1016/S0003-4916(02)00019-2},
  urldate = {2025-11-05},
  copyright = {https://www.elsevier.com/tdm/userlicense/1.0/},
  langid = {english}
}

@misc{wangDecoherenceinduced2025,
  title = {Decoherence-Induced Self-Dual Criticality in Topological States of Matter},
  author = {Wang, Qingyuan and Vasseur, Romain and Trebst, Simon and Ludwig, Andreas W. W. and Zhu, Guo-Yi},
  year = 2025,
  month = mar,
  number = {arXiv:2502.14034},
  eprint = {2502.14034},
  primaryclass = {quant-ph},
  publisher = {arXiv},
  doi = {10.48550/arXiv.2502.14034},
  urldate = {2025-12-25},
  archiveprefix = {arXiv}
}

@misc{weteringZXcalculus2020,
  title = {{{ZX-calculus}} for the Working Quantum Computer Scientist},
  author = {van de Wetering, John},
  year = 2020,
  month = dec,
  number = {arXiv:2012.13966},
  eprint = {2012.13966},
  primaryclass = {quant-ph},
  publisher = {arXiv},
  doi = {10.48550/arXiv.2012.13966},
  urldate = {2025-04-18},
  archiveprefix = {arXiv}
}

@misc{wichettePartition2025,
  title = {A Partition Function Framework for Estimating Logical Error Curves in Stabilizer Codes},
  author = {Wichette, Leon and Hohenfeld, Hans and Mounzer, Elie and {Grans-Samuelsson}, Linnea},
  year = 2025,
  month = jul,
  number = {arXiv:2505.15758},
  eprint = {2505.15758},
  primaryclass = {quant-ph},
  publisher = {arXiv},
  doi = {10.48550/arXiv.2505.15758},
  urldate = {2025-10-07},
  archiveprefix = {arXiv}
}

@misc{williamsonDynamical2025,
  title = {Dynamical Quantum Codes and Logic Gates on a Lattice with Sparse Connectivity},
  author = {Williamson, Dominic J. and Het{\'e}nyi, Bence},
  year = 2025,
  month = oct,
  number = {arXiv:2510.05225},
  eprint = {2510.05225},
  primaryclass = {quant-ph},
  publisher = {arXiv},
  doi = {10.48550/arXiv.2510.05225},
  urldate = {2025-10-08},
  archiveprefix = {arXiv}
}

@misc{wolanskiAutomated2025,
  title = {Automated {{Compilation Including Dropouts}}: {{Tolerating Defective Components}} in {{Stabiliser Codes}}},
  shorttitle = {Automated {{Compilation Including Dropouts}}},
  author = {Wolanski, Stasiu},
  year = 2025,
  month = dec,
  number = {arXiv:2512.01943},
  eprint = {2512.01943},
  primaryclass = {quant-ph},
  publisher = {arXiv},
  doi = {10.48550/arXiv.2512.01943},
  urldate = {2025-12-07},
  archiveprefix = {arXiv}
}

@misc{xuError2025,
  title = {Error Thresholds of Toric Codes with Transversal Logical Gates},
  author = {Xu, Yichen and Zhou, Yiqing and Sethna, James P. and Kim, Eun-Ah},
  year = 2025,
  month = oct,
  number = {arXiv:2510.10835},
  eprint = {2510.10835},
  primaryclass = {quant-ph},
  publisher = {arXiv},
  doi = {10.48550/arXiv.2510.10835},
  urldate = {2025-10-14},
  archiveprefix = {arXiv}
}

@misc{xuFaulttolerant2025,
  title = {Fault-Tolerant Protocols through Spacetime Concatenation},
  author = {Xu, Yichen and Dua, Arpit},
  year = 2025,
  month = apr,
  number = {arXiv:2504.08918},
  eprint = {2504.08918},
  primaryclass = {quant-ph},
  publisher = {arXiv},
  doi = {10.48550/arXiv.2504.08918},
  urldate = {2025-04-15},
  archiveprefix = {arXiv}
}

@misc{xuPhenomenological2025,
  title = {Phenomenological {{Noise Models}} and {{Optimal Thresholds}} of the {{3D Toric Code}}},
  author = {Xu, Ji-Ze and Zhong, Yin and {Martin-Delgado}, Miguel A. and Song, Hao and Liu, Ke},
  year = 2025,
  month = oct,
  number = {arXiv:2510.20489},
  eprint = {2510.20489},
  primaryclass = {quant-ph},
  publisher = {arXiv},
  doi = {10.48550/arXiv.2510.20489},
  urldate = {2025-12-25},
  archiveprefix = {arXiv}
}

@article{yadavalliNoisy2025,
  title = {Noisy Quantum Trees: Infinite Protection without Correction},
  shorttitle = {Noisy Quantum Trees},
  author = {Yadavalli, Shiv Akshar and Marvian, Iman},
  year = 2025,
  month = sep,
  journal = {npj Quantum Information},
  volume = {11},
  number = {1},
  pages = {151},
  issn = {2056-6387},
  doi = {10.1038/s41534-025-00961-x},
  urldate = {2026-01-30},
  langid = {english}
}

@article{zhangXcube2023,
  title = {X-Cube {{Floquet}} Code: {{A}} Dynamical Quantum Error Correcting Code with a Subextensive Number of Logical Qubits},
  shorttitle = {X -Cube {{Floquet}} Code},
  author = {Zhang, Zhehao and Aasen, David and Vijay, Sagar},
  year = 2023,
  month = nov,
  journal = {Physical Review B},
  volume = {108},
  number = {20},
  pages = {205116},
  issn = {2469-9950, 2469-9969},
  doi = {10.1103/PhysRevB.108.205116},
  urldate = {2024-08-28},
  langid = {english}
}

@article{zhuNishimoris2023,
  title = {Nishimori's {{Cat}}: {{Stable Long-Range Entanglement}} from {{Finite-Depth Unitaries}} and {{Weak Measurements}}},
  shorttitle = {Nishimori's {{Cat}}},
  author = {Zhu, Guo-Yi and Tantivasadakarn, Nathanan and Vishwanath, Ashvin and Trebst, Simon and Verresen, Ruben},
  year = 2023,
  month = nov,
  journal = {Physical Review Letters},
  volume = {131},
  number = {20},
  pages = {200201},
  issn = {0031-9007, 1079-7114},
  doi = {10.1103/PhysRevLett.131.200201},
  urldate = {2025-02-06},
  langid = {english}
}

@misc{zhuQubit2023,
  title = {Qubit Fractionalization and Emergent {{Majorana}} Liquid in the Honeycomb {{Floquet}} Code Induced by Coherent Errors and Weak Measurements},
  author = {Zhu, Guo-Yi and Trebst, Simon},
  year = 2023,
  month = nov,
  number = {arXiv:2311.08450},
  eprint = {2311.08450},
  primaryclass = {cond-mat, physics:quant-ph},
  publisher = {arXiv},
  urldate = {2023-11-17},
  archiveprefix = {arXiv}
}

@article{zhuStructured2024,
  title = {Structured Volume-Law Entanglement in an Interacting, Monitored {{Majorana}} Spin Liquid},
  author = {Zhu, Guo-Yi and Tantivasadakarn, Nathanan and Trebst, Simon},
  year = 2024,
  month = dec,
  journal = {Physical Review Research},
  volume = {6},
  number = {4},
  pages = {L042063},
  issn = {2643-1564},
  doi = {10.1103/PhysRevResearch.6.L042063},
  urldate = {2025-12-25},
  langid = {english}
}

\onecolumngrid
\appendix

\counterwithin{table}{section}
\counterwithin{figure}{section}

\section{Additional details}

\subsection{Spacetime codes}
\label{app:spacetimecodes}

In this section, we continue our discussion of spacetime codes.

\paragraph{Stabilizers and logical operators}
To define the spacetime stabilizers and logical operators, we first define
the propagator superoperator $\Pi_{\tau\rightarrow\tau'}$ that takes a
$[Q]_{,\tau}\in\mathcal P_\text{st}$ and multiplies it by gauge operators until
it has support only at half-integer time $\tau'$ (if possible\footnote{In the
absence of mid-circuit measurements, this is always possible by applying
Eq.~\eqref{eq:gauge_propagators}.}). That is, $\Pi_{\tau\rightarrow\tau'}[Q]
\equiv [Q']_{,\tau'}$ for some new $Q' \in \mathcal P_\text{ph}$.

Related to the propagator is the cumulant superoperator,
$\overrightarrow{\cdot}$, which creates a spacetime operator whose action on the
qubits at a half-integer time is the cumulative effect of propagating forward
all operators prior to that time.\footnote{Also known as the ``spackle''
	superoperator.} Specifically, the cumulant of a spacetime operator that has
nontrivial support on only one half-integer timestep is
defined as
\begin{equation}
	\overrightarrow{[Q]_{,\tau}} = \prod_{\tau' = \tau}^{T-0.5}
	\Pi_{\tau\rightarrow \tau'}[Q] \in \mathcal P_{\text{st}}.
\end{equation}
The cumulant is an automorphism of $\mathcal P_{\text{st}}$
\cite{delfosseSpacetime2023}; for $F,G \in \mathcal P_{\text{st}}$, it obeys
$\overrightarrow{FG} = \overrightarrow{F}\overrightarrow{G}$.
For a general $F \in \mathcal P_{\text{st}}$ that acts as $F_\tau \in \mathcal
	P_{\text{ph}}$ at each half-integer time $\tau$---that is, $F =
	\prod_{\tau=0.5}^{T-0.5} [F_{\tau}]_{,\tau}$---we thus have
\begin{equation}
	\overrightarrow F = \prod_{\tau=0.5}^{T-0.5}
	\overrightarrow{[F_\tau]_{,\tau}}.
\end{equation}
The cumulant defines the stabilizers and logical operators of the spacetime
code, as we shall describe now.

For a Pauli measurement $M_\xi\in\mathcal P_\text{ph}$ that occurs at $t_\xi$ in
the circuit, let $o(M_\xi)$ be its measurement outcome when the circuit is
executed in the absence of errors. Depending on the circuit, this outcome may
either be a deterministic value---that is, $M_\xi$ always gives the same outcome
because the system is in its eigenstate when measured---or non-deterministic. If
an outcome is non-deterministic, the product of multiple outcomes may still be
deterministic if the product of the associated $M_\xi$ is an eigenoperator. For
a set of measurement indices $\Xi_k = \{\xi\}$, if the product $\mathcal O_k =
	\prod_{\xi \in \Xi_k} o(M_\xi)$ is deterministic, observing a different outcome
indicates that an error occurred in the support of $\{M_\xi\}$. We use this
property to define the stabilizers of the spacetime code,
\begin{equation}
	S_k = \overrightarrow{M_{(k)}}\mathcal O_k,
\end{equation}
where
\begin{equation}
	M_{(k)} = \prod_{\xi\in\Xi_k} [M_\xi]_{,t_\xi + 0.5}.
\end{equation}
The multiplication by $\mathcal O_k$ ensures that the eigenvalue of $S_k$ is
$+1$ in the absence of errors, and the cumulant ensures that $S_k$ anticommutes
with all errors that would change the outcome of the measurements in $M_{(k)}$.

In a circuit that measures the syndromes of a QEC code, these $M_{(k)}$ take the
form of two consecutive measurements of a stabilizer. These are equivalent to
detector cells \cite{mcewenRelaxing2023, derksDesigning2024,
	gottesmanOpportunities2022}, which are generalizations of stabilizers to
circuits: pairs of stabilizer measurements that signal an error whenever an
outcome changes.
The logical operators of the spacetime code are then the cumulants of the
logical operators of the QEC code.
Specifically, if a logical operator is $L(t)$ at timestep $t$ of a circuit
(where the operator may evolve according to an automorphism in a Floquet code,
for example), then the spacetime logical operator is the membrane
\begin{equation}
	\overrightarrow L = \prod_{t=0}^{T-1} [L(t)]_{,t+0.5}.
	\label{eq:logical_membrane}
\end{equation}
If we have two logical operators $L_1, L_2$ such that $\com{L_1(t)}{L_2(t)} =
	-1$ for all $t$, then $\com{\overrightarrow L_1}{\overrightarrow L_2} = -1$ iff
$T$ is odd \cite{baconSparse2017}. If $T$ is even, then the
	logical operator in Eq.~\eqref{eq:logical_membrane} is in the gauge
	group and the code is trivial. In such a case, we may simply repeat the
	circuit to achieve correct anticommutation.

These logical membranes are the bare logical operators of the spacetime
subsystem code, in that they commute with all stabilizers and the gauge group,
but are not in the gauge group. The dressed logical operators typically take a
simpler structure. In particular, $[L(t_1)]_{,t_1+0.5}$ and
$[L(t_2)]_{,t_2+0.5}$ are gauge-equivalent, as the gauge operators encode
propagations in spacetime. If $T$ is odd, we thus have that $\overrightarrow L$
is gauge-equivalent to $[L(t)]_{,t+0.5}$ for any $t \in [0,T-1]$. That is, the
dressed spacetime logical operators include the original code's logical
operators on a single timeslice.

\paragraph{Subsystem codes}
\label{app:subsystemcodes}
When the Clifford circuit describes the syndrome measurement circuit of a
stabilizer code, the two types of gauge generators discussed in the main text do
indeed generate the spacetime gauge group. However, in a syndrome circuit of a
subsystem code there are additional gauge freedoms not captured. Specifically,
the gauge generators of the static subsystem code at each timestep need to be
determined by considering the stabilizers measured by the circuit; these will
add spacelike spacetime gauge operators \cite{fuSubsystem2025}. One can check if
all such operators have been added---or if they are necessary in the first place
given an unknown Clifford circuit--- by, for example, using Stim to search for
low-weight undetectable errors \cite{gidneyStim2021}.

\subsection{SM mapping}
\label{app:SM_mapping}

The SM model ultimately aims to relate the partition function of some
quenched-disordered classical random-bond model to the probability of a class of
logical errors. Specifically, define the logical coset
\begin{equation}
	\overline E = E\mathcal G
\end{equation}
for some Pauli error $E$, and $\mathcal G$ the gauge group (the following
analysis proceeds identically if we replace $\mathcal G$ by the stabilizer
group). We wish to compute the probability 
\begin{equation}
	\mathbb P(\overline E) = \sum_{\epsilon \in E\mathcal G} \mathbb P(\epsilon).
\end{equation}
If $\epsilon_i \in \mathcal P_i = \{I, X, Y, Z\}$ is a single-qubit Pauli error
on qubit $i$, then we can write its probability in the form
\begin{equation}
	\begin{aligned}
		\mathbb P(\epsilon_i)
		 & = \left[
			\left(\frac{\mathbb P(I)\mathbb P(X)}{\mathbb P(Y)\mathbb
				P(Z)}\right)^{\com{\epsilon_i}{X}}
			\left(\frac{\mathbb P(I)\mathbb P(Y)}{\mathbb P(X)\mathbb
				P(Z)}\right)^{\com{\epsilon_i}{Y}}
			\left(\frac{\mathbb P(I)\mathbb P(Z)}{\mathbb P(X)\mathbb
				P(Y)}\right)^{\com{\epsilon_i}{Z}}
			\mathbb P(I)\mathbb P(X)\mathbb P(Y)\mathbb P(Z)
		\right]^{1/4}                                                        \\
		 & = \exp\left[\frac14 \sum_{\alpha \in \mathcal P_i}
			\sum_{\tau \in \mathcal P_i} \ln \left[\mathbb P(\tau)\right]
		\com{\alpha}{\tau} \com{\epsilon_i}{\alpha} \right]                  \\
		 & \equiv \exp\left[\sum_{\alpha \in \mathcal P_i}  K_i(\alpha)
			\com{\epsilon_i}{\alpha}\right];\quad K_i(\alpha) = \frac14 \sum_{\tau
			\in \mathcal P_i} \ln [\mathbb P(\tau)] \com{\alpha}{\tau}.
	\end{aligned}
\end{equation}
For a generic multi-qubit error $\epsilon = \otimes_i \epsilon_i$, we therefore
have
\begin{equation}
	\mathbb P(\epsilon) = \prod_i \mathbb P(\epsilon_i) = \exp\left[ \sum_i
		\sum_{\alpha \in \mathcal P_i} K_i(\alpha)\com{\epsilon}{\alpha}\right]
		\equiv e^{-H_0},
\end{equation}
where
\begin{equation}
	H_0 = -\sum_i \sum_{\alpha \in \mathcal P_i} K_i(\alpha)
	\com{\epsilon}{\alpha}.
\end{equation}
Now, any errors $\epsilon, \epsilon' \in E\mathcal G$ are related by a product
of gauge operators, $\epsilon' = g\epsilon$, $g \in \mathcal G$. If $\mathcal G
= \braket{g_1, g_2,\ldots, g_m}$ is generated by $m$ gauge operators
$g_k$
then we can represent $E\mathcal G$ by $m$-bit binary vectors $\textsf b
\in \textsf B_m = \{0,1\}^m$, as
\begin{equation}
	\epsilon_{\textsf b} = E \prod_{k=1}^m g_k^{\textsf b_k}.
\end{equation}
For some Pauli $\alpha \in \mathcal P$, the scalar commutator acts on this
representation as
\begin{equation}
	\begin{aligned}
		\com{\epsilon_\textsf{b}}{\alpha}
		 & = \com{E}{\alpha}\prod_{k=1}^m \com{g_k}{\alpha}^{\textsf b_k} \\
		 & = \com{E}{\alpha}\prod_{k:\com{g_k}{\alpha} =-1}(-1)^{\textsf b_k}.
	\end{aligned}
\end{equation}
Therefore,
\begin{equation}
	\mathbb P(\overline E) = \sum_{\textsf b \in \textsf B_m} 
	e^{- H_E(\textsf b)},
\end{equation}
where
\begin{equation}
	H_E({\textsf b}) = -\sum_i \sum_{\alpha \in \mathcal P_i}K_i(\alpha)
	\com{E}{\alpha} \prod_{k : \com{g_k}{\alpha}=-1}(-1)^{\textsf b_k}.
\end{equation}
Identifying classical Ising spins $\sigma_k = (-1)^{\textsf b_k}$ thus brings us
to the Hamiltonian in Eq.~\eqref{eq:H_ES_gauge}.

\subsubsection{Discrete Walsh transform}
\label{app:walsh-hadamard}

In this section we detail how spins can be integrated out of the Hamiltonian in
order to simplify the model. We wish to integrate out spin $\sigma_0$ from a
Hamiltonian of the form 
\begin{equation}
	H = H_0 - \sigma_0 P(\bm\Sigma)
\end{equation}
where $H_0$ is independent of $\sigma_0$, and $P$ encodes the
$\sigma_0$-dependent interactions
\begin{equation}
	P(\bm\Sigma) = \sum_{i=1}^r K_i
    \Sigma_i
\end{equation}
where $\bm\Sigma = [\Sigma_1,\,\ldots,\Sigma_r] \in \{-1,1\}^r$ is a product of
spins and interactions signs. For example, if $r=1$ (spin $\sigma_0$ is
involved in only one interaction, i.e. it is degree-$1$), 
\begin{equation}
	\bm\Sigma = [\eta_1\sigma_1], \quad P(\bm\Sigma) = K_1\eta_1\sigma_1,
\end{equation}
or if $r=2$,
\begin{equation}
	\bm\Sigma
	= [\eta_1\sigma_1, \, \eta_2\sigma_2],\quad
	P(\bm\Sigma) = K_1\eta_1\sigma_1 + K_2\eta_2\sigma_2.
	\label{eq:P_sigma_example}
\end{equation}
Tracing out $\sigma_0$, we get an effective Hamiltonian
\begin{equation}
\begin{aligned}
	\sum_{\sigma_0 \in \{-1,1\}} e^{-H} 
	&= e^{- H_0} \left(e^{P(\bm\Sigma)} + e^{- P(\bm\Sigma)}\right) \\ 
	&= 2
    e^{-H_0} \cosh\left[P(\bm\Sigma)\right] \\ 
	&= e^{- H_0 + \ln\cosh\left[ P(\bm\Sigma)\right] + \ln 2} \\ 
	H_\text{eff} &= H_0 - f(\bm\Sigma)
\end{aligned}
\end{equation}
where
\begin{equation}
	f(\bm\Sigma) = \ln \cosh\left[ P(\bm\Sigma)\right],
\end{equation}
and we ignore the constant contribution. When $r=1$, $\cosh[P(\bm\Sigma)] =
\cosh|K_1\eta_1\sigma_1|$ is independent of $\eta_1\sigma=\pm1$ and we are left
with $H_\text{eff} = H_0$ up to constants. In this way, degree-$1$ spins and
their sole interaction term can be removed from the Hamiltonian without
affecting the partition function (up to an unimportant pre-factor).

For $r > 1$, we cannot remove $f(\bm\Sigma)$ and we instead aim to write
$f(\bm\Sigma)$ as a linear combination of $\Sigma_i$. To do so, we define Walsh
coefficients
\begin{equation}
	F(S) = \frac1{2^r} \sum_{\bm\theta \in \{-1,1\}^r} f(\bm\theta)
	\prod_{i \in S}\theta_i, \quad S \subseteq \{1,2,\ldots,r\}
\end{equation}
where $S$ is an index set that determines which components $\theta_i$ of
$\bm\theta$
are included. A summation over all non-empty $S$ recovers $f(\bm{\Sigma})$
(up to constant terms) \cite{berloffExact2025}
\begin{equation}
	H_\text{eff} = H_0 - \sum_{S \neq \emptyset} F(S) \prod_{i\in
	S}\Sigma_i.
\end{equation}
For example, taking $r=2$ and using $P$ from Eq.~\eqref{eq:P_sigma_example}, we
get 
\begin{align}
	f(\{-1,-1\}) &= f(\{1,1\}) = \ln\cosh(K_1+K_2), \\
	f(\{-1,1\}) &= f(\{1,-1\}) = \ln\cosh|K_1-K_2| \\ 
	F(\{1\}) &= F(\{2\}) = 0 \\ 
	F(\{1,2\}) &= \frac12 \ln\cosh(K_1+K_2) - \frac12 \ln\cosh|K_1-K_2|,
\end{align}
which results in an effective Hamiltonian 
\begin{equation}
	H_\text{eff} = H_0 - \frac1{2}
	\ln\left(\frac{\cosh(K_1+K_2)}{\cosh|K_1-K_2|}\right)
	\eta_1\eta_2 \sigma_1\sigma_2.
\end{equation}
Now, assume that 
\begin{equation}
	K_i = \frac12 \ln \left(\frac{1-p_i}{p_i}\right) \equiv \frac12 \ln
	\left(\frac{1+\pi_i}{1-\pi_i}\right),
	\end{equation}
for some $\pi_i \in (0,1)$ and $p_i = \frac12(1-\pi_i)$. Then,
\begin{align}
	\frac{\cosh(K_1+K_2)}{\cosh|K_1-K_2|} 
	&= \frac{e^{K_1}e^{K_2} + e^{-K_1}e^{-K_2}}{e^{K_1}e^{-K_2} + e^{-K_1}e^{K_2}}
	\\ 
	&= \frac{\sqrt{\frac{1+\pi_1}{1-\pi_1} \frac{1+\pi_2}{1-\pi_2}} +
		\sqrt{\frac{1-\pi_1}{1+\pi_1}\frac{1-\pi_2}{1+\pi_2}}}{\sqrt{\frac{1+\pi_1}{1-\pi_1}\frac{1-\pi_2}{1+\pi_2}}
	+ \sqrt{\frac{1-\pi_1}{1+\pi_1}\frac{1+\pi_2}{1-\pi_2}}} \\ 
	&= \frac{(1+\pi_1)(1+\pi_2) +
	(1-\pi_1)(1-\pi_2)}{(1+\pi_1)(1-\pi_2) +
	(1-\pi_1)(1+\pi_2)} \\ 
	&= \frac{1+\pi_1\pi_2}{1 -\pi_1\pi_2} \\ 
	&\equiv \frac{1-p_{1,2}}{p_{1,2}}, \qquad p_{1,2} =
	\frac12\left(1-\pi_1\pi_2\right).
\end{align}
If $\pi_i = (1-2p_X)^{x_i}(1-2p_Z)^{z_i}$ for some
non-negative integers $x_i,z_i$, then this recovers the effective probability
parameters from Eq.~\eqref{eq:peff}. Substituting this back into the effective
Hamiltonian brings us to the form of the effective interaction strength from
Eq.~\eqref{eq:Keff}.

When $r \geq 3$, the number of nonzero $F(S)$, and therefore the number of
interactions in the effective Hamiltonian, is generally larger than $r$. We thus
do not consider integrating out spins involved in more than two interactions;
this also limits this procedure to independent $X$-$Z$ noise channels.
Otherwise, contribution from $K(Y)$ terms increase the number of interactions
each spin undergoes, and restrict this procedure to single-qubit gauge
operators.

\subsection{Additional spin diagram simplifications}
\label{app:spin_diagrams}

In addition to the two simplification rules discussed in the main text, cf.
Fig~\ref{fig:spins_rules}, a third rule can also be described that involves
changing the basis of gauge generators. That is, for two gauge generators $g_1,
g_2$, the mapping $g_2 \mapsto g_2' = g_1g_2$ preserves $\mathcal G$.
Diagrammatically, we draw this by updating the interactions associated with the
spin for $g_2$:
\begin{center}
	\includegraphics{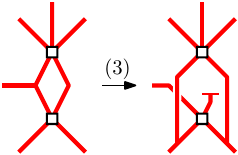}
\end{center}
This rule may be useful for re-arranging bonds to reveal simpler or more
informative spin diagrams. For example, using this rule we can investigate the
differences between a \texttt{SWAP} gate implemented natively (or equivalently,
as a phenomenological representation) as in Fig.~\ref{fig:spins}, and a
\texttt{SWAP} gate implemented via three \texttt{CNOT} gates:
\begin{center}
	\includegraphics{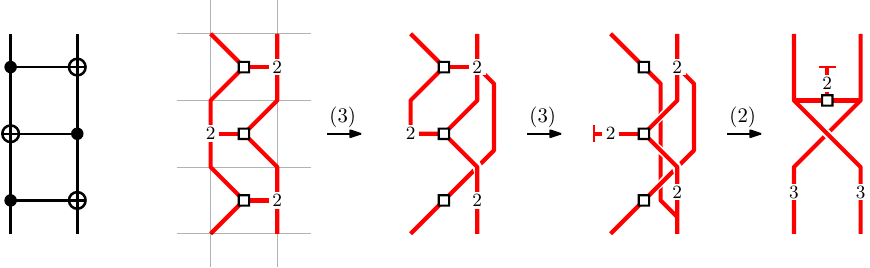}
\end{center}
(The simplifying rules used in each step are shown in parentheses.) Unlike the
native implementation, this spin diagram includes an additional spin and
interaction that highlights the potential for hook errors to spread between the
two wires. Spin diagrams, in conjunction with the three simplification rules,
thus serve as a useful tool for understanding the propagation of errors and the
differences between implementations of circuit operations.

\subsection{Additional figures}

\begin{figure}[h]
	\centering
	\includegraphics{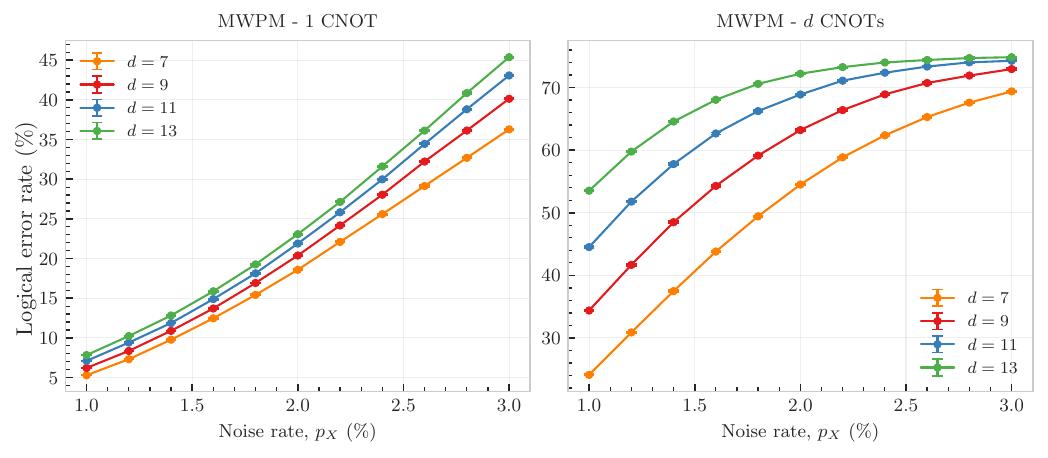}
	\caption{MWPM logical error rates for two repetition codes with standard
		syndrome extraction circuits. (Left) Subject to one transversal
		\texttt{CNOT}, with $N=4d-2$, $T=4d-1$. Unlike Fig.~\ref{fig:rep_cnot} with
		$T=16d-1$, no threshold is observed at this reduced circuit depth. 
		(Right) Subject to $d$ \texttt{CNOTs}, with $N=4d-2$ and $T=16d-1$. Unlike
	the ML decoder in Fig.~\ref{fig:rep_cnot_d}, the MWPM decoder has no threshold.}
	\label{fig:rep_cnot_appendix}
\end{figure}

\section{Numerical methods}
\label{app:methods}

\subsection{Monte Carlo}

Monte Carlo simulations are used to estimate the success (or failure)
probabilities of a maximum likelihood decoder for each quantum circuit. This
amounts to computing the free-energy differences between systems with quenched
disorders corresponding to distinct logical classes. By
Eq.~\eqref{eq:ML-decoding} and Eq.~\eqref{eq:free-energy}, the success
probability of an ML decoder is given by
\begin{equation}
	\mathbb P(\text{success}) =
	\left\langle{\frac{\max_L\{e^{-F_{C_\textsf{s}L}}\}}{\sum_L
			e^{-F_{C_\textsf{s}L}}}}\right\rangle_\eta 
			\equiv 
			\left\langle\frac{\max_L\{e^{-(F_{C_\textsf{s}L}-
				F_{C_\textsf{s}})}\}}{\sum_L
		e^{-(F_{C_\textsf{s}L} - F_{C_\textsf{s}})}}\right\rangle_\eta 
\end{equation}
where $\braket{\cdot}_\eta$ is the average over quenched disorders according to
the probability distribution of $\bm\eta$. The latter formula is used when
computing free-energy differences is preferable. Each free energy is computable
from
\begin{equation}
	F_{E} = -\ln \mathcal Z_E = -\ln \braket{e^{-H_E}}_\beta
\end{equation}
where $\mathcal Z_E$ is the partition function and $\braket{\cdot}_\beta$ is the
thermal average at a given quenched disorder.

We employ two approaches to estimate these free energies. 

\paragraph{Parallel tempering.} Parallel tempering (or replica exchange Monte
Carlo) improves upon the standard Metropolis-Hastings algorithm with reduced
critical slowing down near phase transitions. For a given disorder realization,
we maintain $R$ replicas (typically $10-100$) of the system at inverse
temperatures $\beta_1=1,\ldots,\beta_R$. All replicas share the same quenched
bond disorder but evolve independently via single-spin Metropolis updates. After
each sweep of the system (order $N$ updates, where $N$ is the total number of
spins), we attempt replica exchanges between replicas of adjacent temperatures
$i,i+1$ using the Metropolis criterion 
\begin{equation}
	\mathbb P(\text{accept}) = \min \left\{1, e^{(\beta_{i+1}-\beta_i)(\mathcal
	E_{i+1}-\mathcal E_i)} \right\}
\end{equation}
where $\mathcal E_i$ is the energy of replica $i$. This maintains detailed
balance of the simulation while enabling configurations to diffuse
through temperature space. Colder replicas borrow entropy from hotter
replicas and more readily escape local energy minima.

The free energy $F_{C_\textsf{s}L}$ is slow to compute due to the exponential
function being dominated by high-energy but improbable configurations. We
instead use the multi-step Bennett acceptance ratio \cite{bennettEfficient1976,
voterMonte1985}: to estimate $F_{C_\textsf{s}L} - F_{C_\textsf{s}}$, we
decompose $L$ into $M$ ``components'' $Q_1,Q_2,\ldots,Q_M$, where each $Q_\mu
\in \mathcal P_\text{st}$ is a (small) Pauli operator (e.g., a single-qubit
Pauli). They satisfy
\begin{equation}
	L_m = \prod_{\mu=1}^m Q_\mu, \quad L_0 = I, \quad L_M = L.
\end{equation}
Then, we reduce the free-energy estimation to a product of
smaller ratios
\begin{equation}
	F_{C_\textsf{s}L} - F_{C_\textsf{s}} = -\ln \frac{\mathcal
	Z_{C_\textsf{s}L}}{\mathcal Z_{C_{\textsf s}}} = - \ln \left[ \prod_{m=1}^M \frac{\mathcal Z_{C_\textsf{s} L_m}}{\mathcal Z_{C_\textsf{s} L_{m-1}}} \right].
\end{equation}
We simulate $(M+1)$ $\beta$-ladders of $R$ replicas each with quenched disorder
given by $C_{\textsf s},\, C_{\textsf s} L_1, \,\ldots, C_{\textsf s} L$ and
compute
\begin{equation}
	\frac{\mathcal Z_{C_{\textsf s}L_m}}{\mathcal Z_{C_{\textsf s}L_{m-1}}} =
	\frac{\braket{f({C_{\textsf s}L_{m}}; C_{\textsf
			s}L_{m-1})}_\beta} {\braket{f({C_{\textsf s}L_{m-1}}; C_{\textsf
		s}L_{m})}_\beta}, \quad f(E;E') = \min\left\{1, e^{-\Delta \mathcal
E(E;E')}\right\}
\end{equation}
where $\Delta \mathcal E(E;E')$ is the energy difference acquired by placing the
spins from the replica evolving under quenched disorder $E'$ (at temperature
$\beta_1$) onto the Hamiltonian with quenched disorder $E$ (unlike replica
exchange, this swap does not follow through; we record only the energy
difference if the swap were to happen).

\paragraph{Population annealing.} An alternative to parallel tempering,
population annealing maintains a population of $R$ replicas (typically $10^2$ to
$10^4$) all at the same temperature and same quenched disorder. The temperature
is gradually decreased from $\beta_\text{min}$ to $1$ via $n_\text{steps}$
intermediate values. At each
temperature step from $\beta_k$ to $\beta_{k+1}$, every replica $r$ is assigned
a weight 
\begin{equation}
	\tau_r(k) = \exp \left[-(\beta_{k+1}-\beta_k)\mathcal E_r\right]
\end{equation}
and the population is resampled (with replacement) according to normalized
weights $\tilde\tau_r$. The new population is thermalized over several
Metropolis updates at the new temperature. The free energy for a given quenched
disorder is then directly computed from the cumulative weights
\begin{equation}
	F_E = - \sum_{k=1}^{n_\text{steps}} \ln \left[\frac1R \sum_{r=1}^R
	\tau_r(k)\right].
\end{equation}
Population annealing naturally amplifies low-energy configurations while culling
high-energy configurations, making it particularly effective for disordered
systems with rough energy landscapes, such as lattice gauge models.

\subsection{Additional simulation details}
\label{app:additional_details}

Parallel tempering simulations were typically executed with $1000-2000$
thermalization steps, $5000$ to $20\,000$ sweeps (over which the Bennett
acceptance ratios were computed), and $10\,000$ to $50\,000$ quenched disorder
realizations. $10-20$ temperature steps were used for the replicas. 
Population annealing typically used $200-1000$ thermalization steps, $100-800$
population size, and $5 000$ to $20\,000$ disorder realizations. 

Errors were computed using jackknifing: for a nonlinear estimator $\hat{\theta}$
computed from $n$ samples, we compute leave-one-out estimates $\hat{\theta}_{-i}$
by excluding each sample in turn. The jackknife variance is $\sigma^2 =
\frac{n-1}{n}\sum_i (\hat{\theta}_{-i} - \bar{\theta})^2$. Error bars display the
estimated $95\%$ confidence intervals, indicating $\pm 2\sigma$.

Finally, intersection points of logical error rates (for thresholds) were
computed by taking sample data within the vicinity of the observed crossing
point. Weighted least-squares regression (weighted by the sample errors)
approximates each distance curve as a straight line, $y = m_i x + b_i$. The
intersection point $(x_c,y_c)$ can be found by the overdetermined equation
\begin{equation}
	\mathfrak E [\bm p] = \begin{pmatrix} \bm m & \bm{-1}\end{pmatrix} \bm p
	+\bm b= \bm 0, \quad \bm p = 
	\begin{pmatrix} x \\ y\end{pmatrix}
\end{equation}
and using least squares to minimize residuals: $\hat{\bm p}_c =
\operatorname{arg min}_{\bm p} ||\mathfrak E[\bm p]||^2$.
Error bars are estimated using bootstrapping, varying each sample data point
using the jackknife variance and then recalculating the estimator.

\end{document}